\documentclass[twocolumn]{aastex63}

\newcommand{\sgr}{\mbox{SGR\,J1935+2154~}}

\newcommand{\tbb}{\mbox{$T_{\rm bb}$~}}
\usepackage{rotating}
\usepackage{bm}
\usepackage{xcolor}
\usepackage{todonotes}
\usepackage{soul}
\usepackage{multirow}

\shortauthors{Yang et al.}
\begin{document}

\title{Bursts before Burst: A Comparative Study on FRB 200428-associated and FRB-absent X-Ray Bursts from SGR J1935+2154}
\correspondingauthor{Bin-Bin Zhang}
\email{bbzhang@nju.edu.cn}

\author[0000-0003-0691-6688]{Yu-Han Yang}
\affiliation{School of Astronomy and Space Science, Nanjing University, Nanjing 210093, China}
\affiliation{Key Laboratory of Modern Astronomy and Astrophysics (Nanjing University), Ministry of Education, China}

\author[0000-0003-4111-5958]{Bin-Bin Zhang}
\affiliation{School of Astronomy and Space Science, Nanjing University, Nanjing 210093, China}
\affiliation{Key Laboratory of Modern Astronomy and Astrophysics (Nanjing University), Ministry of Education, China}
\affiliation{Department of Physics and Astronomy, University of Nevada Las Vegas, NV 89154, USA}

\author[0000-0002-0633-5325]{Lin Lin}
\affiliation{Department of Astronomy, Beijing Normal University, Beijing 100875, China}

\author[0000-0002-9725-2524]{Bing Zhang}
\affiliation{Department of Physics and Astronomy, University of Nevada Las Vegas, NV 89154, USA}

\author[0000-0001-6545-4802]{Guo-Qiang Zhang}
\affiliation{School of Astronomy and Space Science, Nanjing University, Nanjing 210093, China}
\affiliation{Key Laboratory of Modern Astronomy and Astrophysics (Nanjing University), Ministry of Education, China}

\author[0000-0002-7555-0790]{Yi-Si Yang}
\affiliation{School of Astronomy and Space Science, Nanjing University, Nanjing 210093, China}
\affiliation{Key Laboratory of Modern Astronomy and Astrophysics (Nanjing University), Ministry of Education, China}

\author[0000-0001-6606-4347]{Zuo-Lin Tu}
\affiliation{School of Astronomy and Space Science, Nanjing University, Nanjing 210093, China}
\affiliation{Key Laboratory of Modern Astronomy and Astrophysics (Nanjing University), Ministry of Education, China}

\author{Jin-Hang Zou}
\affiliation{College of Physics, Hebei Normal University, Shijiazhuang 050024, China}

\author{Hao-Yang Ye}
\affiliation{School of Astronomy and Space Science, Nanjing University, Nanjing 210093, China}

\author{Fa-Yin Wang}
\affiliation{School of Astronomy and Space Science, Nanjing University, Nanjing 210093, China}
\affiliation{Key Laboratory of Modern Astronomy and Astrophysics (Nanjing University), Ministry of Education, China}

\author{Zi-Gao Dai}
\affiliation{School of Astronomy and Space Science, Nanjing University, Nanjing 210093, China}
\affiliation{Key Laboratory of Modern Astronomy and Astrophysics (Nanjing University), Ministry of Education, China}

\begin{abstract}

 Accompanied by an X-ray burst, the fast radio burst (FRB) FRB 200428 was recently confirmed as originating from the Galactic magnetar soft gamma repeater (SGR) SGR J1935+2154. Just before and after FRB 200428 was detected, the Five-hundred-meter Aperture Spherical radio Telescope (FAST) had been monitoring SGR J1935+2154 for eight hours. From UTC 2020 April 27 23:55:00 to 2020 April 28 00:50:37, FAST detected no pulsed radio  emission from SGR J1935+2154, while Fermi/Gamma-ray Burst Monitor registered 34 bursts in the X/soft $\gamma$-ray band, forming a unique sample of X-ray bursts in the absence of FRBs. After a comprehensive analysis on light curves, time-integrated, and time-resolved spectral properties of these FRB-absent X-ray bursts, we compare this sample with the FRB-associated X-ray burst detected by Insight-HXMT, INTEGRAL, and Konus-Wind. The FRB-associated burst distinguishes itself from other X-ray bursts by its nonthermal spectrum and a higher spectral peak energy, but otherwise is not atypical. We also compare the cumulative energy distribution of our X-ray burst sample with that of first repeating FRB source, FRB 121102, with the calibration of FRB 200428-X-ray burst association. We find a similarity between the two, offering indirect support of the magnetar origin of cosmological FRBs. The event rate density of magnetar bursts is about $\sim 150$ times higher than the FRB event rate density at the energy of FRB 200428. This again suggests that, if all FRBs originate from magnetars, only a small fraction of X-ray bursts are associated with FRBs.

\end{abstract}


\section{Introduction} \label{sec:intro}

Magnetars are observed with persistent X-ray emissions, conventional bursts, random outbursts, and sparse giant flares (GFs), consistent with being powered by the decay of their extreme magnetic fields \citep{1995Thompson,1996Thompson}. The soft gamma repeater (SGR) SGR J1935+2154 was first discovered through its short magnetar-like bursts by the Burst Alert Telescope (BAT) on board the Neil Gehrels Swift Observatory \citep[Swift;][]{2014Stamatikos}. Its magnetar nature has been confirmed by the follow-up observations, which revealed the spin period $P = 3.24$ s, spin-down rate $\dot{P}=1.43 \times 10^{-11}$ s s$^{-1}$, and an inferred surface dipole-magnetic field $B \sim 2.2 \times 10^{14}$ G \citep{2016Israel}. These follow-up observations revealed that this magnetar has had several active episodes, making it by far the most prolific magnetar \citep{2017Younes,2020Borghese,2020Lina,2020Linb}. The distance of SGR J1935+2154 is $4.4^{+2.8}_{-1.3}$ kpc, which is estimated from a bright expanding dust-scattering X-ray ring observed by the X-ray Telescope \citep{2020Mereghetti} on board Swift. Meanwhile, its associated supernova remnant (SNR) G57.2+0.8 was inferred to be at $D = 6.6 \pm 0.7$ kpc \citep{2020Zhou}. By analyzing the contributions of the dispersion measure, \citet{2020Zhong} obtained the distance of SGR J1935+2154, $D=9.0\pm2.5$ kpc.

Recently, the fast radio burst (FRB) FRB 200428 was reported from the direction of \sgr \citep{2020Bochenek,2020CHIME}. Interestingly, it is associated with an X-ray burst of SGR J1935+2154, which was detected by Insight-HXMT \citep{2020Li}, AGILE \citep{2020Tavani}, INTEGRAL \citep{2020Mereghetti}, and Konus-Wind \citep{2020Ridnaia}. The properties of the FRB-associated X-ray burst obtained with different telescopes are shown in Table \ref{table:FRB-XRB}. Compared to other FRBs from extragalactic origins, this unique Galactic FRB is roughly 25 times less energetic than the weakest of the FRB population \citep{2020Bochenek}. The FRB emission could be caused by various mechanisms \citep[see][for a recent review]{2020Zhang} such as a disturbance propagating from the magnetar crust to the magnetosphere \citep{2016Cordes,2016Katz,2020Katz,2017Kumar,2018Lu,2018Yang,2020Kumar,2020Lu,2020Lyubarsky,2020Wang,2020YangYP}, an interaction between an asteroid and the magnetosphere \citep{2020Dai,2020Geng}, as well as a model invoking a relativistic shock outside the magnetosphere\citep{2014Lyubarsky,2016Murase,2017Beloborodov,2020Beloborodov,2017Waxman,2019Metzger,2020Margalit,2020Wu,2020Yu}.

Before and after the detection of FRB 200428, four targeted observations with a total exposure of eight hours on SGR J1935+2154 were undertaken by the Five-hundred-meter Aperture Spherical radio Telescope (FAST) from UTC 2020 April 15 21:54:00 to 2020 April 28 23:35:00 \citep{2020Linc}. The FAST observation yielded no dispersed pulsed emission at all, which placed a stringent flux and fluence upper limit for the radio emission of SGR J1935+2154. During the FRB-absent period, a series of X-ray bursts were detected by Fermi/Gamma-ray Burst Monitor (GBM) about 14 hr before FRB 200428 occurred. Therefore, \citet{2020Linc} concluded that the association with an observable FRB of short magnetar X-ray bursts is rather rare. \citet{2020Younes} found 24 X-ray bursts simultaneously observed by NASA's NICER and Fermi/GBM about 13 hr prior to FRB 200428, which largely overlap with our sample (34 bursts) in this Letter. They claimed that these bursts are temporally similar to but spectrally different from the FRB-associated X-ray burst. The unusual hardness of the energy spectrum for the FRB-associated X-ray burst has been noticed since its discovery \citep{2020Mereghetti,2020Ridnaia}. This work aims to provide a comprehensive analysis of all the FRB-absent magnetar bursts observed by Fermi/GBM, with a focus on their detailed timing and spectral properties as well as similarities and differences with the FRB-associated burst.

In Section \ref{sec:analysis}, we describe the data reduction and burst search procedure followed by the results of temporal, time-integrated and time-resolved analyses of FRB-absent magnetar bursts.In Section \ref{sec:special}, we compare our FRB-absent sample to the FRB-associated burst in order to determine how unique the latter is. In Section \ref{sec:frb}, assuming that all FRBs are originated from magnetars, we compare our X-ray burst sample with the radio burst sample from the first repeating FRB source, FRB 121102, in terms of the cumulative energy distribution and event rate. In Section \ref{sec:sum}, we briefly summarize the results.

\section{Observation and Data Analysis}\label{sec:analysis}

\subsection{Bursts Identification} \label{sec:obs}

\begin{figure}
\includegraphics[angle=0,width=0.195\textwidth]{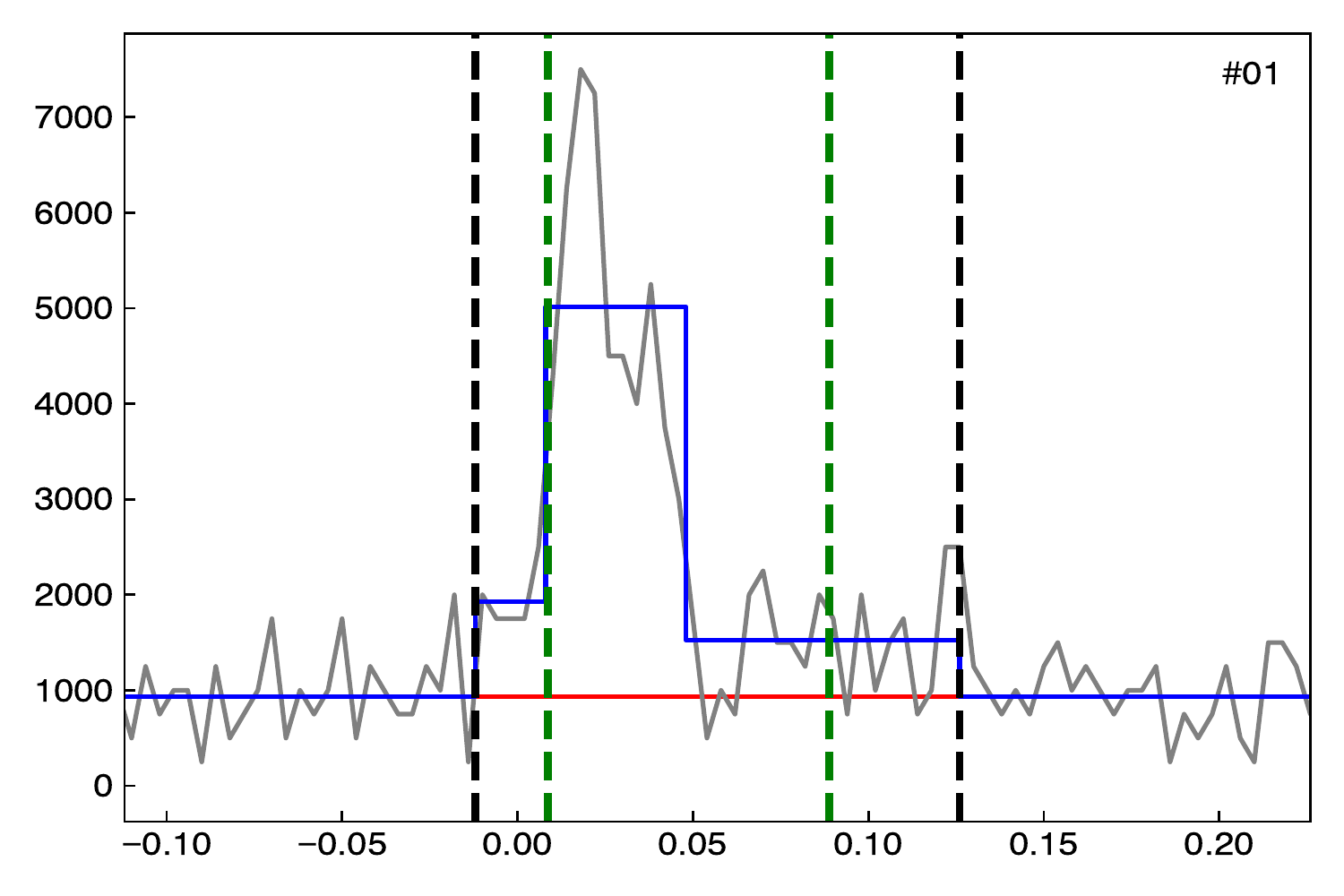}
\includegraphics[angle=0,width=0.195\textwidth]{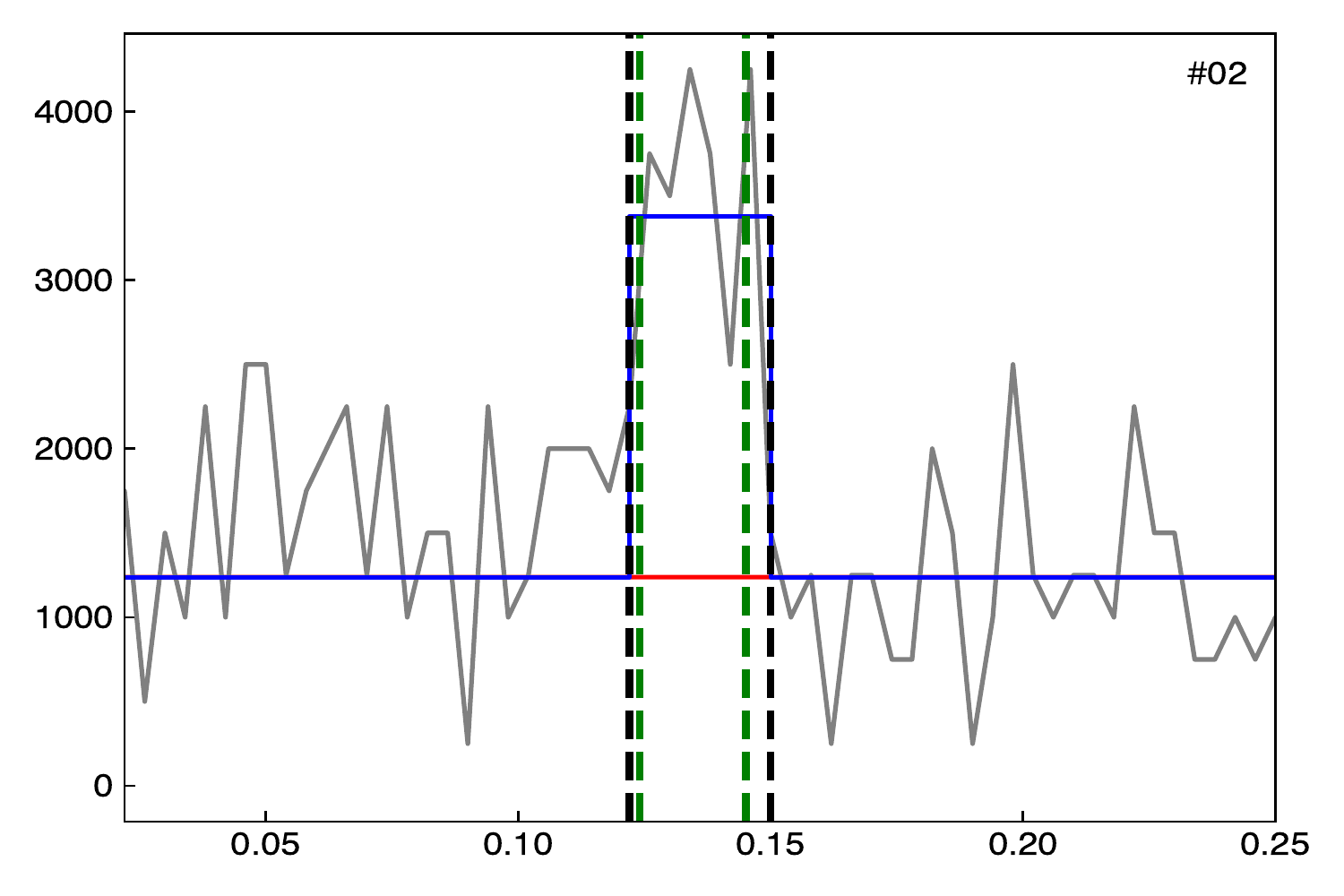}
\includegraphics[angle=0,width=0.195\textwidth]{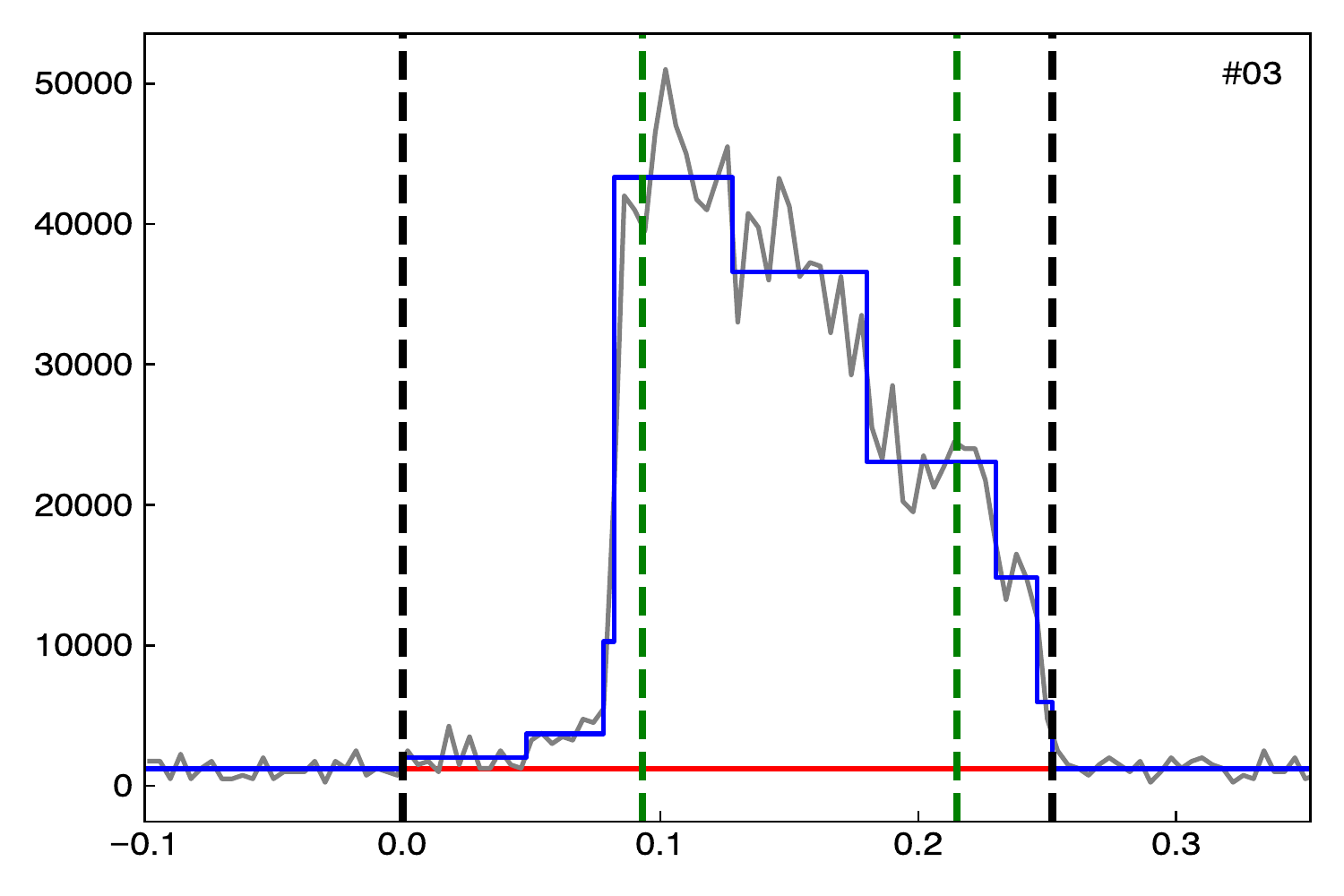}
\includegraphics[angle=0,width=0.195\textwidth]{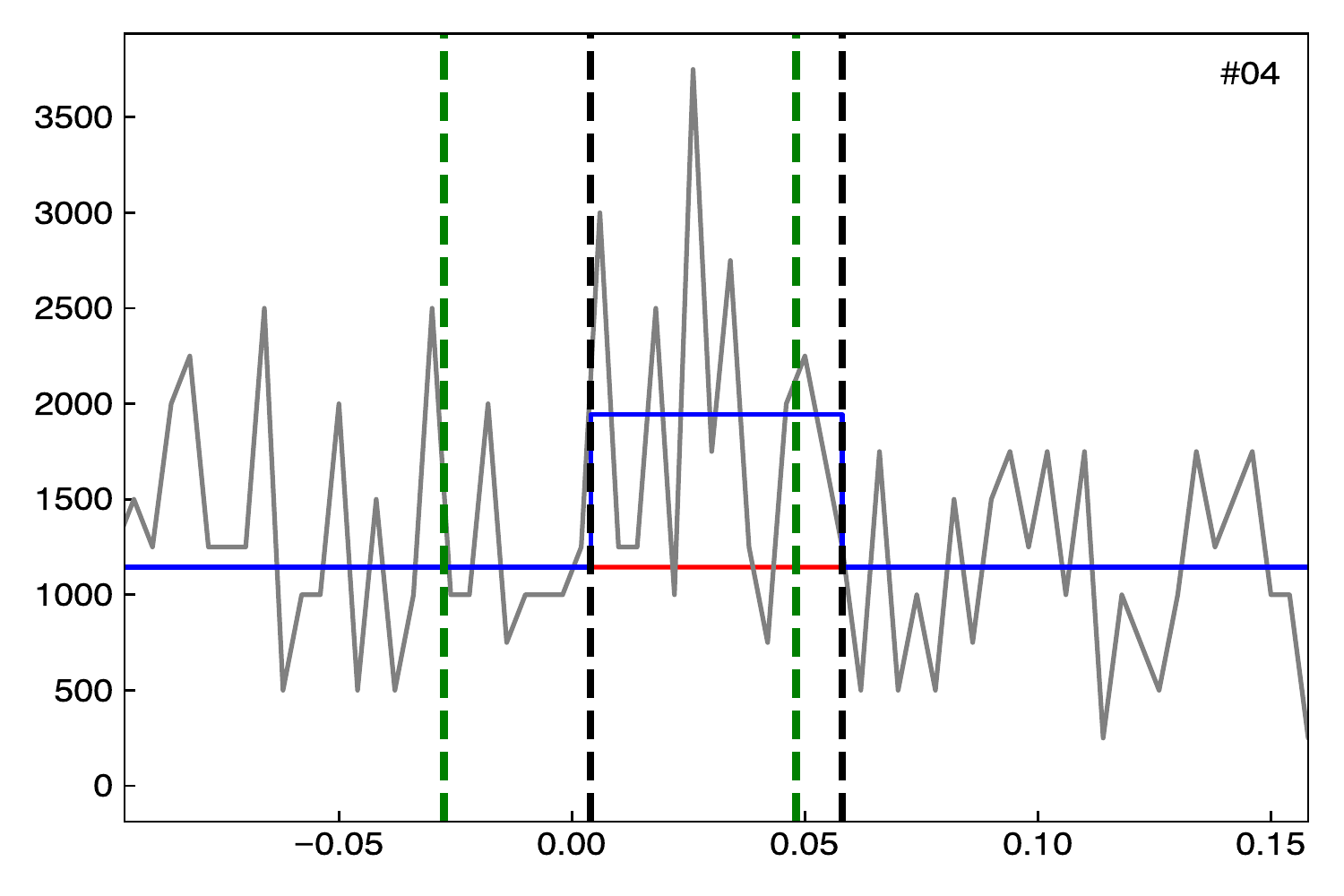}
\includegraphics[angle=0,width=0.195\textwidth]{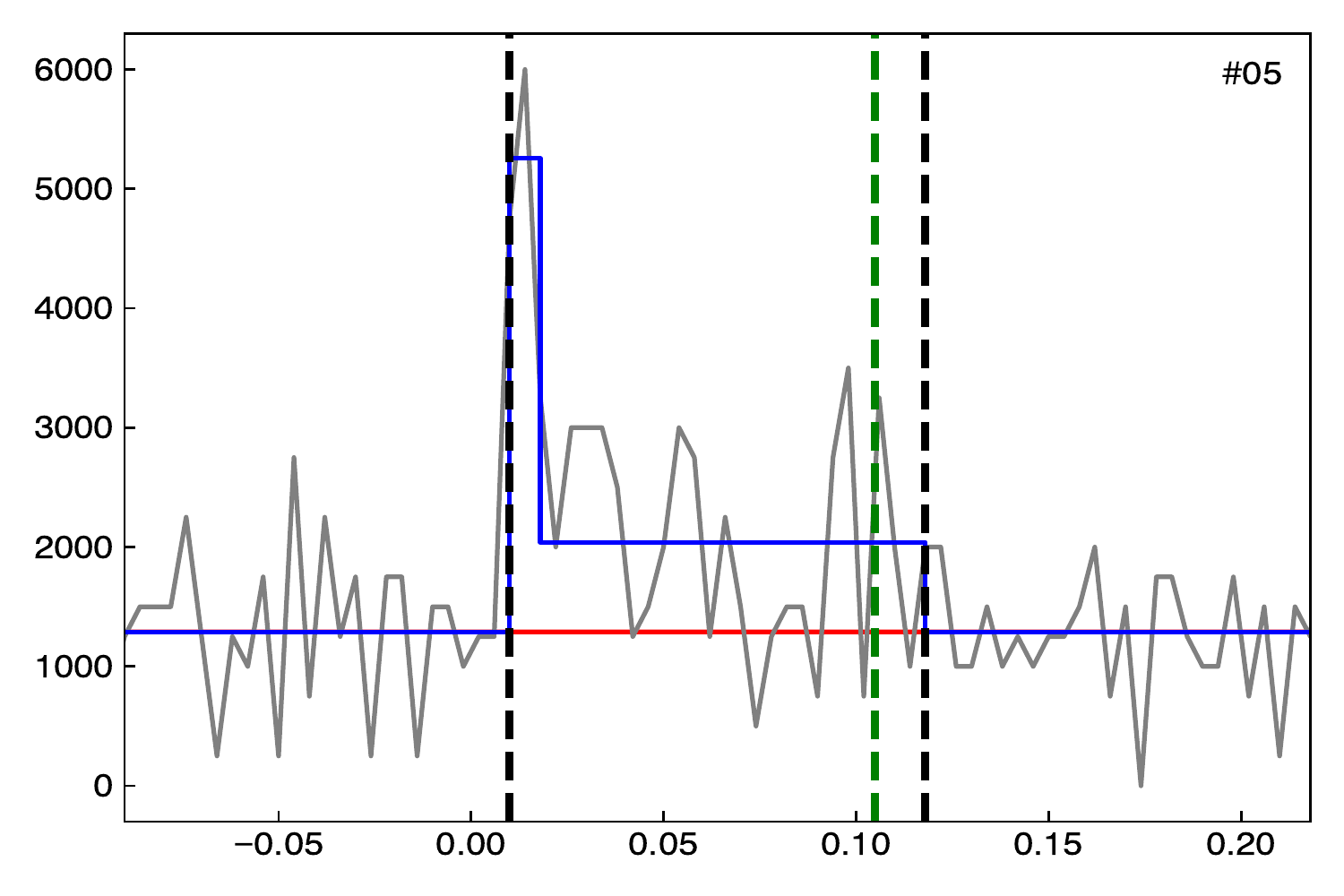}
\includegraphics[angle=0,width=0.195\textwidth]{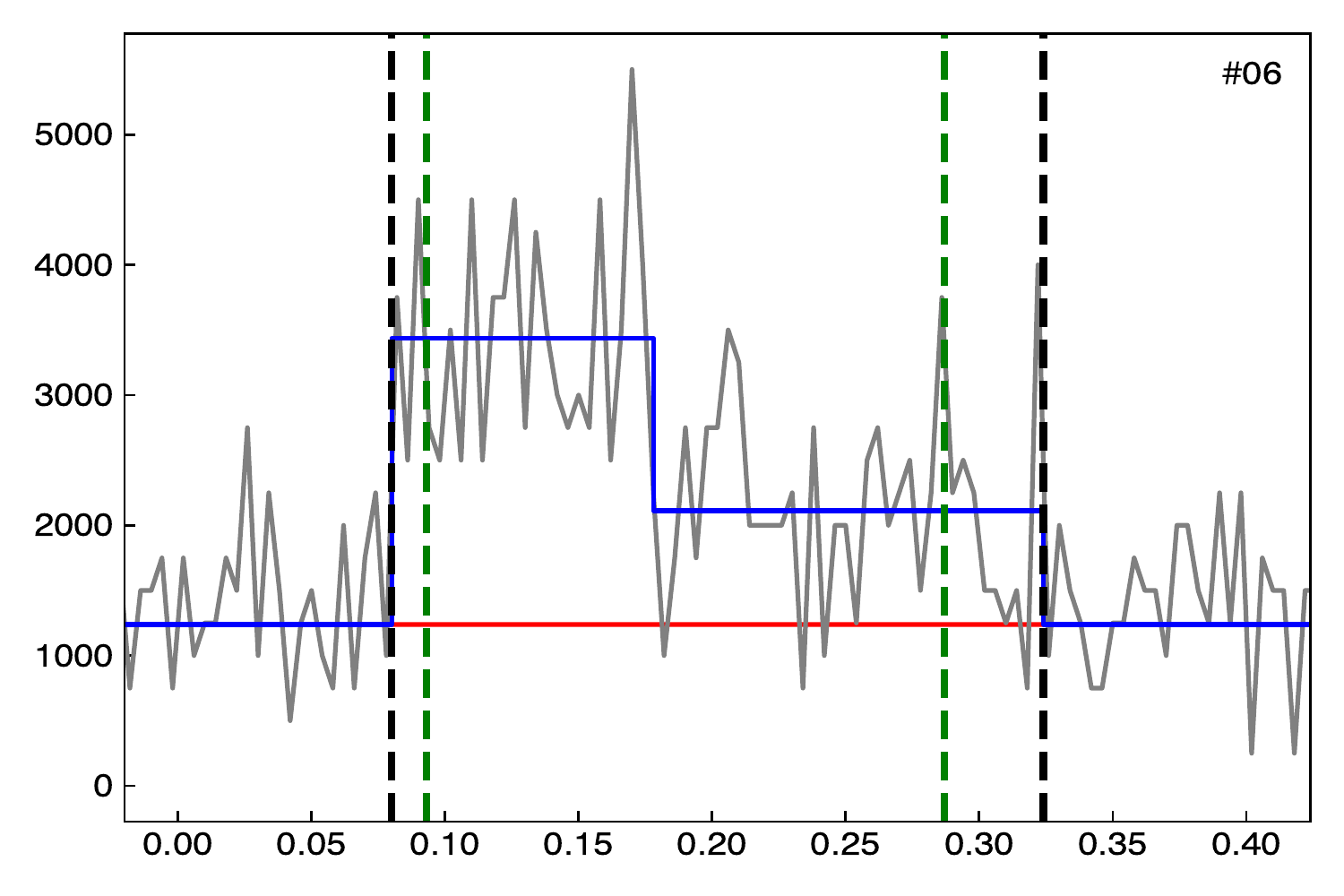}
\includegraphics[angle=0,width=0.195\textwidth]{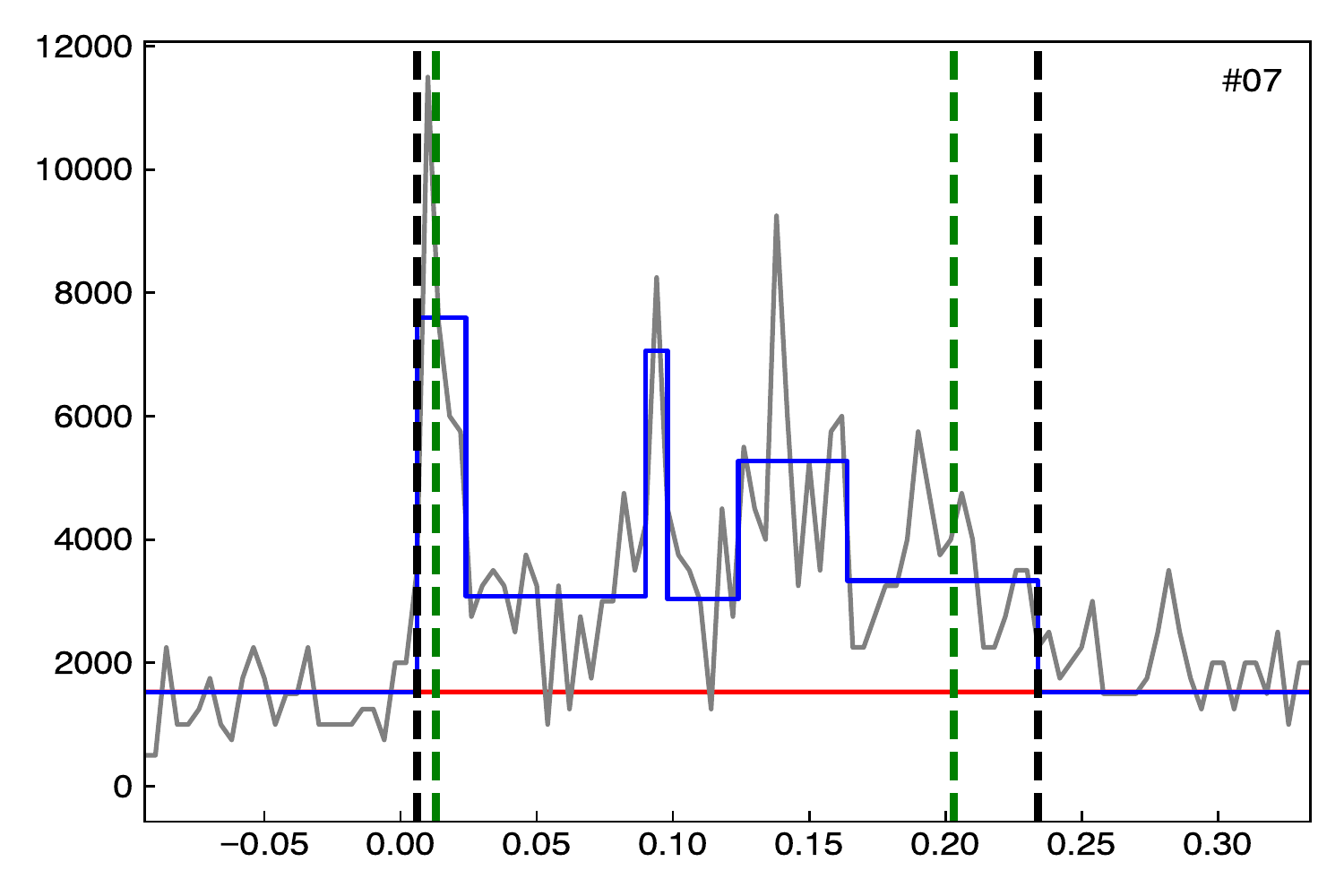}
\includegraphics[angle=0,width=0.195\textwidth]{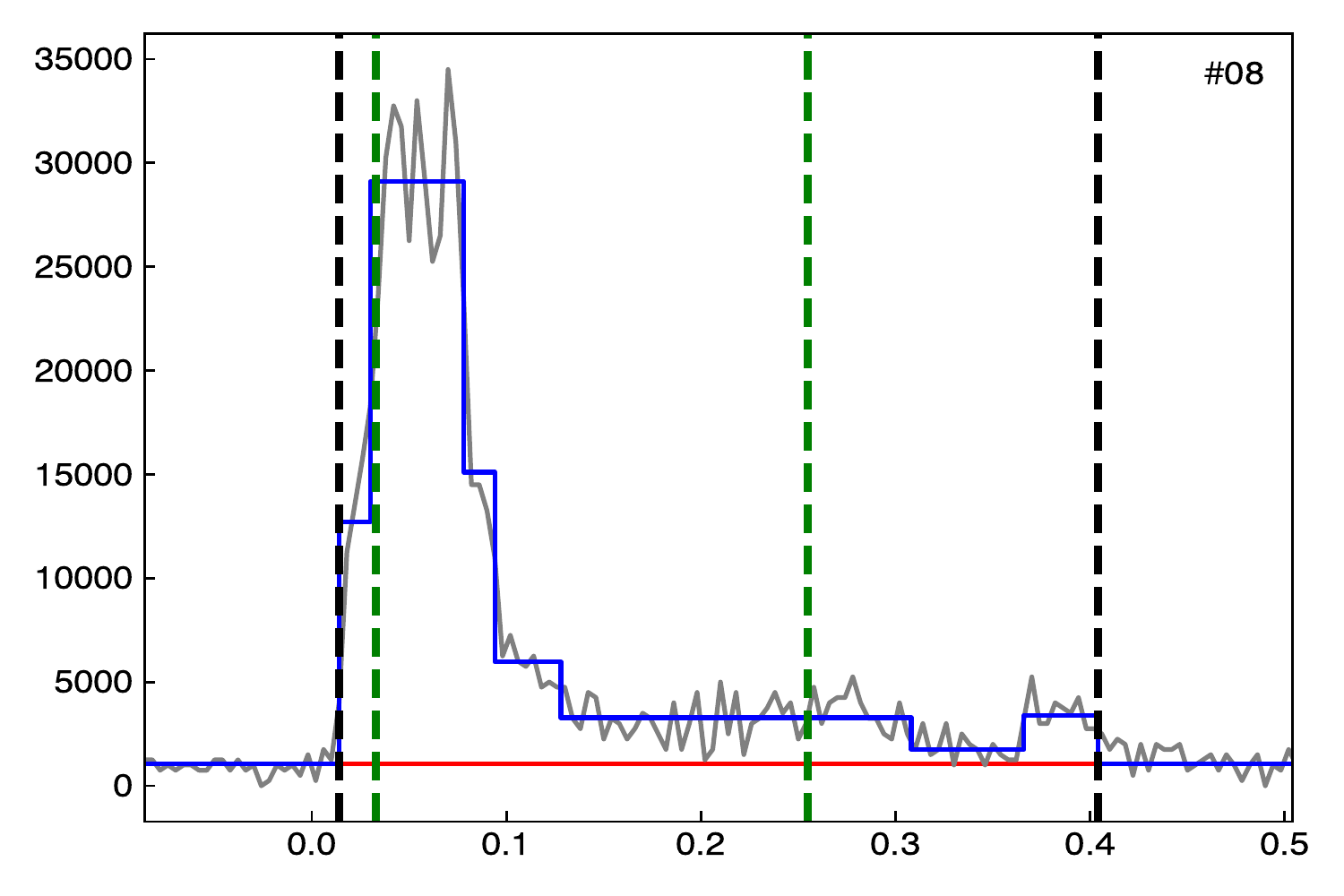}
\includegraphics[angle=0,width=0.195\textwidth]{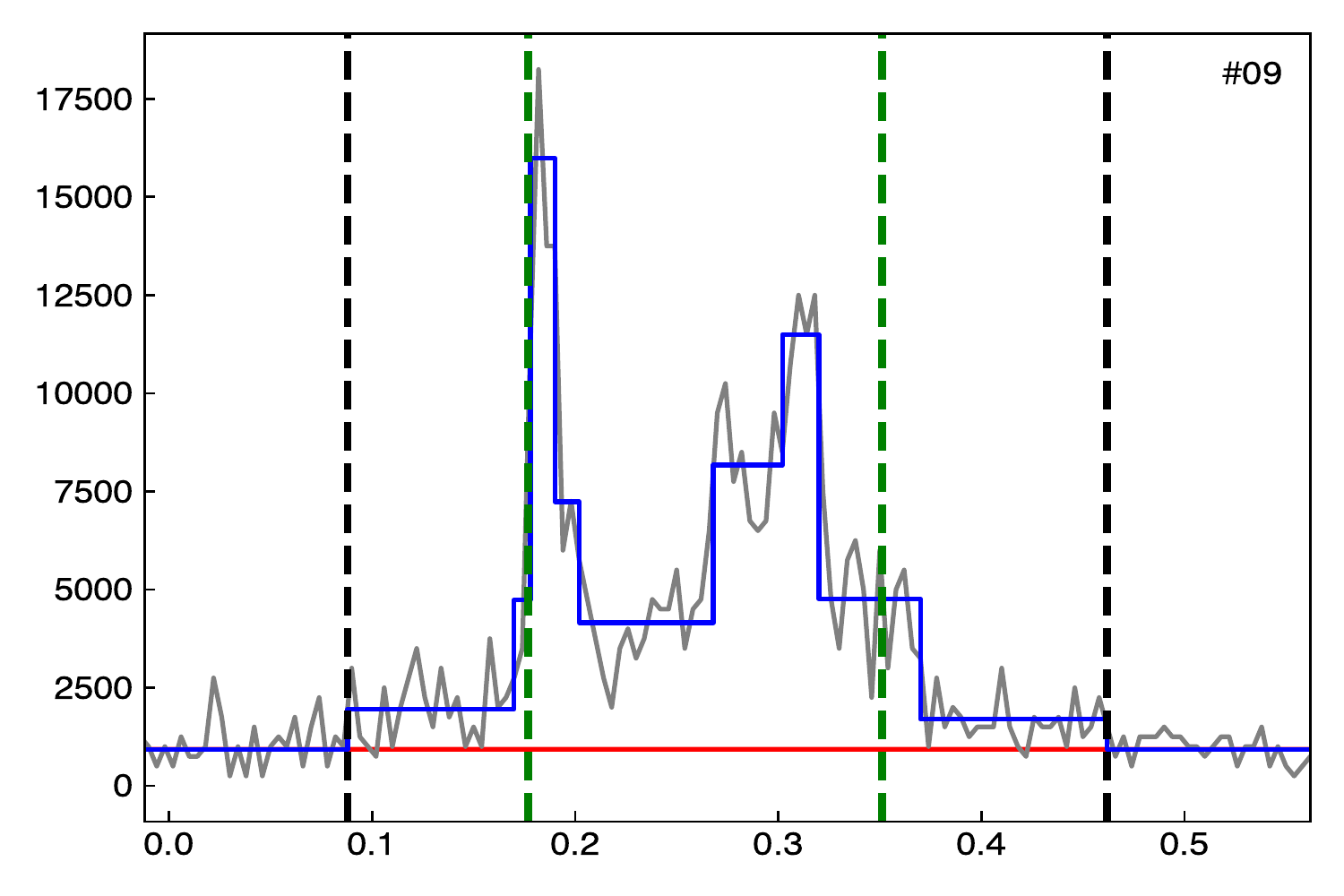}
\includegraphics[angle=0,width=0.195\textwidth]{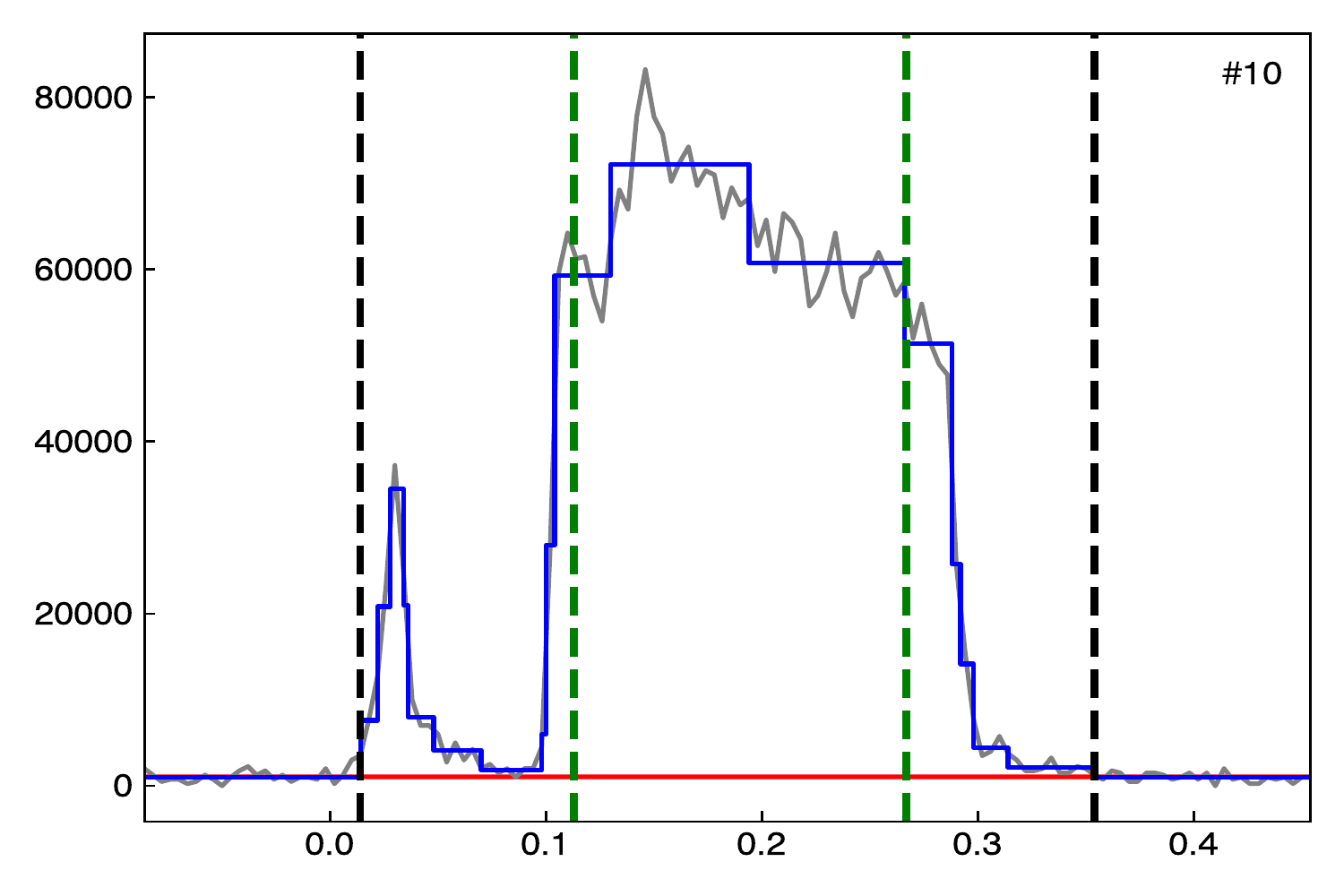}
\includegraphics[angle=0,width=0.195\textwidth]{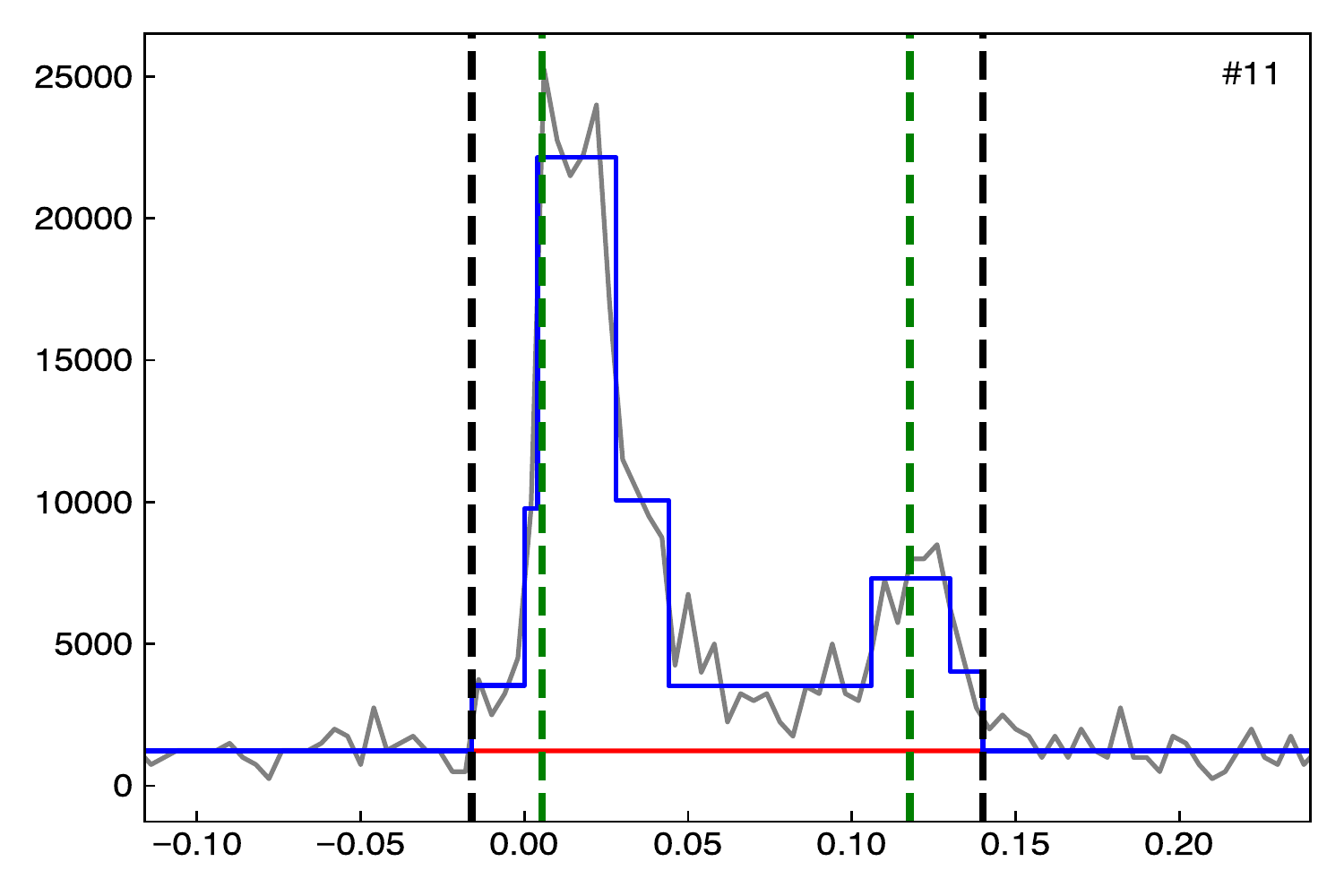}
\includegraphics[angle=0,width=0.195\textwidth]{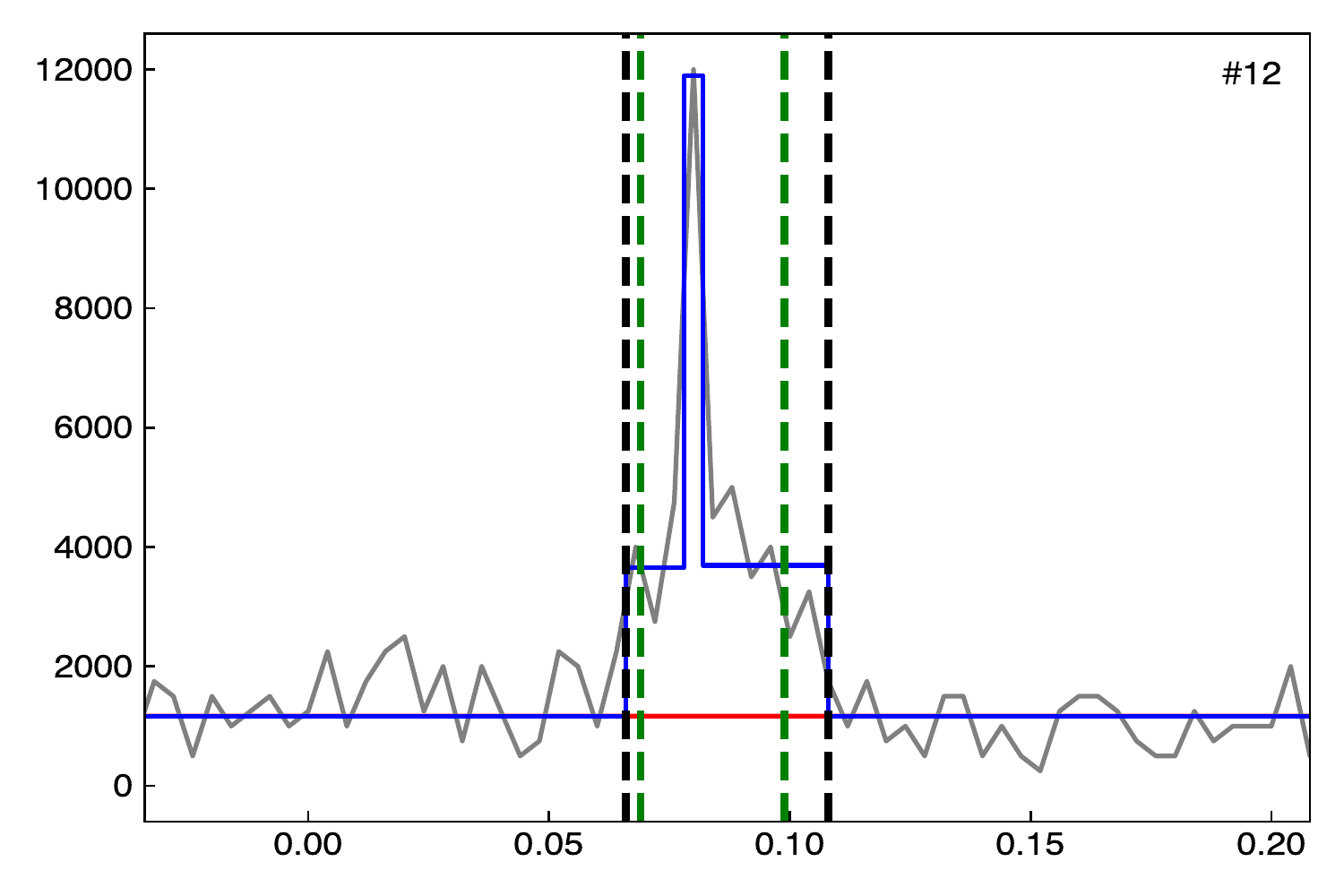}
\includegraphics[angle=0,width=0.195\textwidth]{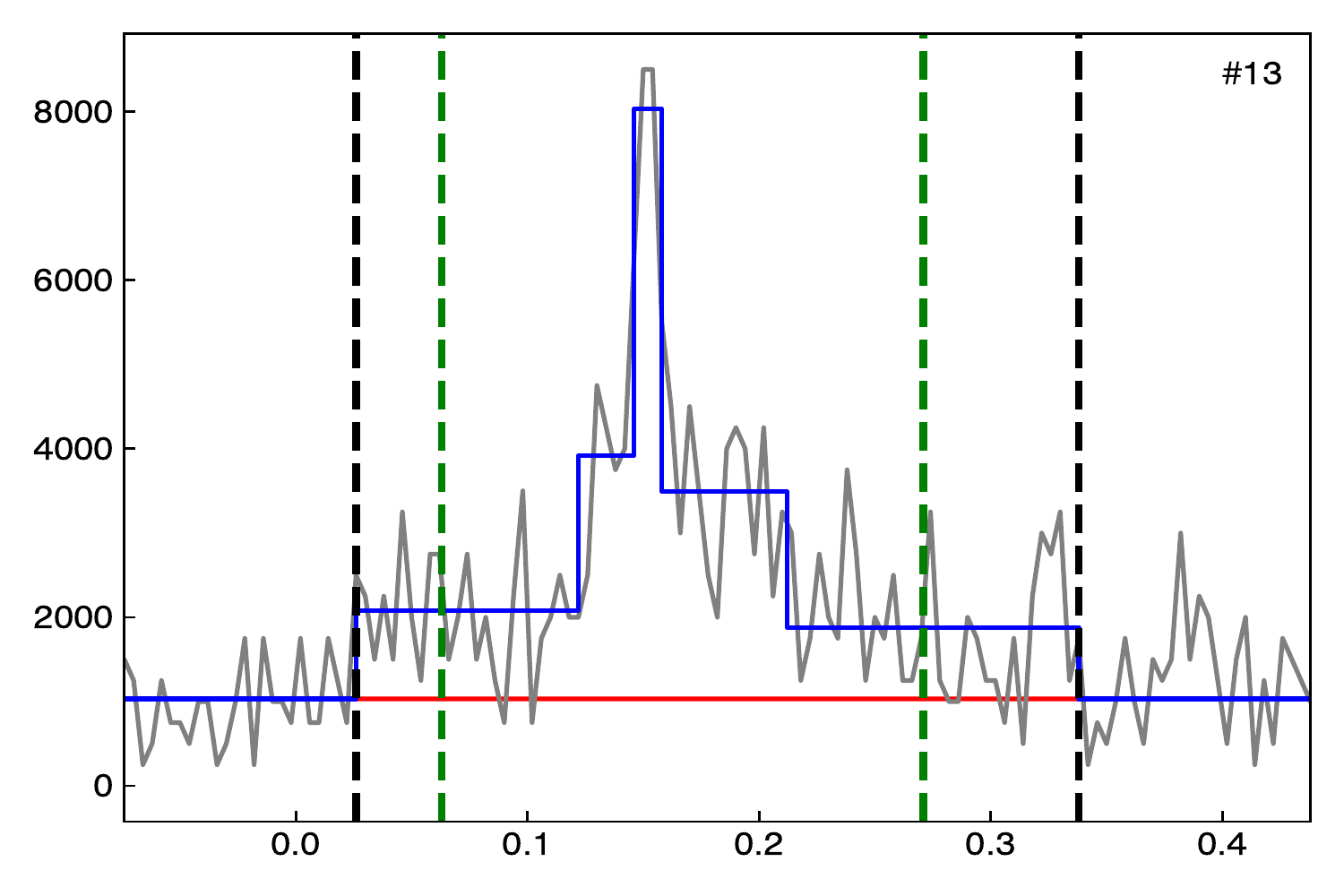}
\includegraphics[angle=0,width=0.195\textwidth]{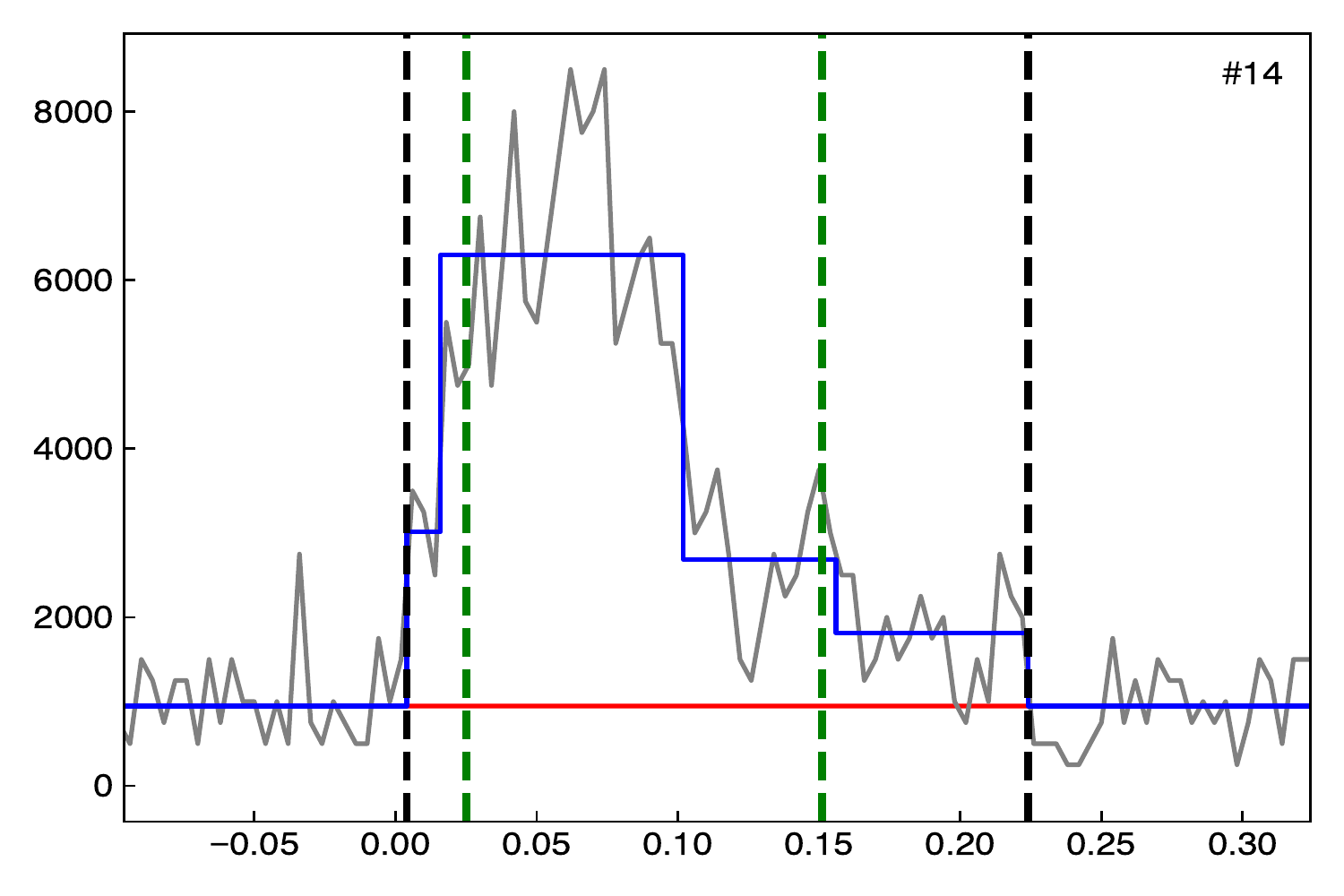}
\includegraphics[angle=0,width=0.195\textwidth]{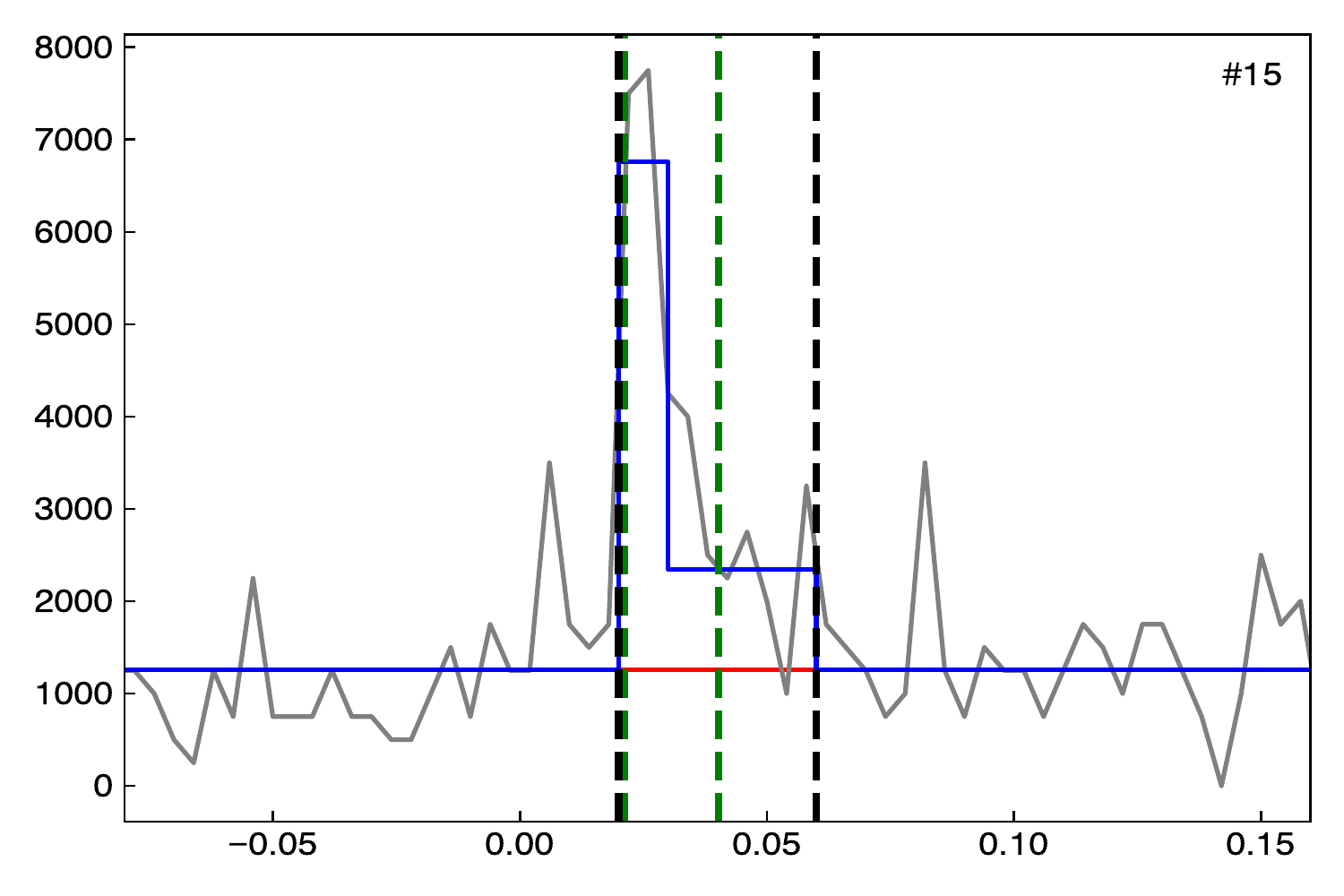}
\includegraphics[angle=0,width=0.195\textwidth]{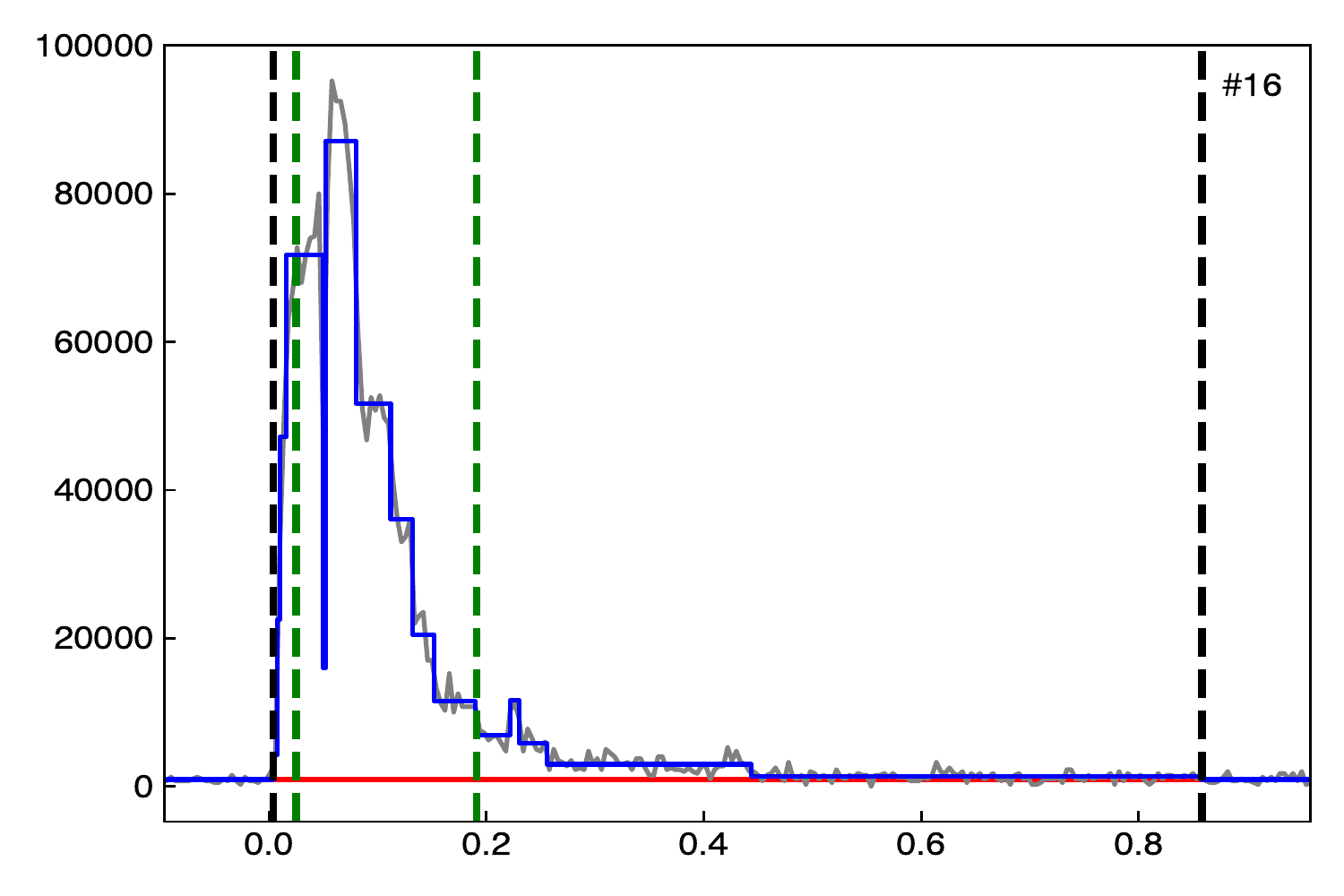}
\includegraphics[angle=0,width=0.195\textwidth]{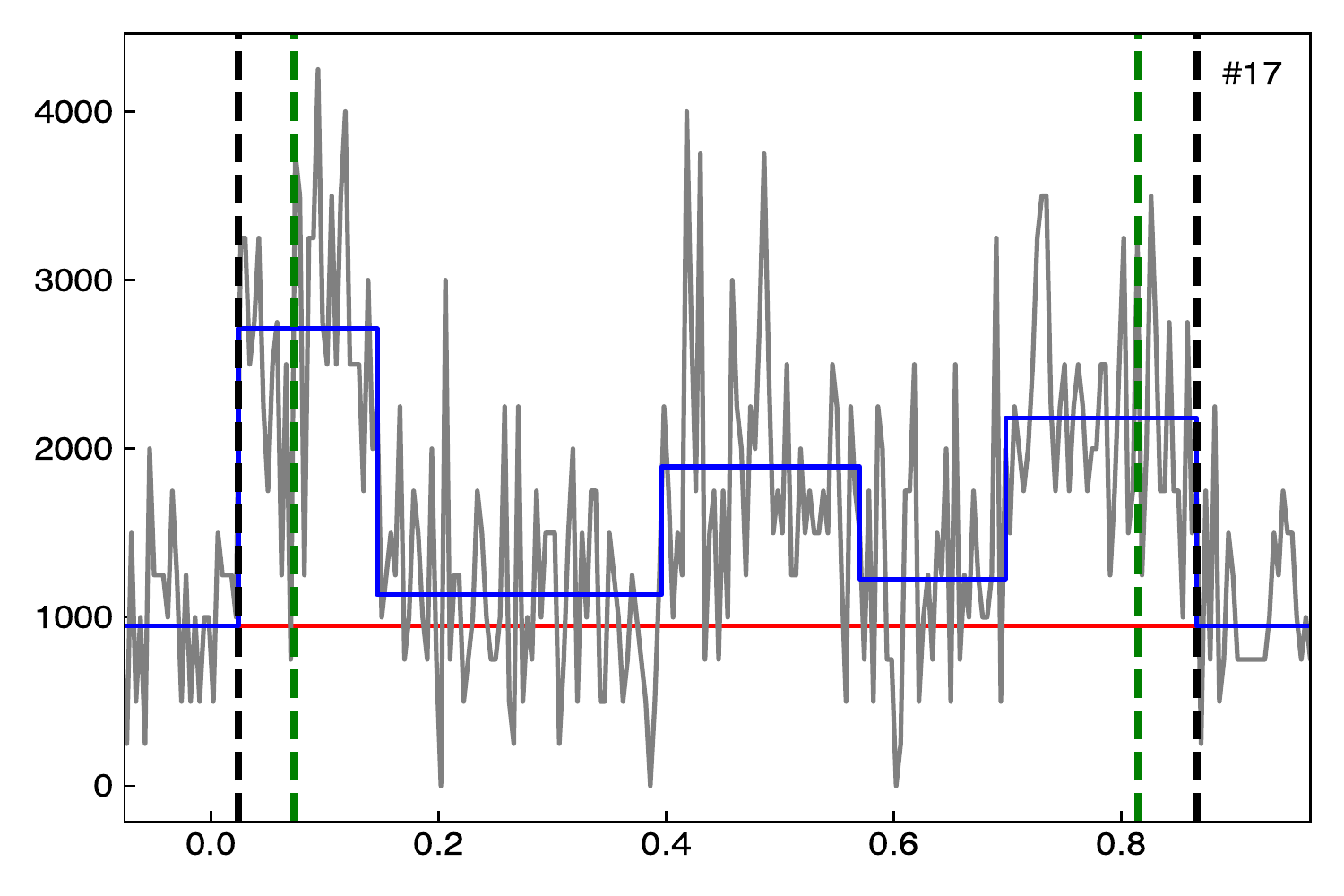}
\includegraphics[angle=0,width=0.195\textwidth]{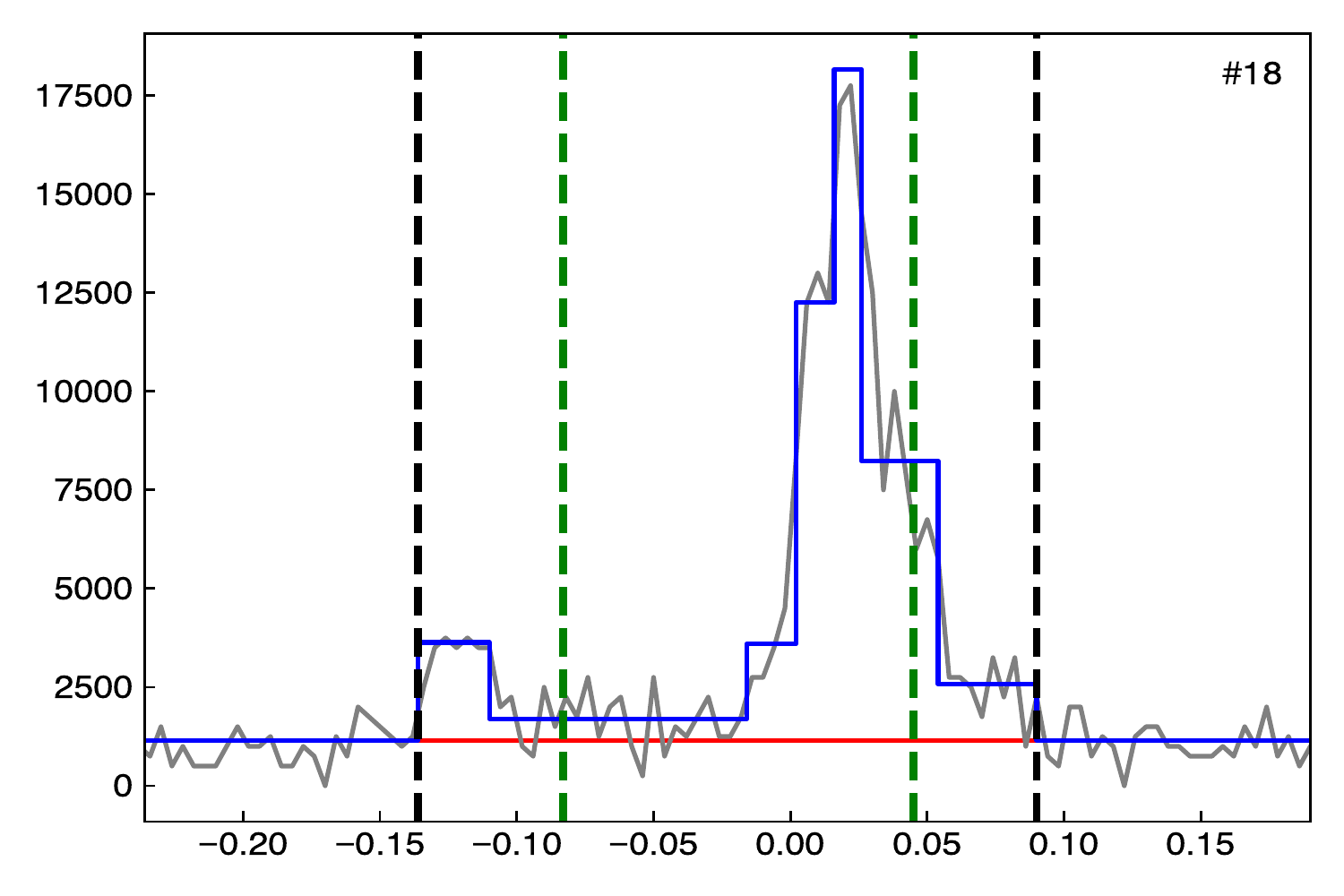}
\includegraphics[angle=0,width=0.195\textwidth]{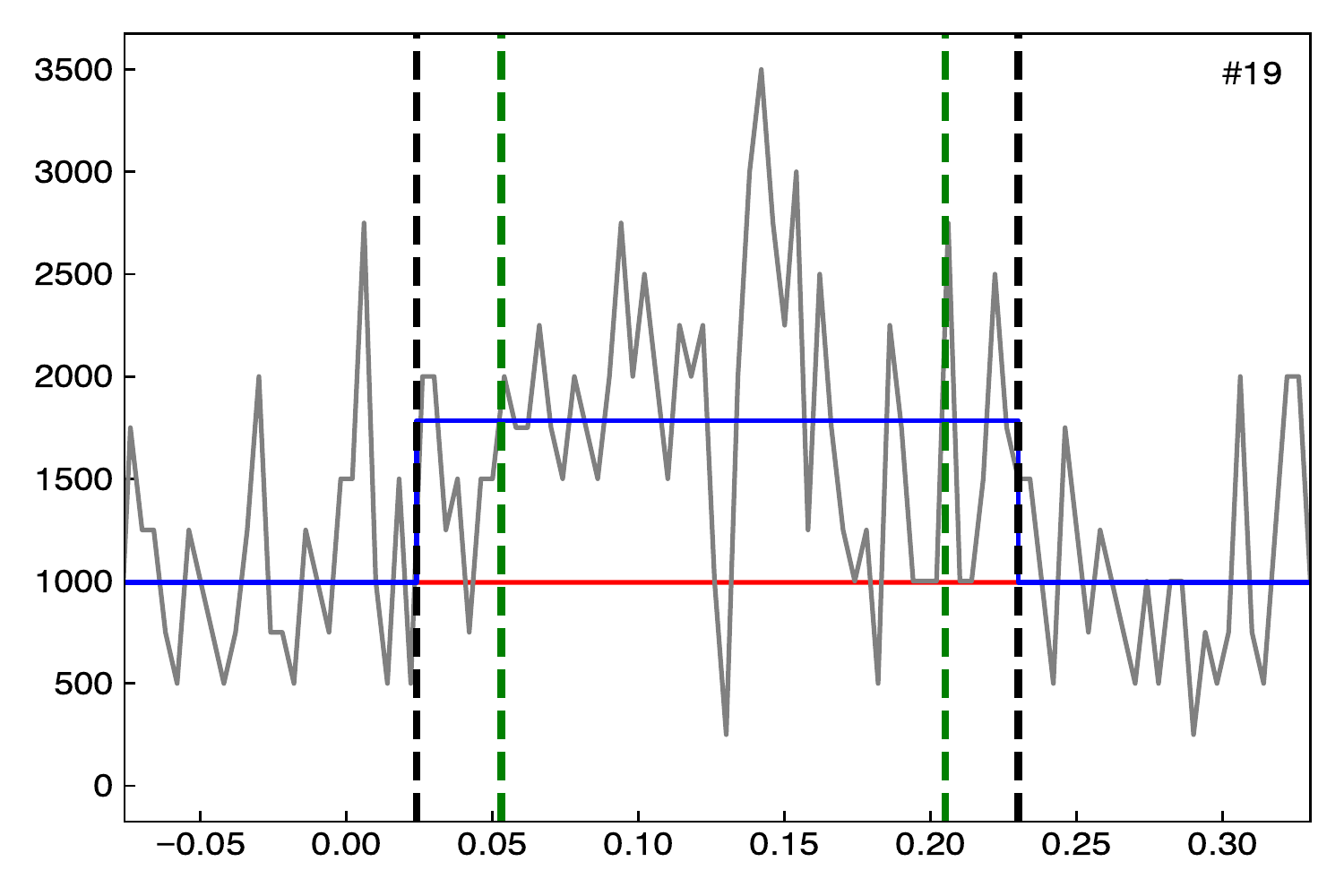}
\includegraphics[angle=0,width=0.195\textwidth]{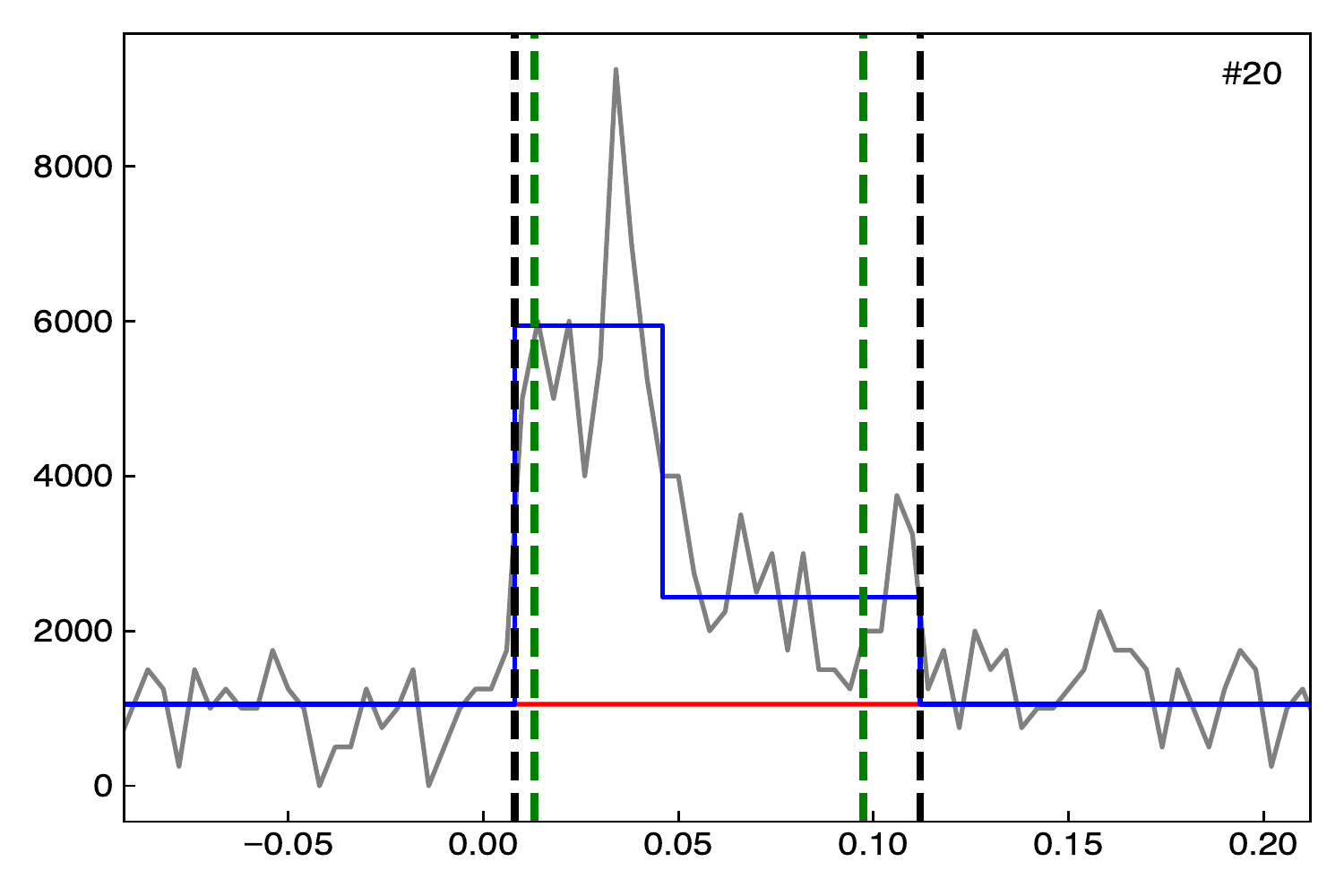}
\includegraphics[angle=0,width=0.195\textwidth]{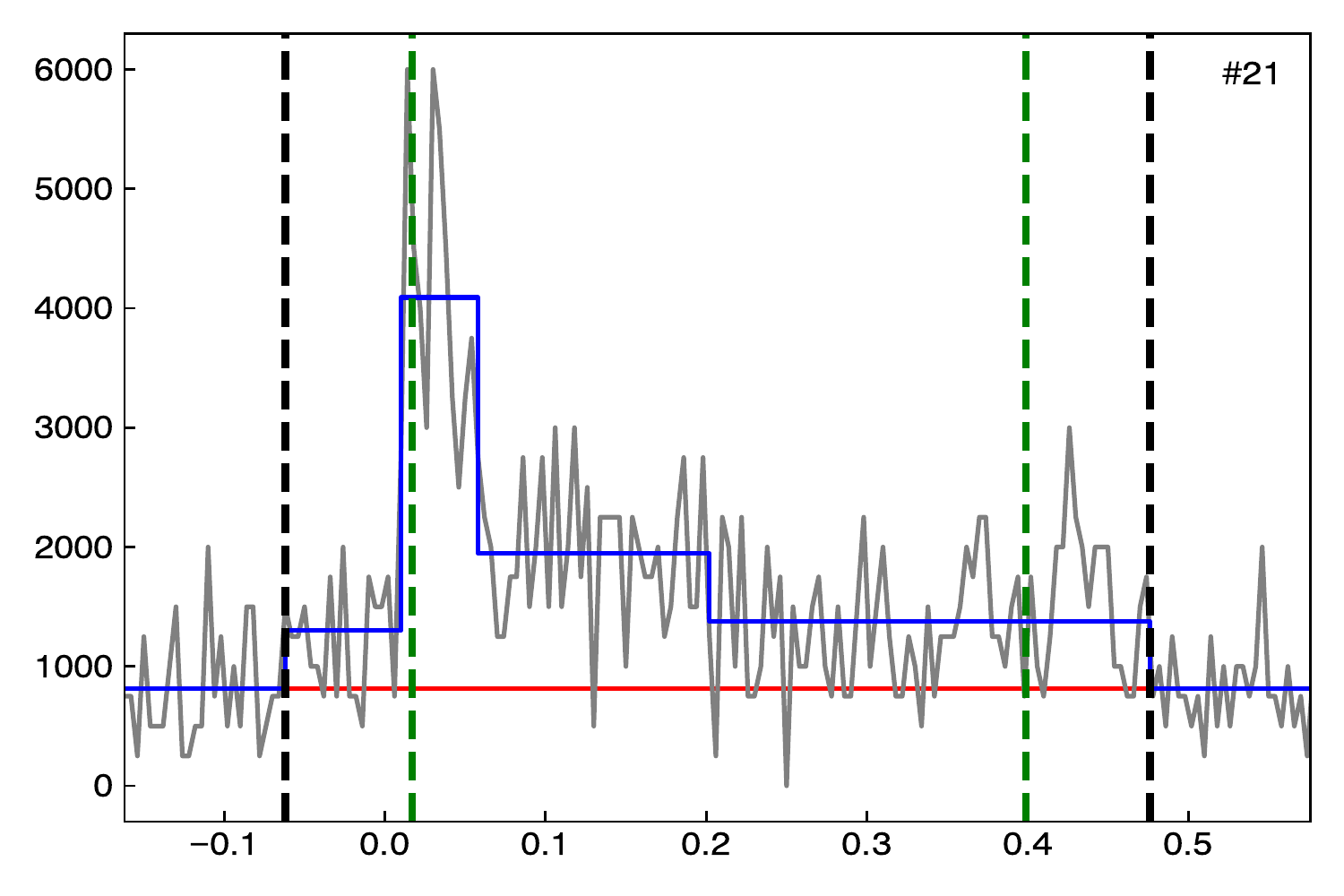}
\includegraphics[angle=0,width=0.195\textwidth]{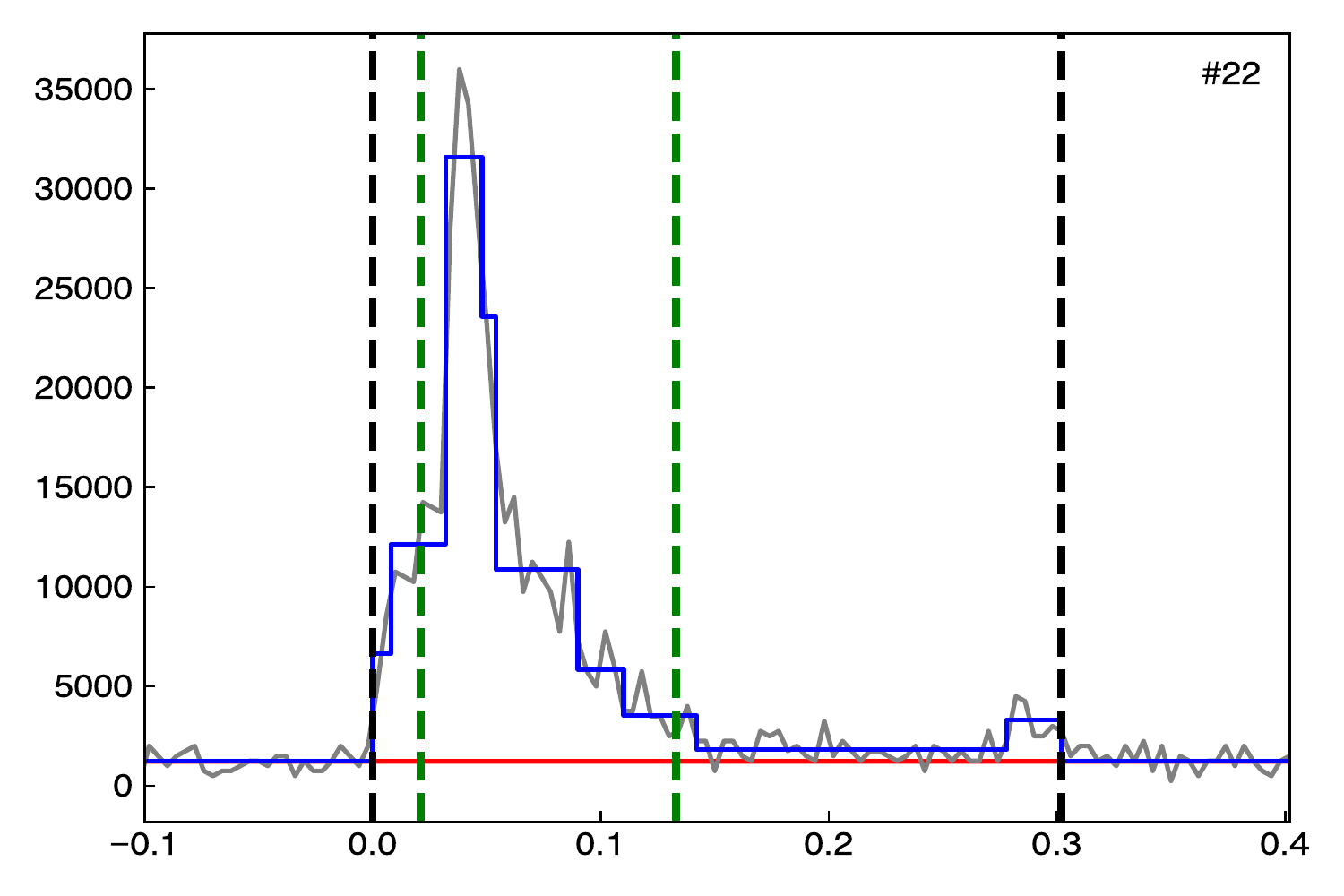}
\includegraphics[angle=0,width=0.195\textwidth]{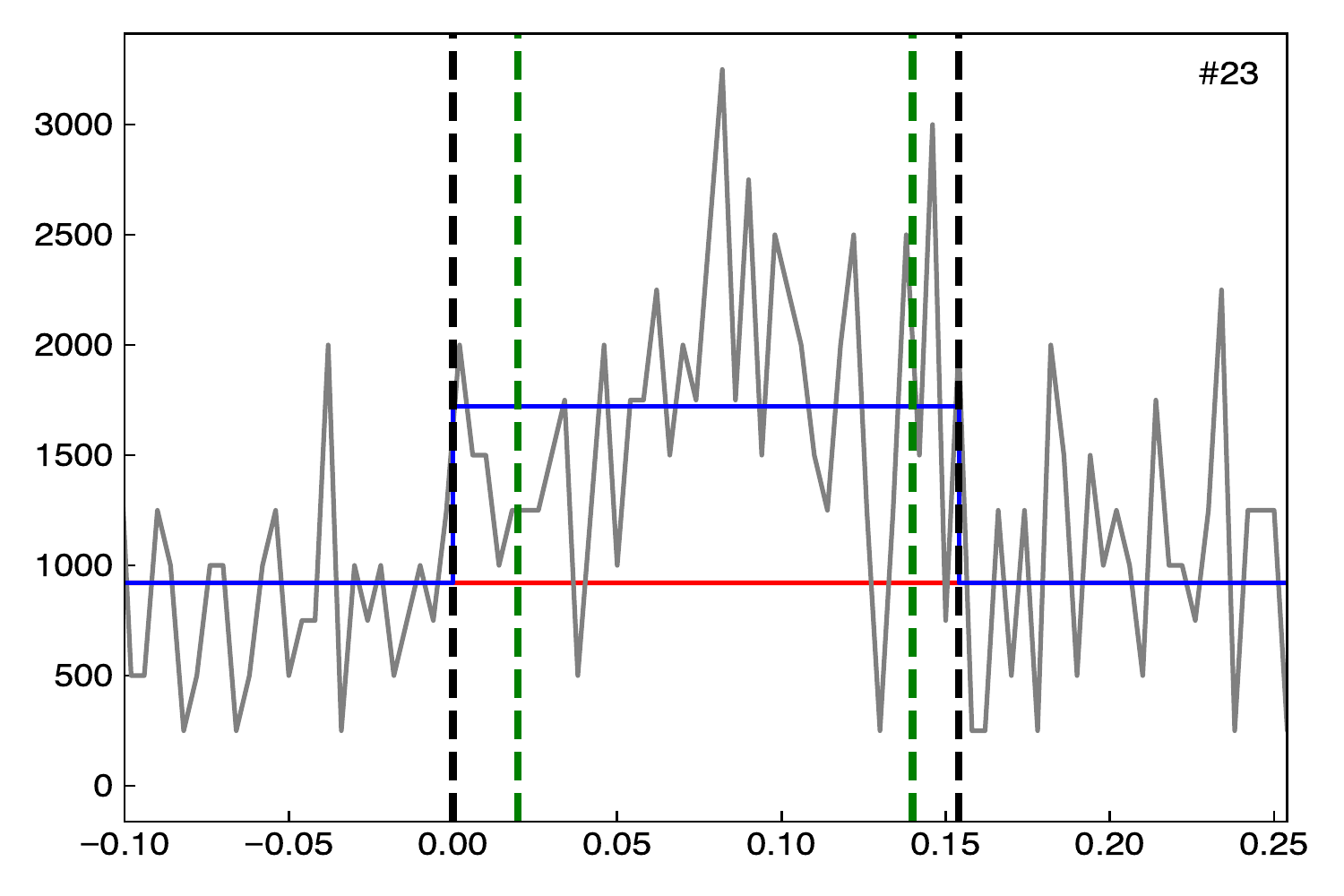}
\includegraphics[angle=0,width=0.195\textwidth]{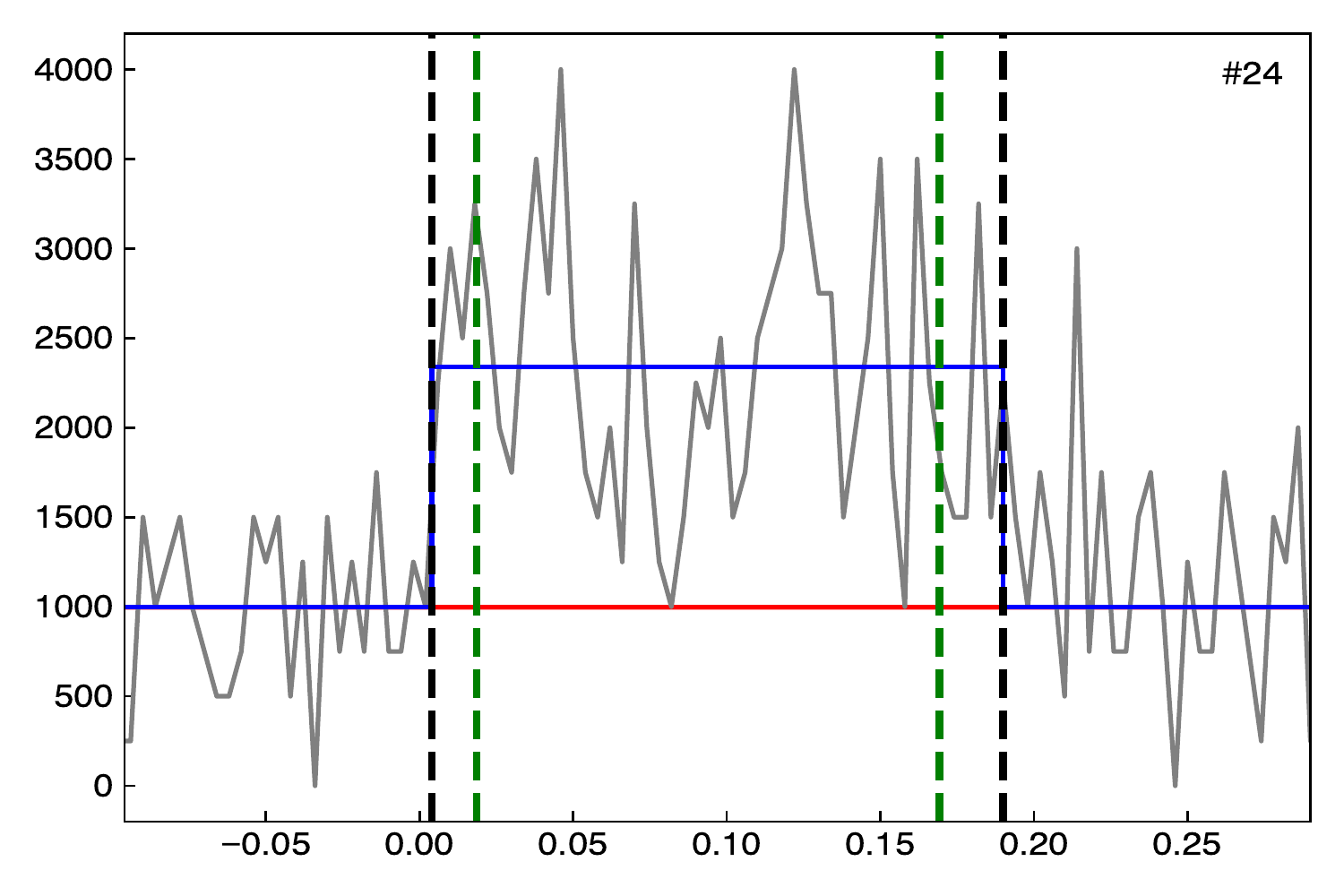}
\includegraphics[angle=0,width=0.195\textwidth]{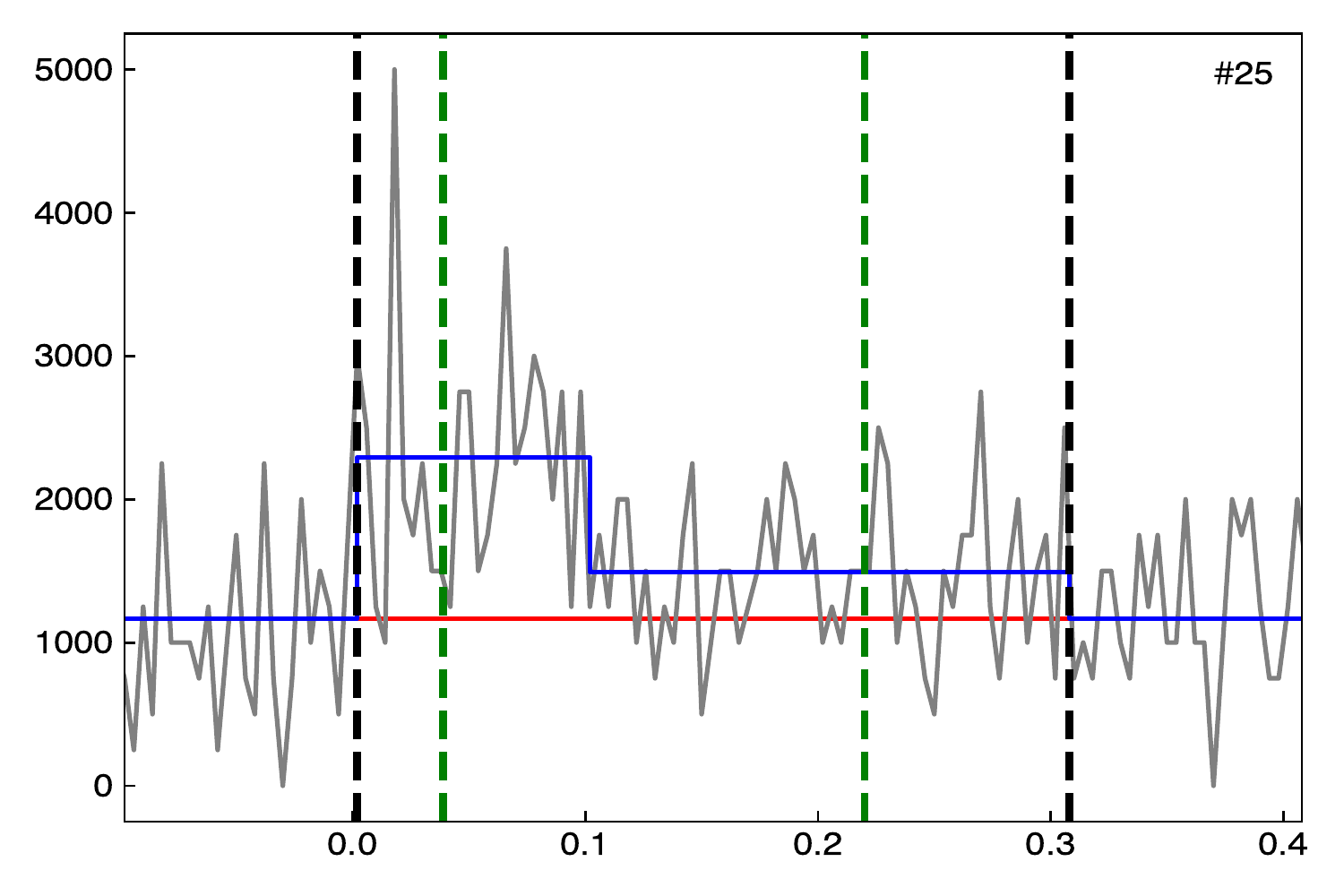}
\includegraphics[angle=0,width=0.195\textwidth]{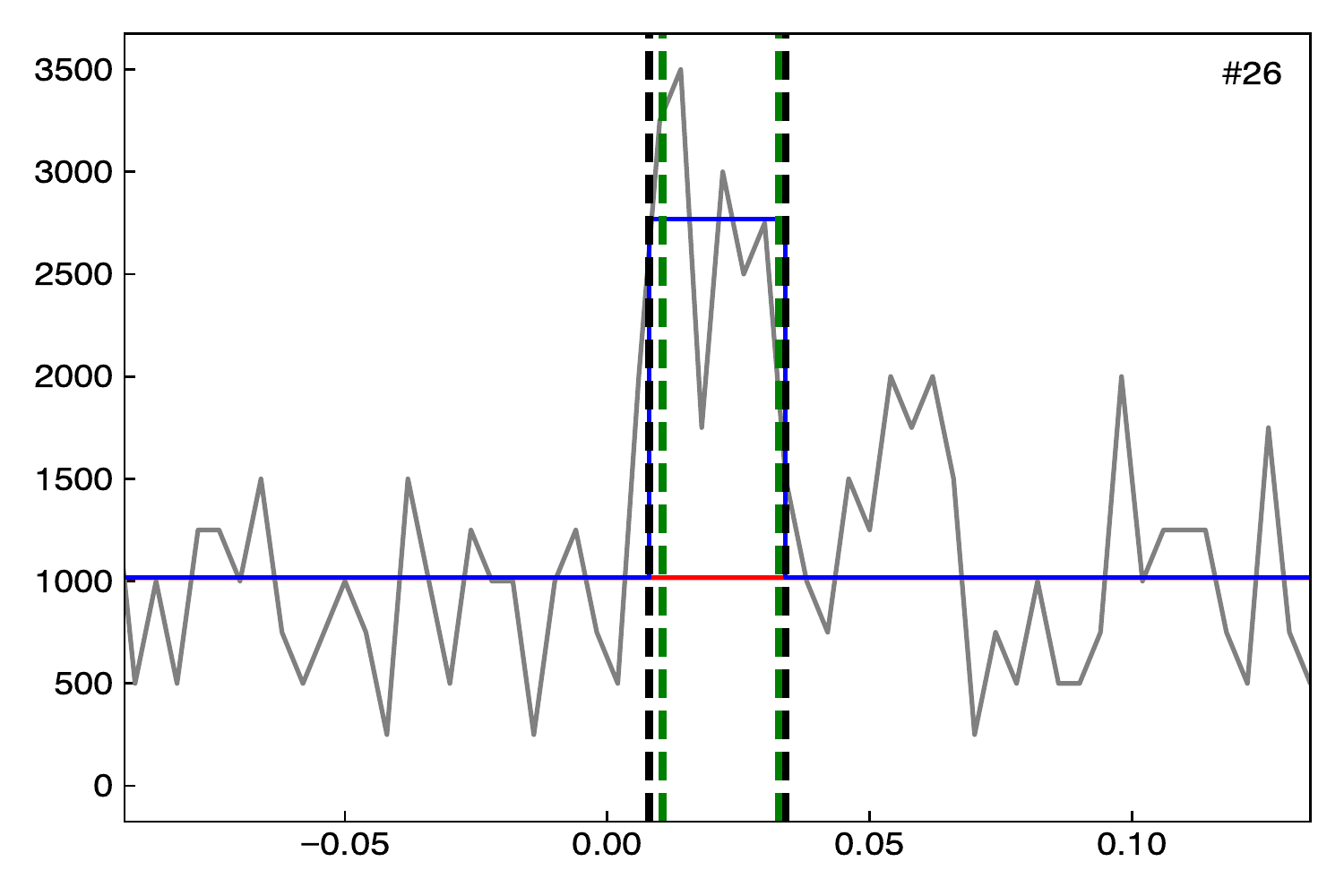}
\includegraphics[angle=0,width=0.195\textwidth]{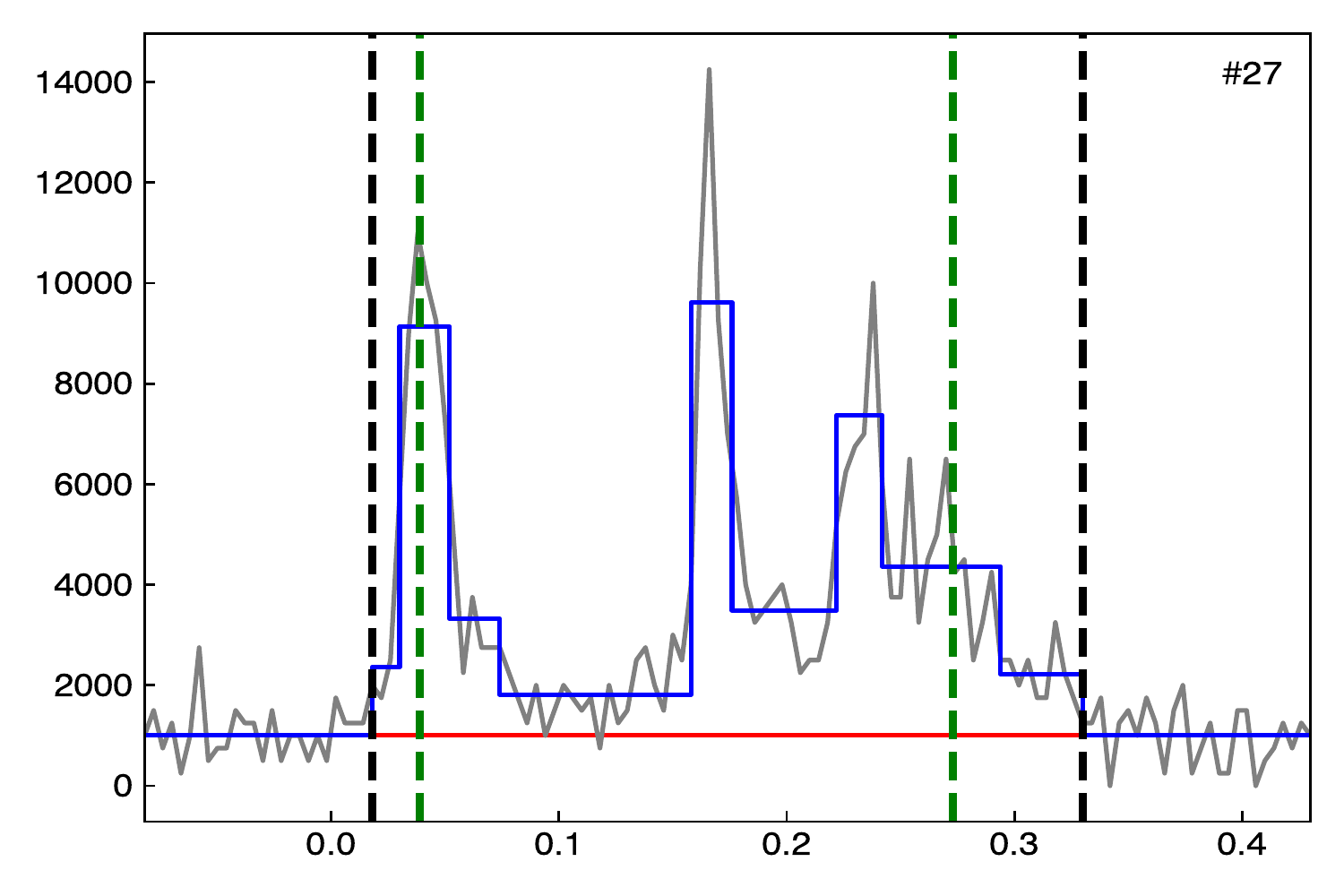}
\includegraphics[angle=0,width=0.195\textwidth]{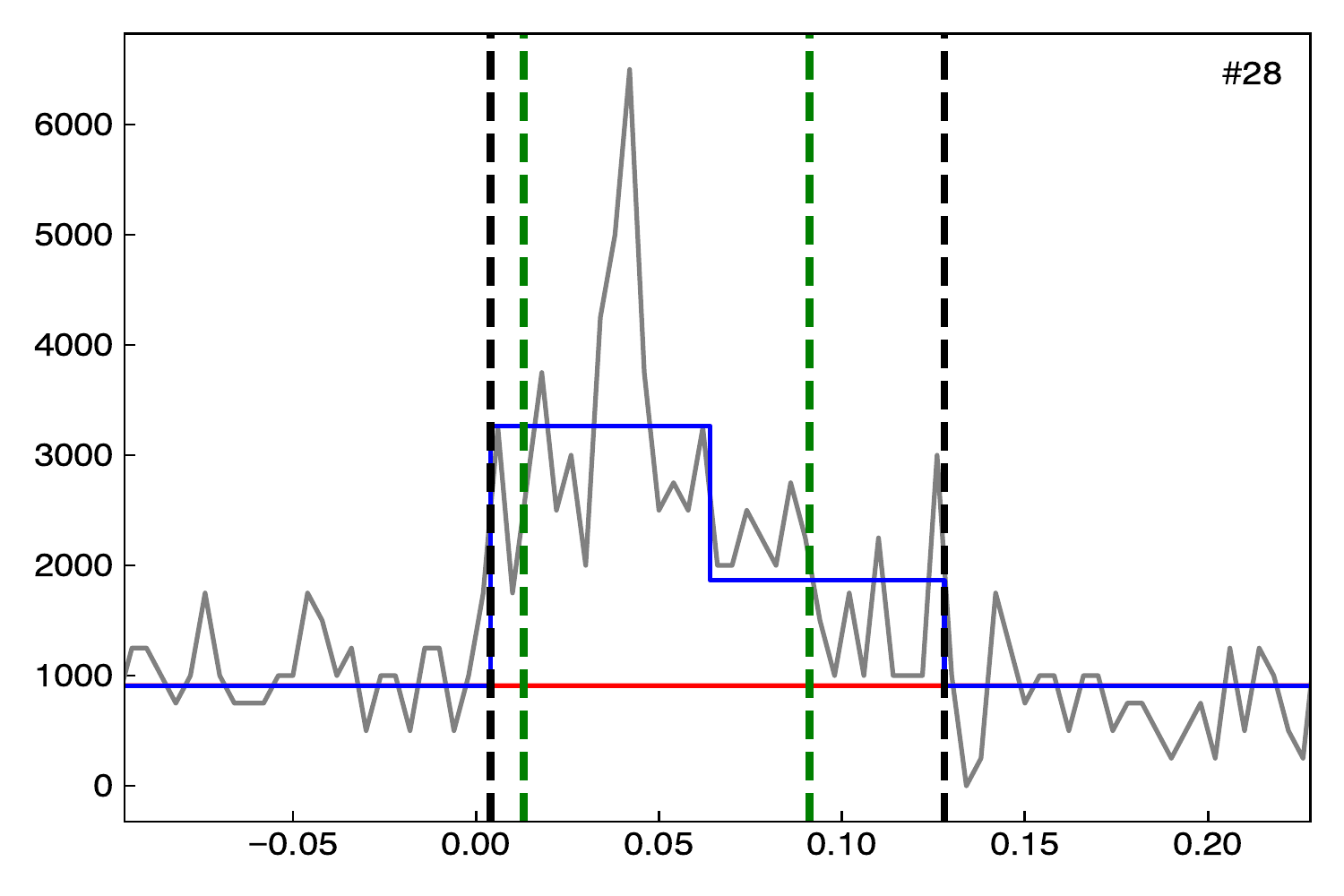}
\includegraphics[angle=0,width=0.195\textwidth]{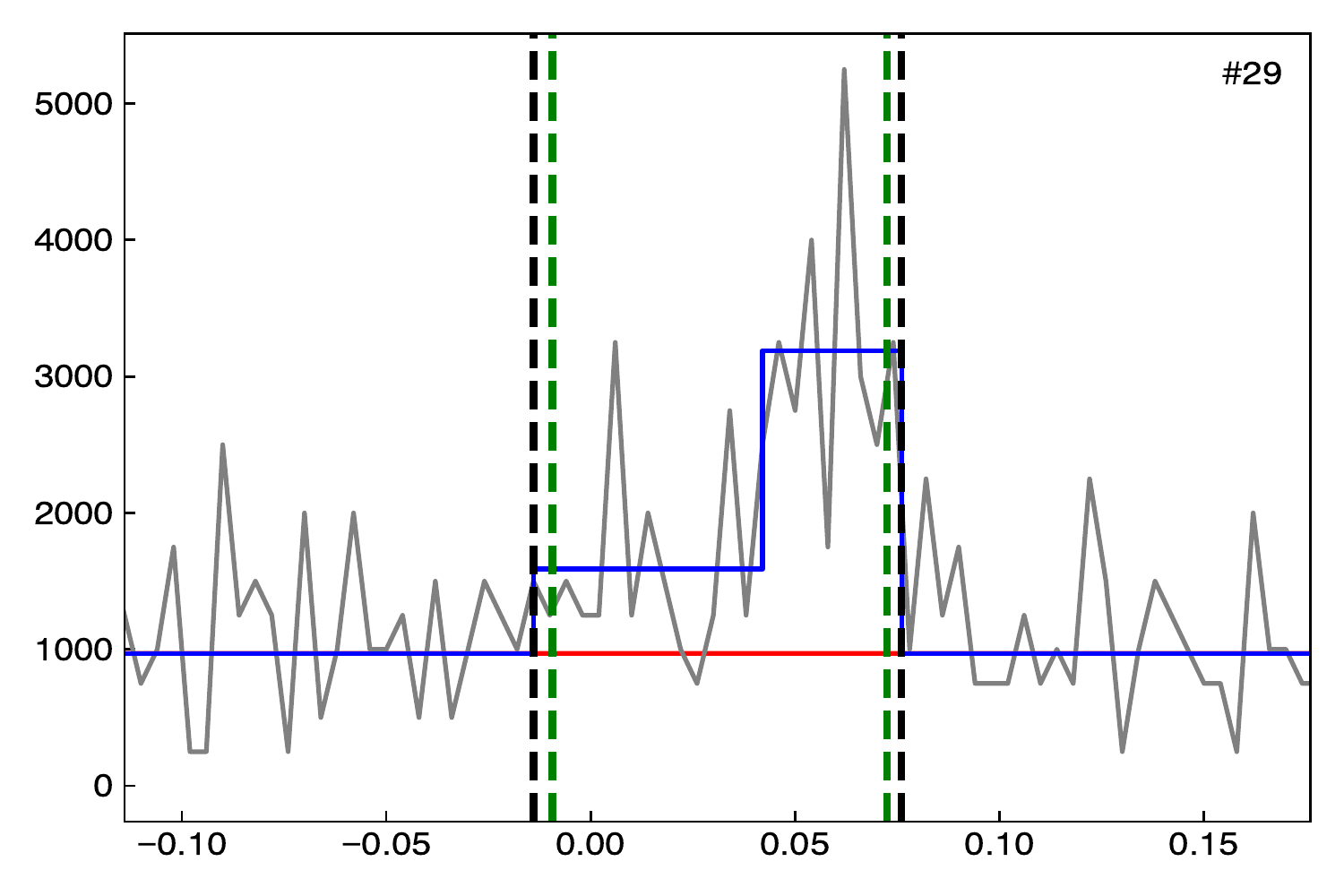}
\includegraphics[angle=0,width=0.195\textwidth]{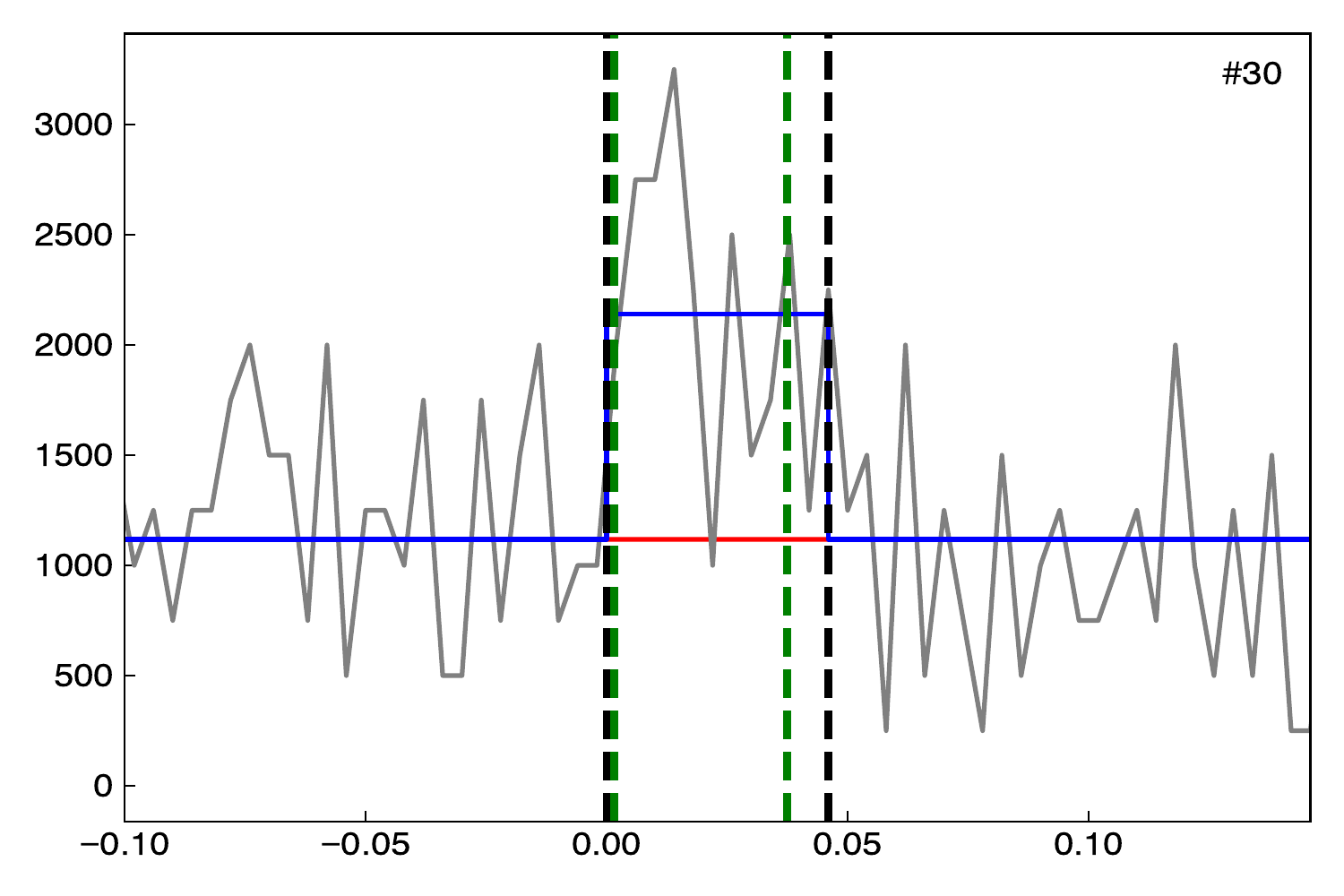}
\includegraphics[angle=0,width=0.195\textwidth]{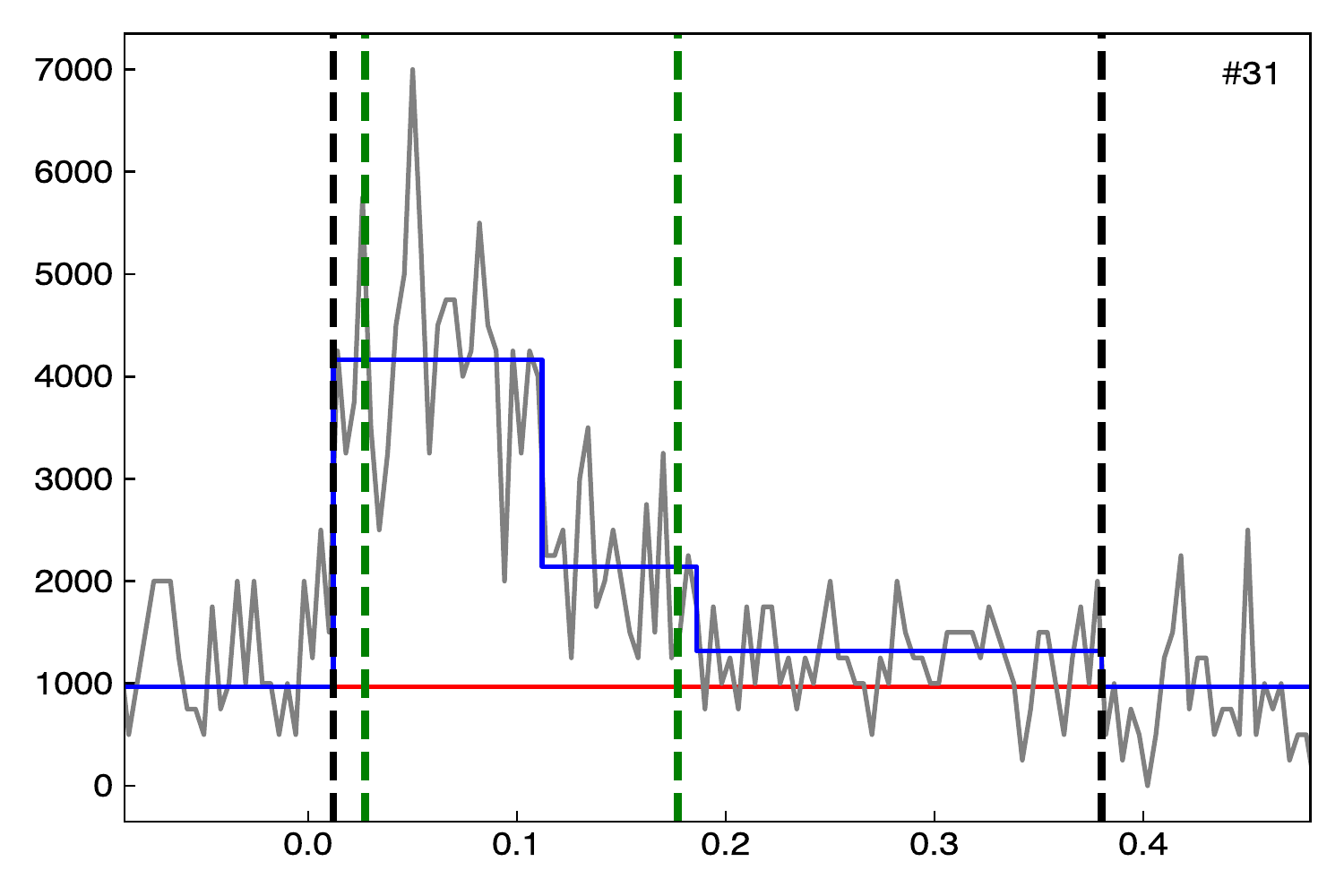}
\includegraphics[angle=0,width=0.195\textwidth]{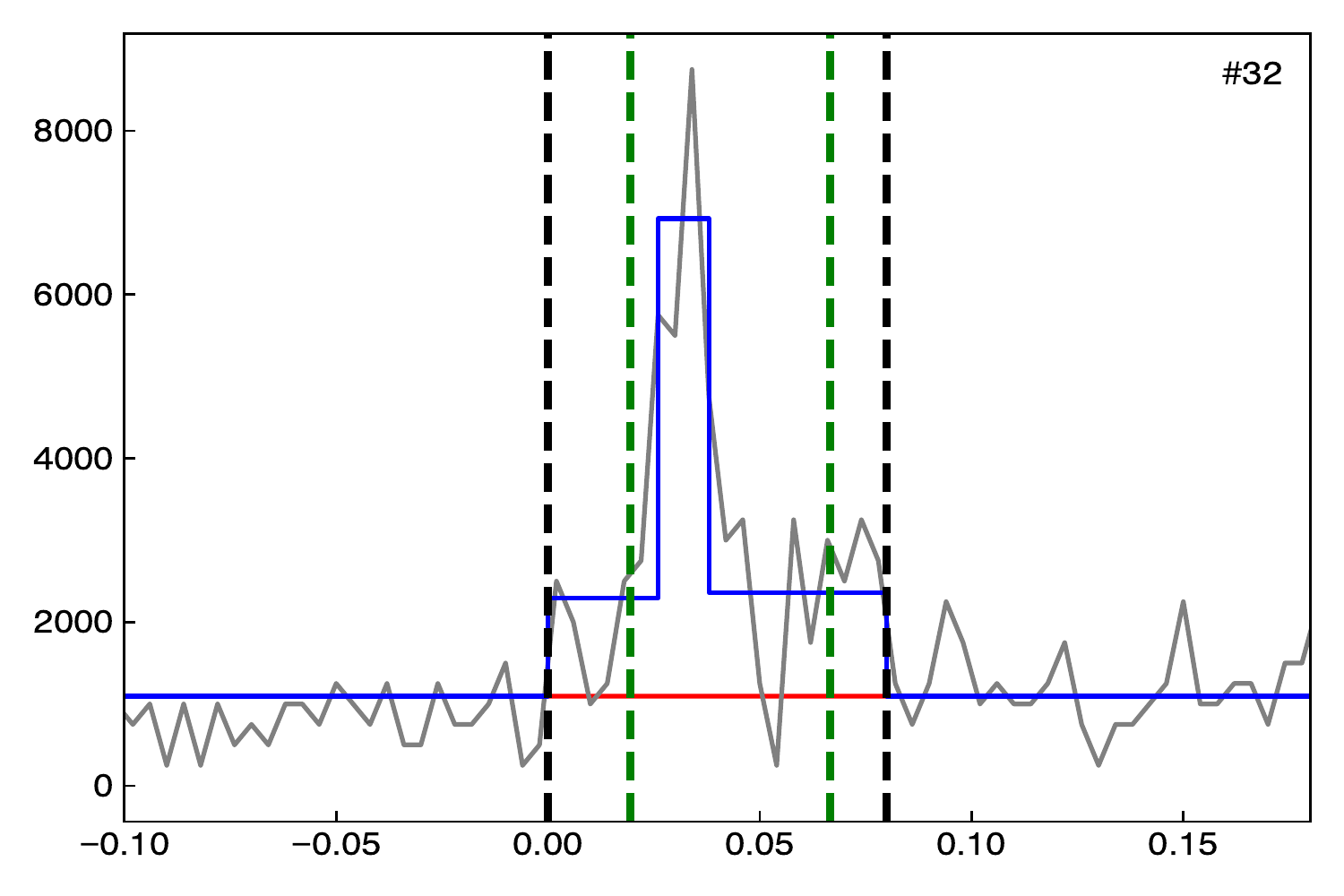}
\includegraphics[angle=0,width=0.195\textwidth]{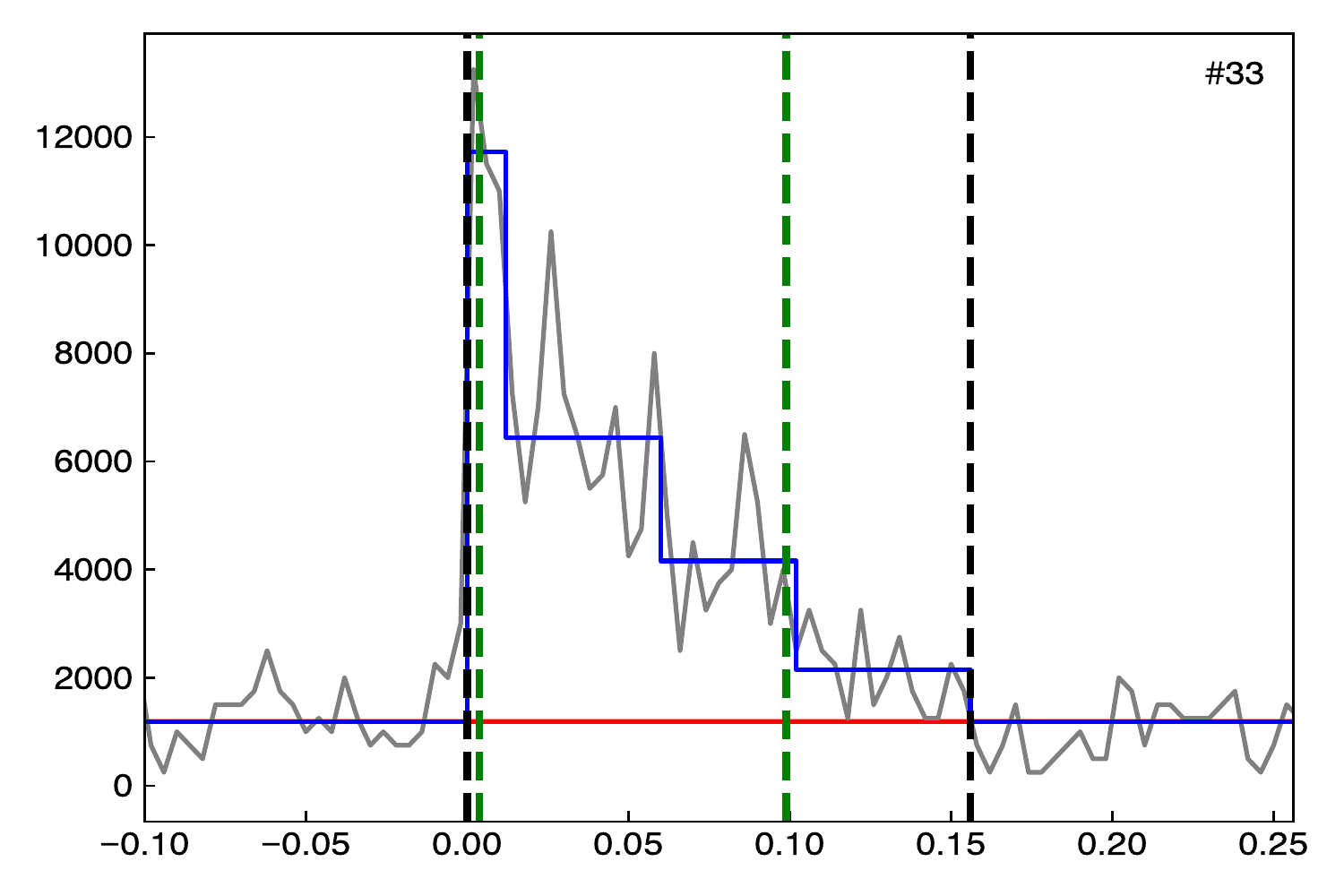}
\includegraphics[angle=0,width=0.195\textwidth]{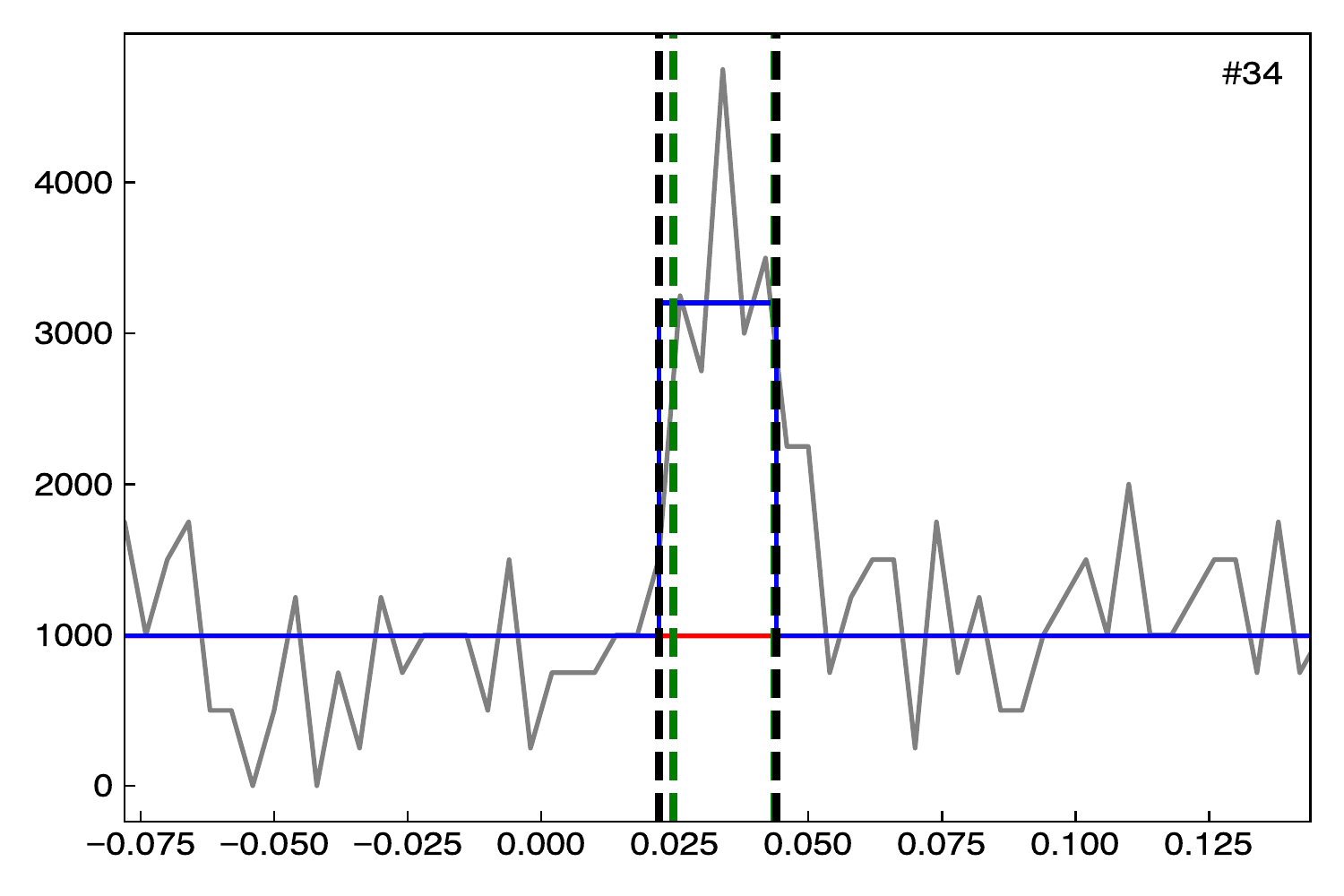}
\includegraphics[angle=0,width=0.195\textwidth]{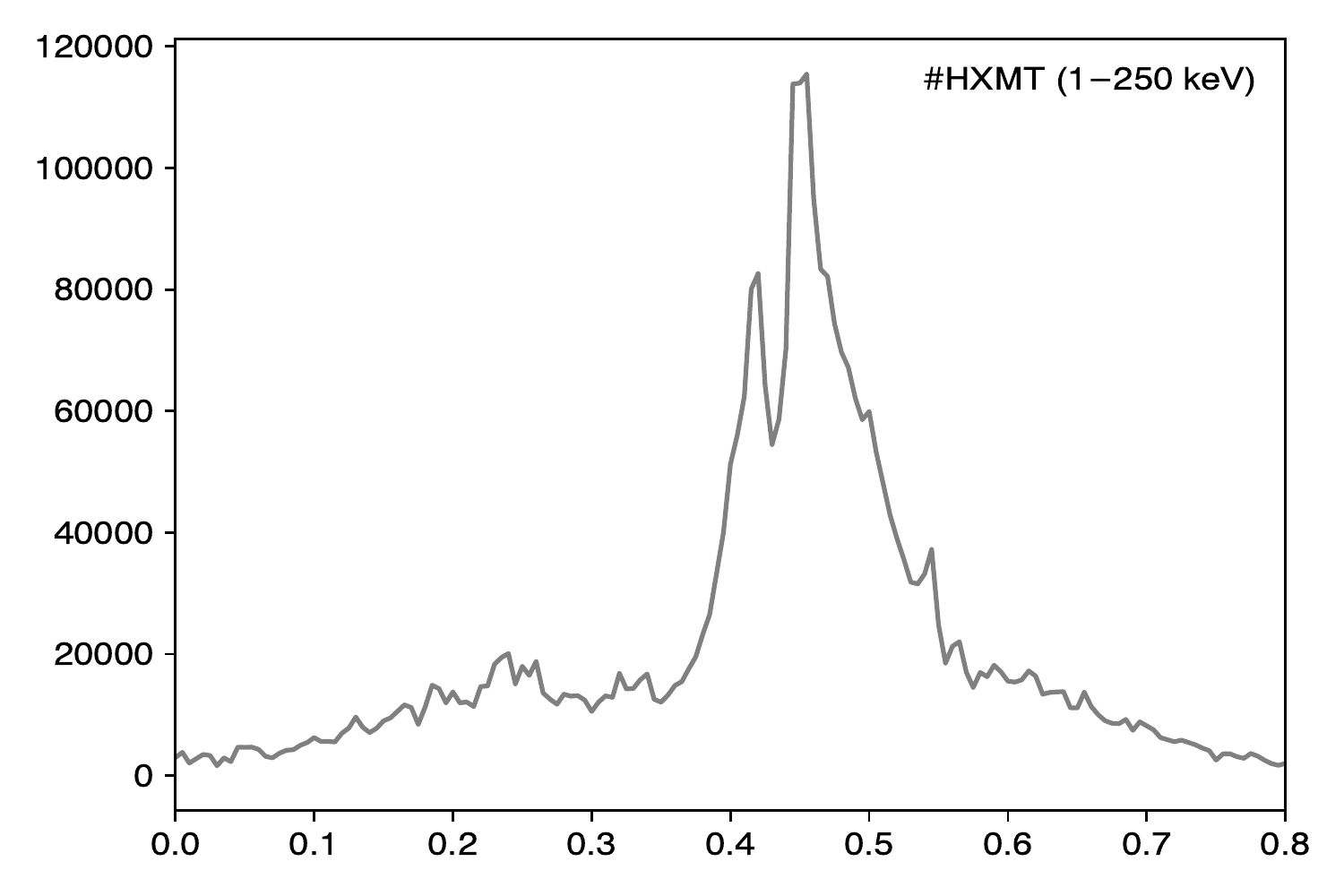}
\caption{Light curves of SGR J1935+2154. For each burst, only the detector with the smallest viewing angle to the source direction is used for its light curve plot. The energy range is 8--200 keV. The gray, blue, and red lines denote the count rates light curve, bayesian blocks, and background, respectively. The black and green vertical dashed lines show time intervals of $T_{\rm bb}$ and $T_{90}$. Bottom right: the light curve of the FRB-associated burst observed by HXMT in the energy range of 1--250 keV \citep{2020Li}.}
\label{fig:LC}
\end{figure}

Not all SGR bursts can trigger Fermi/GBM. As \cite{2020Lina} suggested, an untriggered search is needed to identify all potential bursts throughout the data available, such that a complete burst sample can be picked up. We perform such an untriggered burst search by applying the Bayesian Block method (\citealt{2013Scargle}; Figure \ref{fig:LC}) on the unbinned GBM continuous time-tagged event (TTE) data within the range of 8 keV--1 MeV energy. The GBM data are acquired during the time period of the third session of FAST observations, which correspond to the time interval between UTC 2020 April 27 23:55:00 and 2020 April 28 00:50:37. We apply the Bayesian Block method to find the best ``blocks" of time-series data over which the underlying signal is constant within the observational errors. In our work, the background block can be straightforwardly determined by the lowest constant blocks around the burst (red lines in Figure \ref{fig:LC}). A burst is defined in such a way that within its starting time ($T_{\rm bb,1}$) and ending time ($T_{\rm bb,2}$), all blocks are continuously above the background. Using such an approach, we identified 34 bursts from SGR J1935+2154 observed within the time period mentioned above, as listed in Table \ref{table:time}. We notice that some bursts in our sample are separated by only sub-seconds, which are combined into one event in \citet{2020Linc}.

\subsection{Temporal Analysis}\label{subsec:tem}

The light curves of all 34 bursts are shown in Figure \ref{fig:LC}. All light curves are plotted with a uniform bin size of 4\,ms. The Bayesian blocks are also plotted on top. In addition to the Bayesian block duration (\tbb$=T_{\rm bb,2}-T_{\rm bb,1}$), we also use $T_{90}$ to describe the burst duration. $T_{90}$ is the time interval within which the cumulative counts of the burst increases from 5\% to 95\% of the total counts \citep{1993Kouveliotou}. Both \tbb and $T_{90}$ are calculated in 8-200 keV and 2\,ms temporal resolution. The starting and ending times of $T_{\rm bb}$ and $T_{90}$ are shown in black and green vertical dashed lines in each panel of Figure \ref{fig:LC}. In addition, we obtained the minimum time variability (MTV) of each burst, which is defined as the length of the shortest block. We list the burst time, \tbb, $T_{90}$, and MTV of each burst event in Table \ref{table:time}.

\begin{figure}
\centering
\includegraphics[angle=0,width=0.48\textwidth]{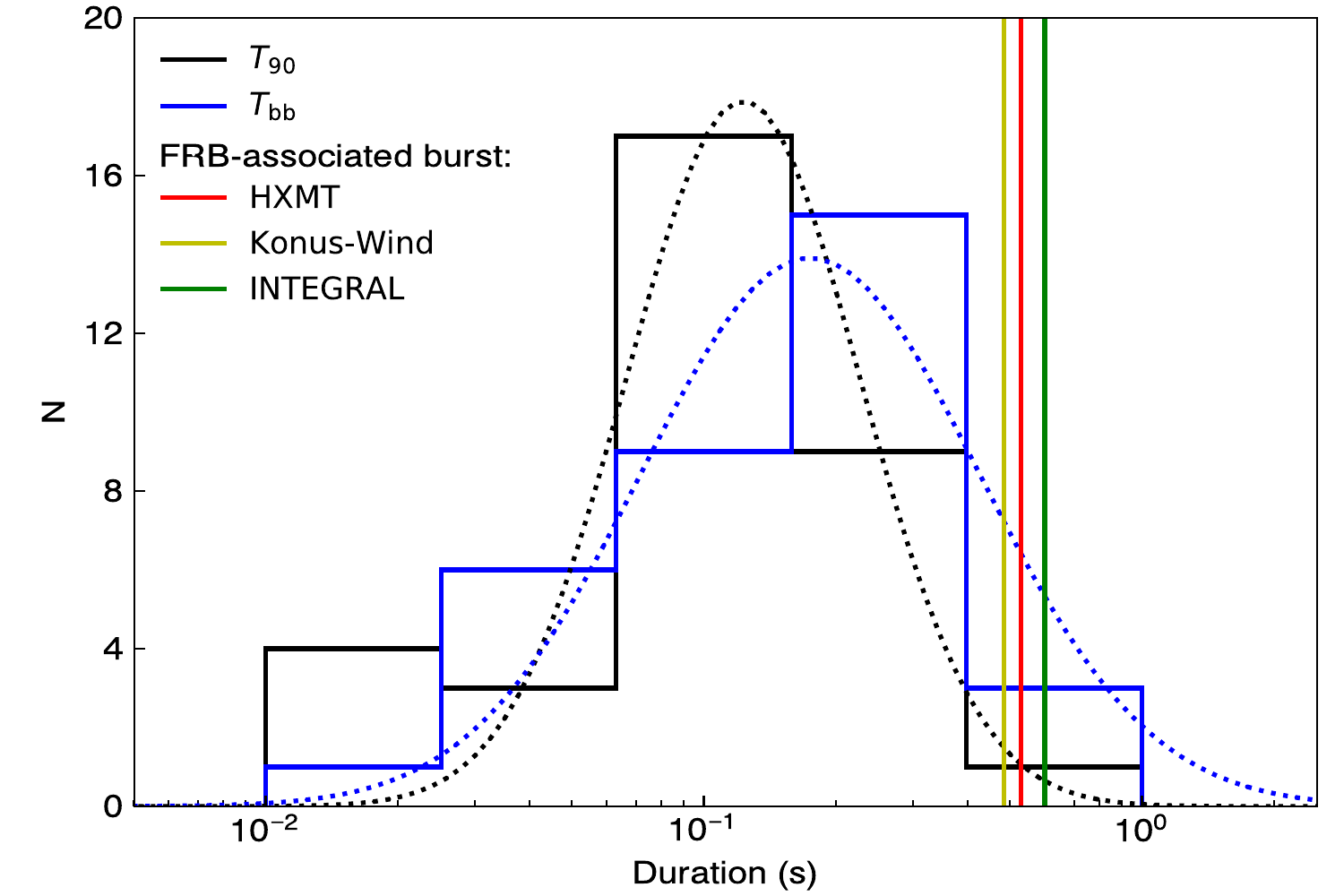}
\includegraphics[angle=0,width=0.48\textwidth]{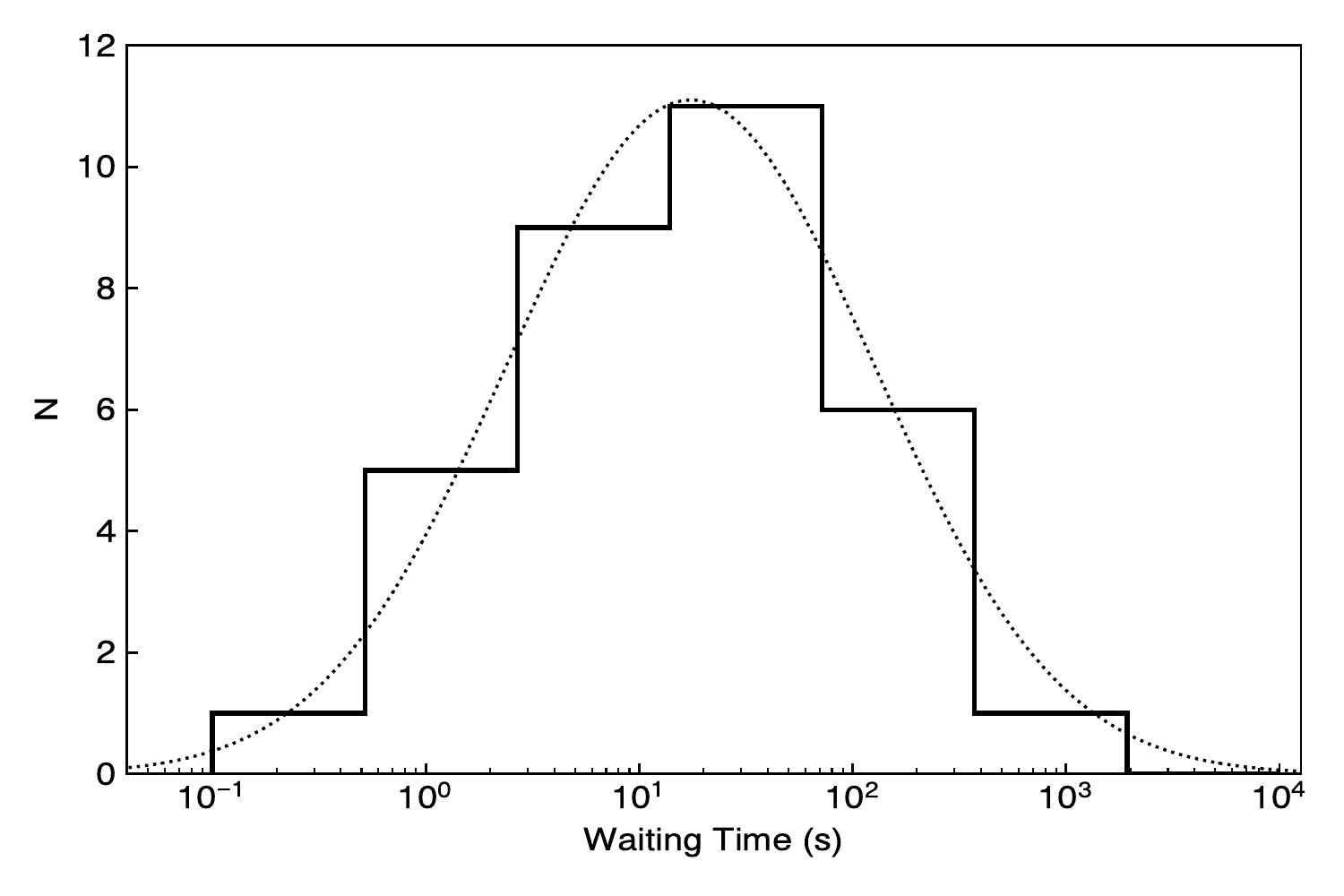}
\includegraphics[angle=0,width=0.48\textwidth]{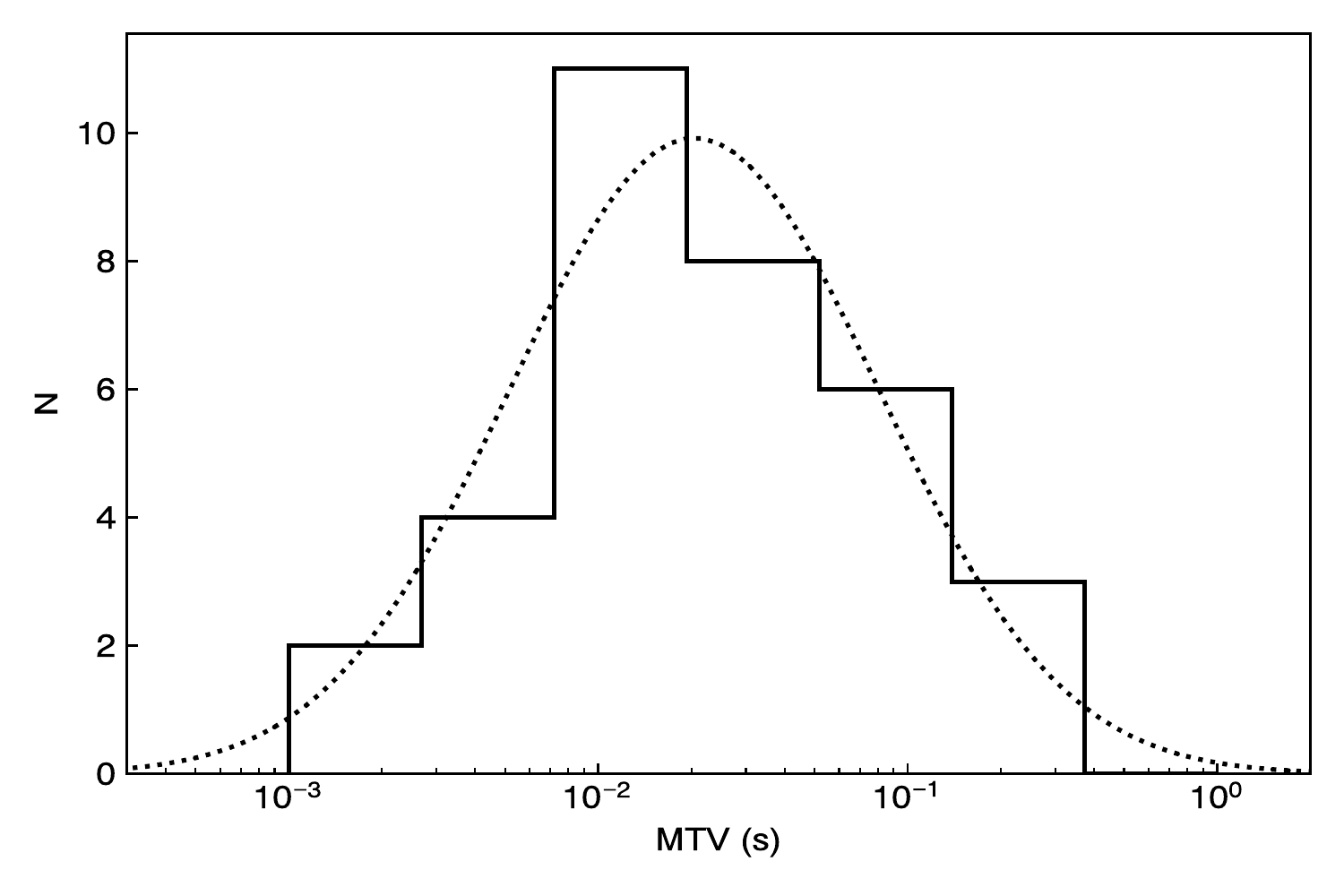}
\includegraphics[angle=0,width=0.48\textwidth]{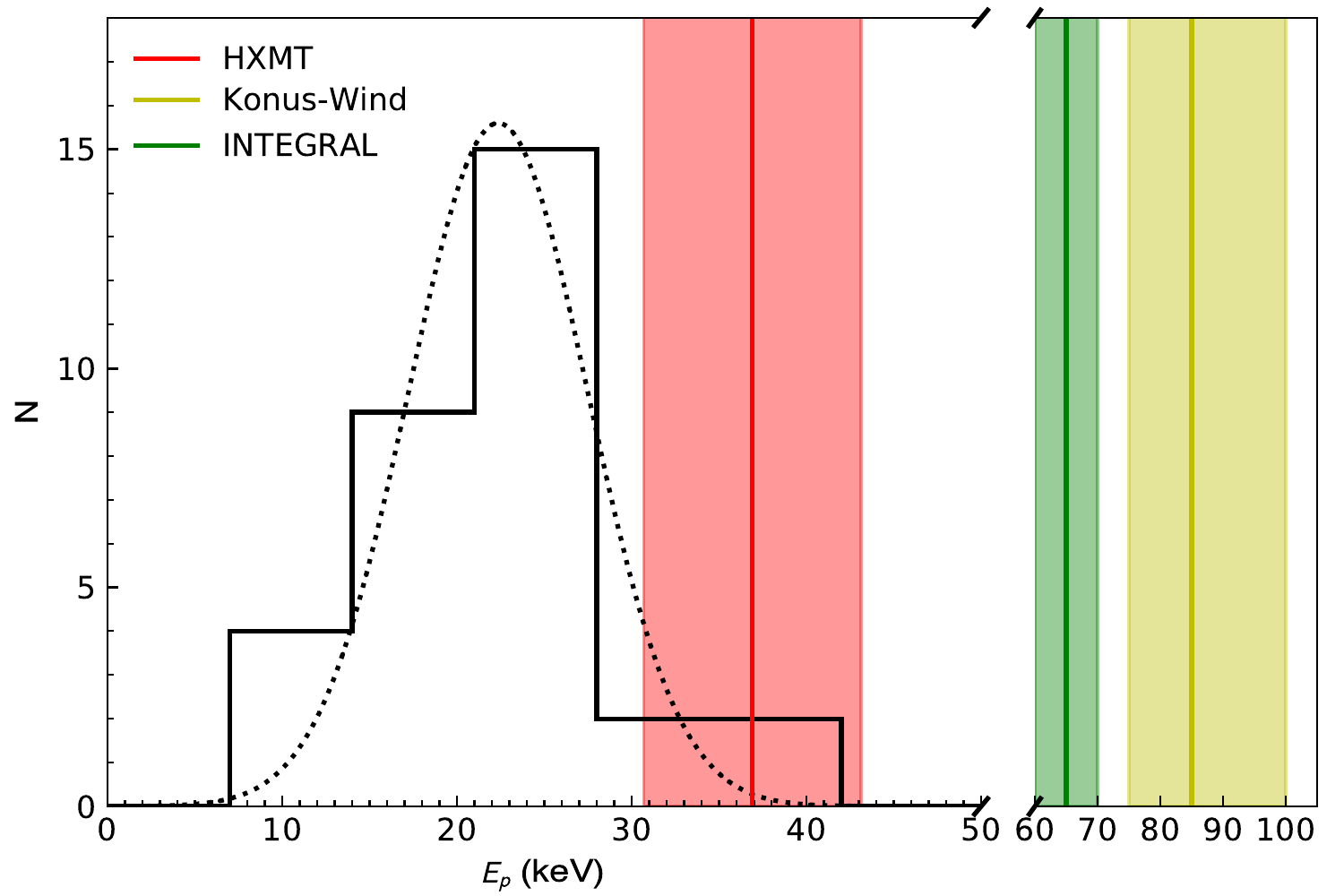}
\includegraphics[angle=0,width=0.48\textwidth]{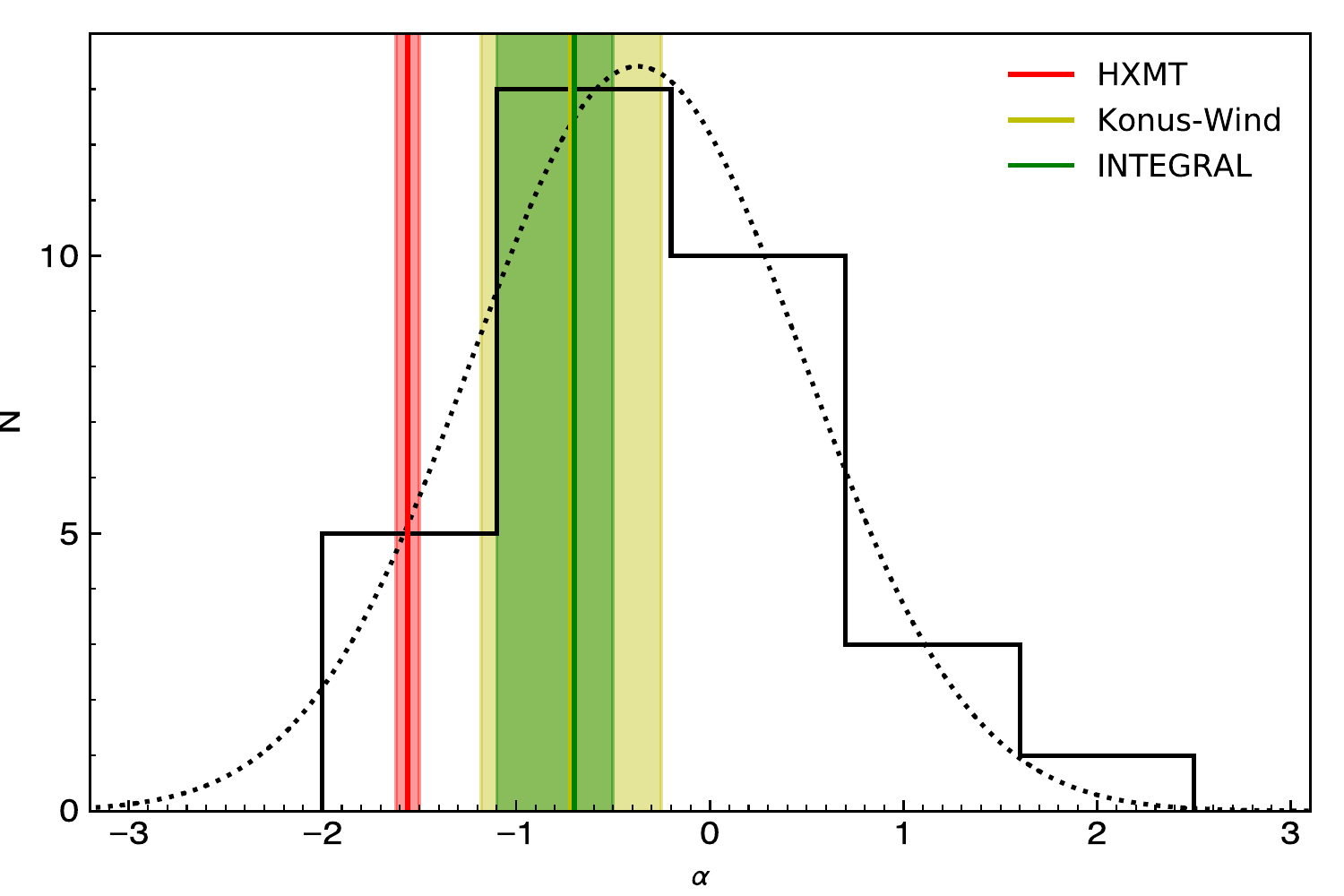}
\includegraphics[angle=0,width=0.48\textwidth]{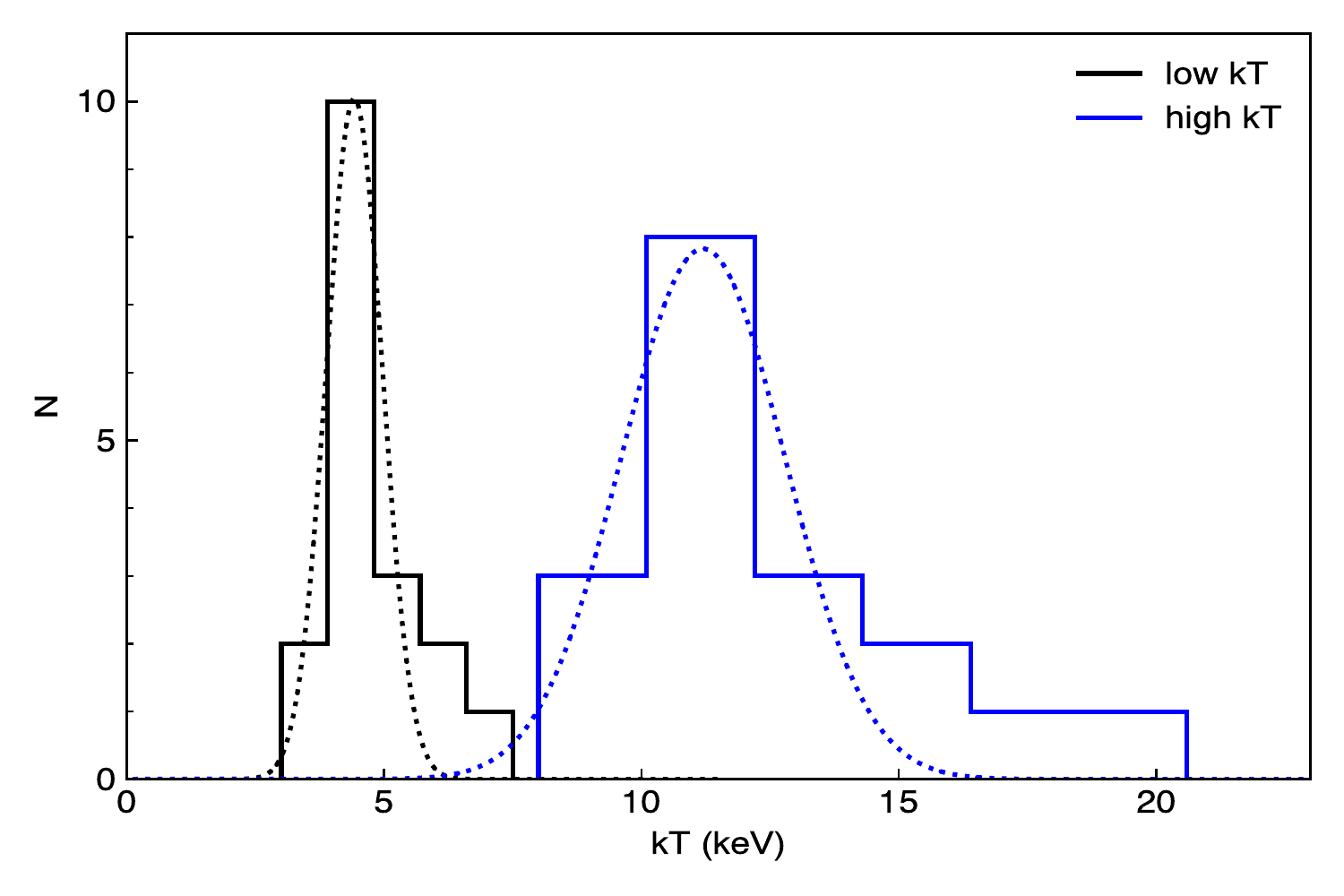}
\includegraphics[angle=0,width=0.48\textwidth]{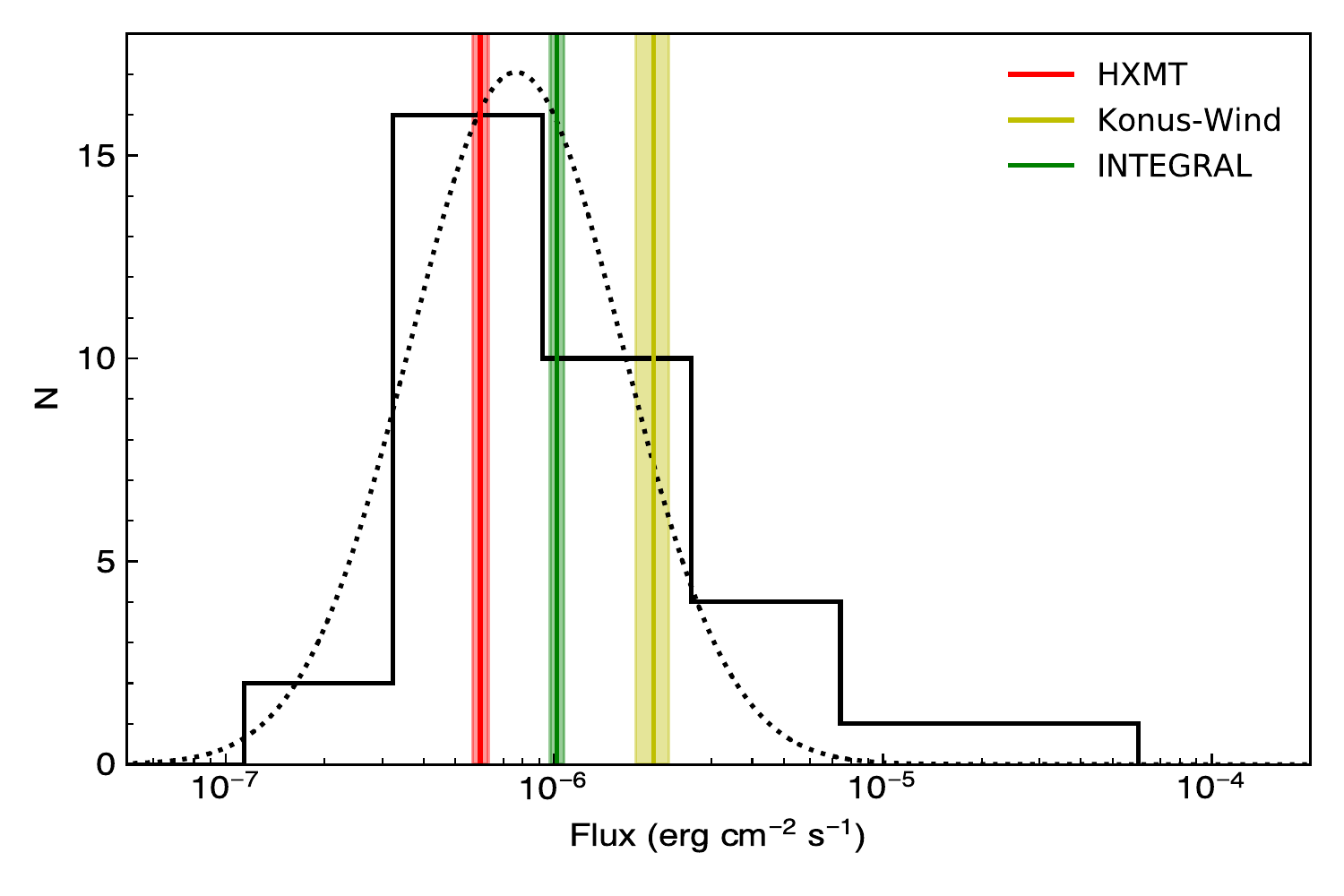}
\includegraphics[angle=0,width=0.48\textwidth]{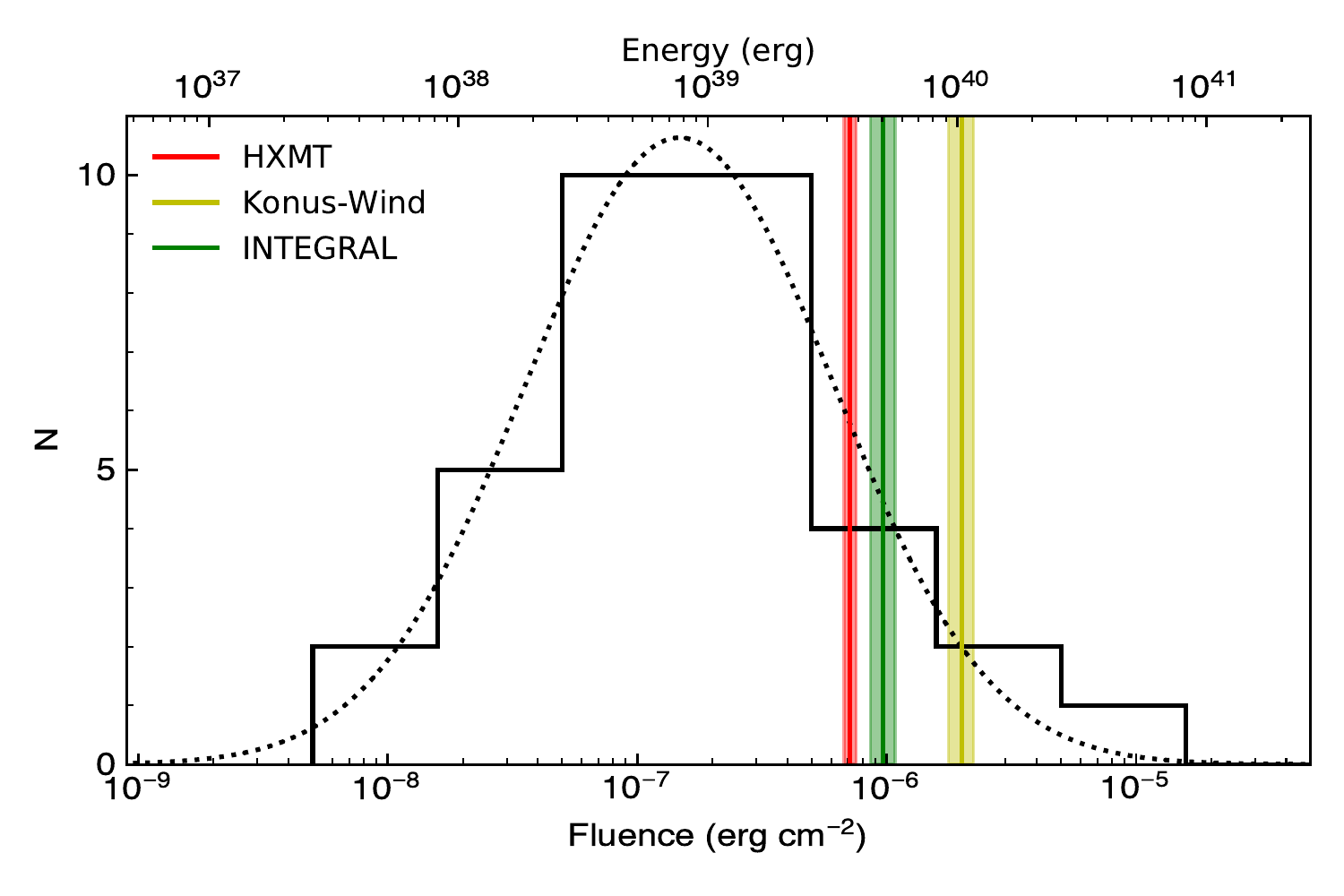}
\caption{Distributions of characteristic timescales, time-integrated spectral fitting parameters, and other derived parameters. Flux, fluence, and energy are calculated in the energy band of 8--200 keV. The dotted curves are the best Gaussian fits to the histograms. The red/green/yellow vertical lines and corresponding shadow areas represent the parameter values and errors of the FRB-associated burst obtained with HXMT/INTEGRAL/Konus-Wind \citep{2020Li,2020Mereghetti,2020Ridnaia}}.
\label{fig:dis}
\end{figure}

\begin{figure}
\centering
\includegraphics[angle=0,width=0.48\textwidth]{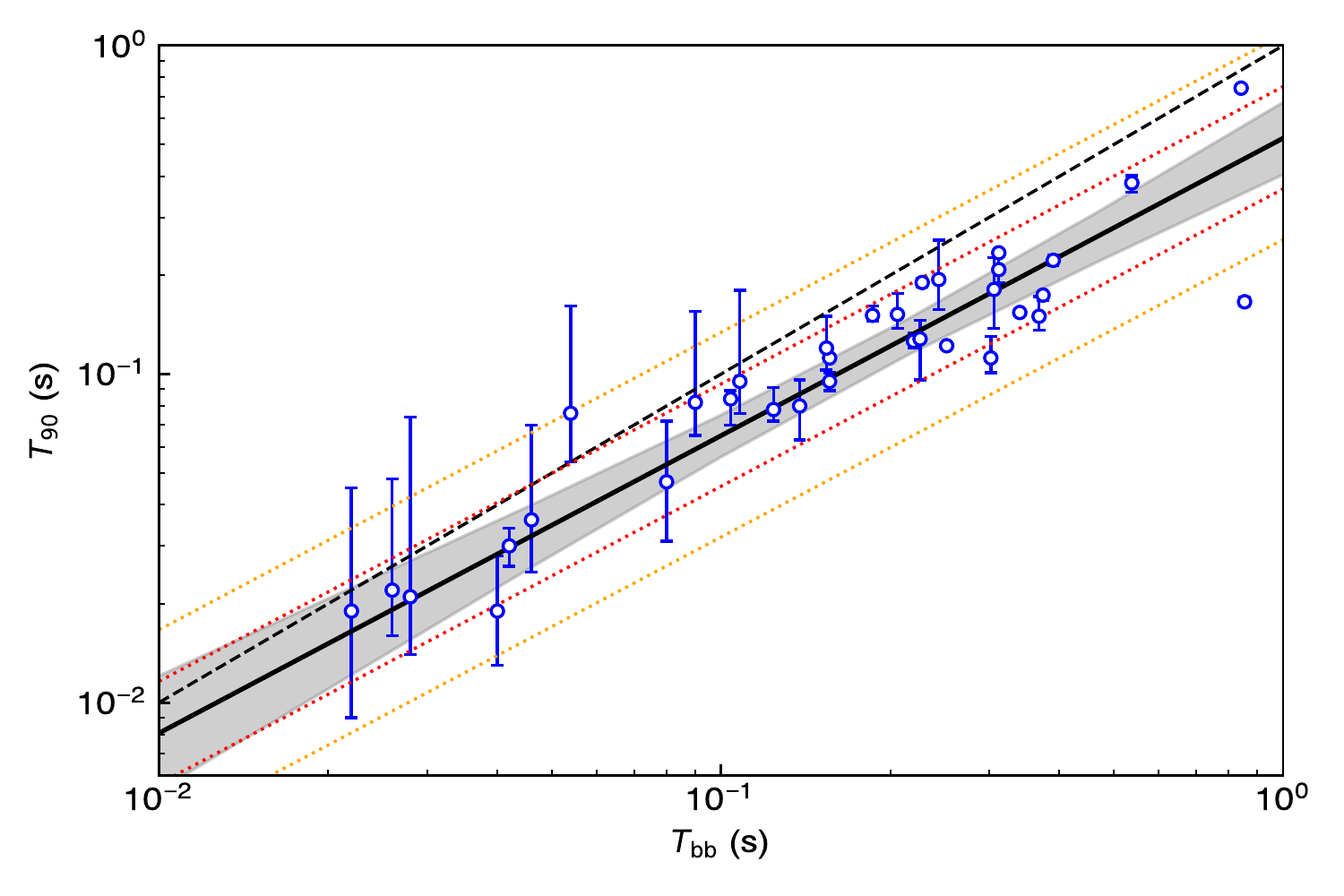}
\includegraphics[angle=0,width=0.48\textwidth]{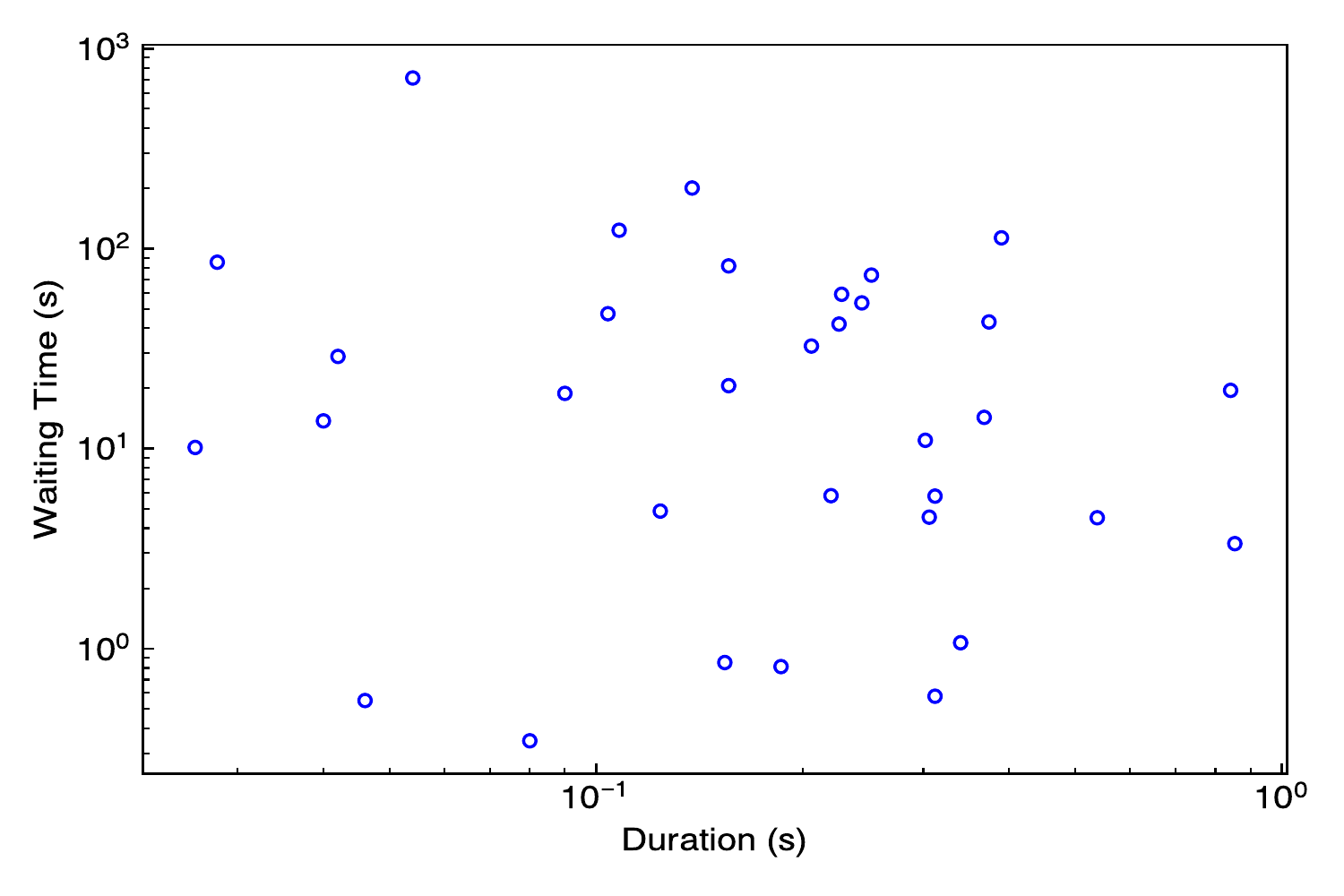}
\includegraphics[angle=0,width=0.48\textwidth]{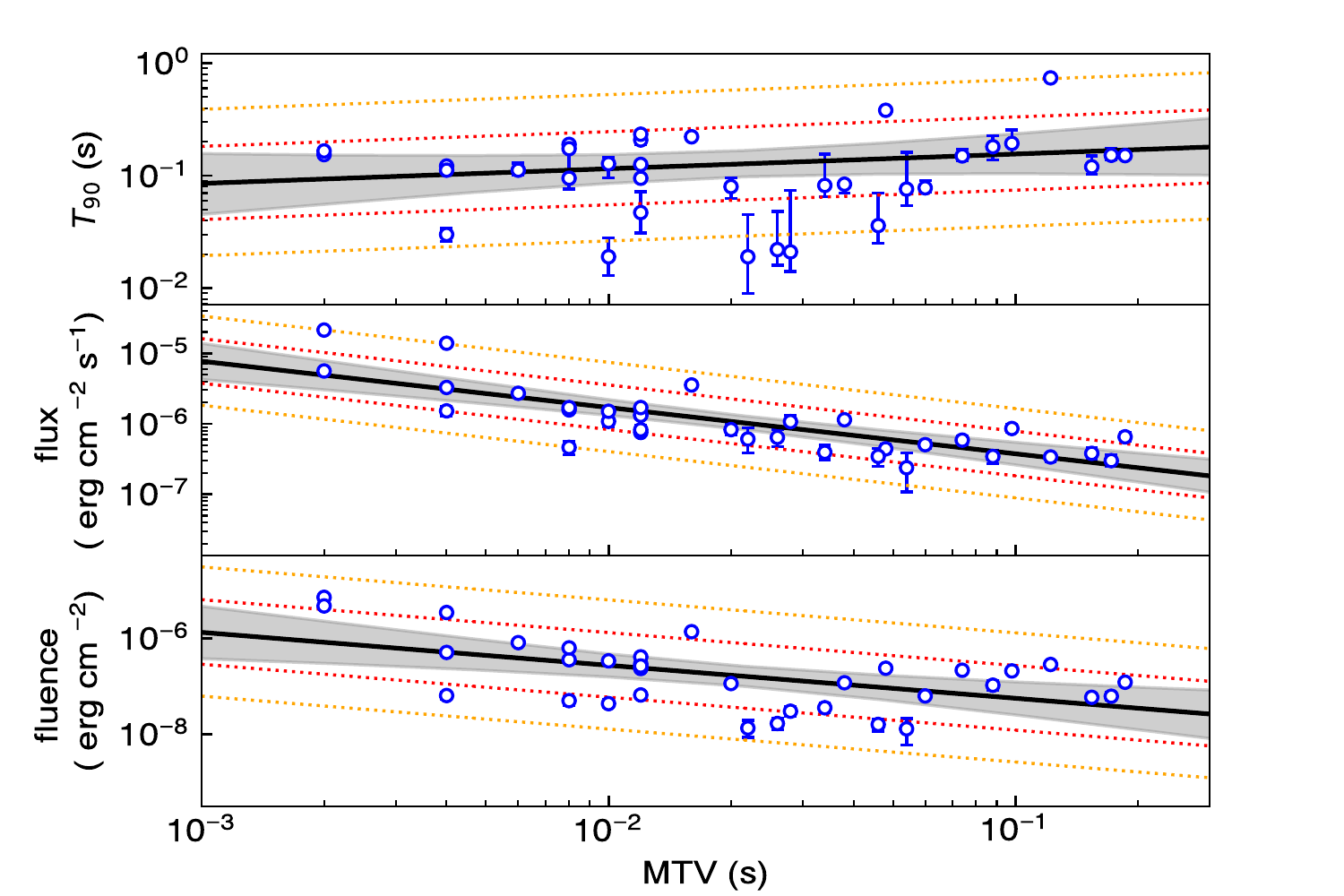}
\includegraphics[angle=0,width=0.48\textwidth]{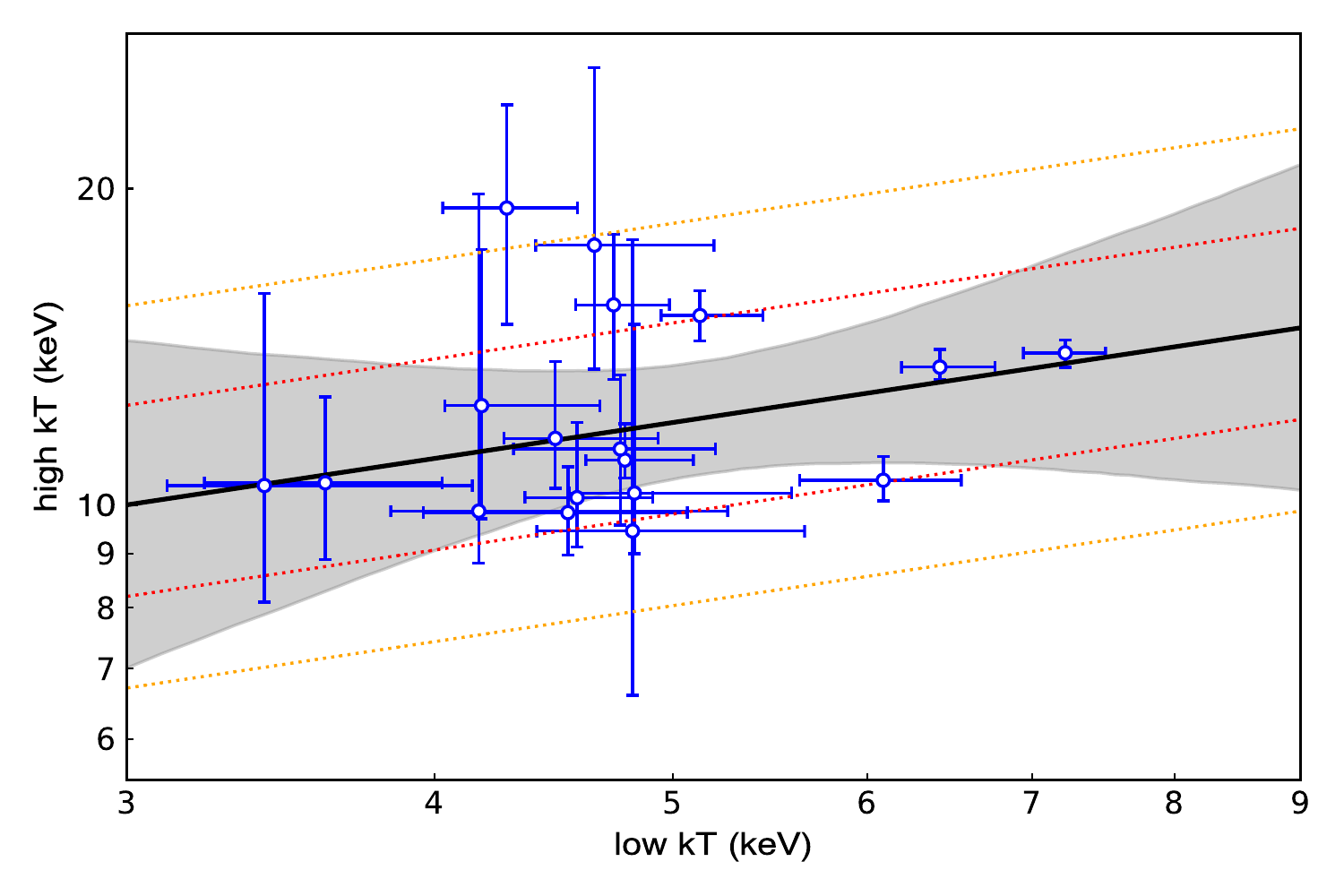}
\includegraphics[angle=0,width=0.48\textwidth]{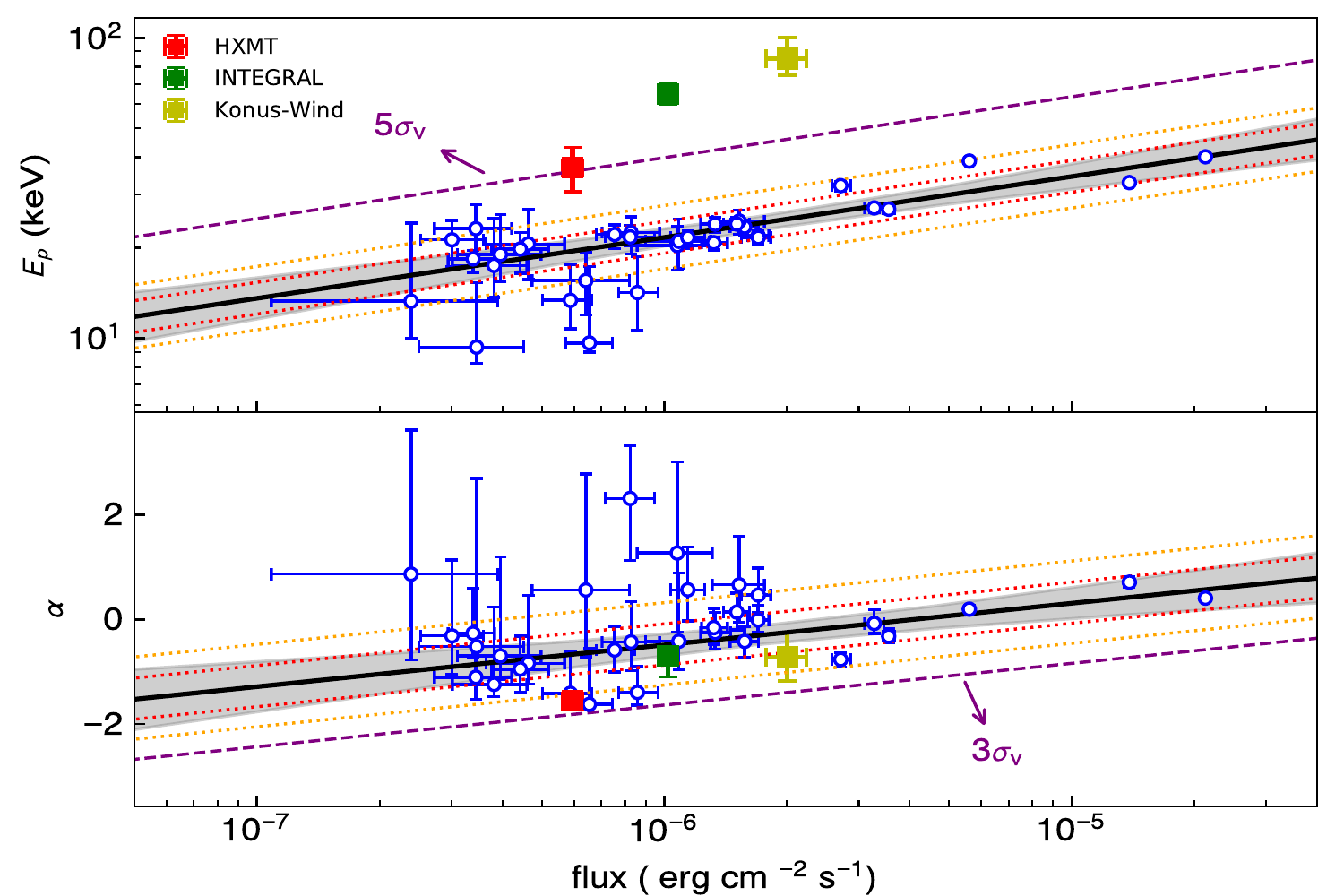}
\includegraphics[angle=0,width=0.48\textwidth]{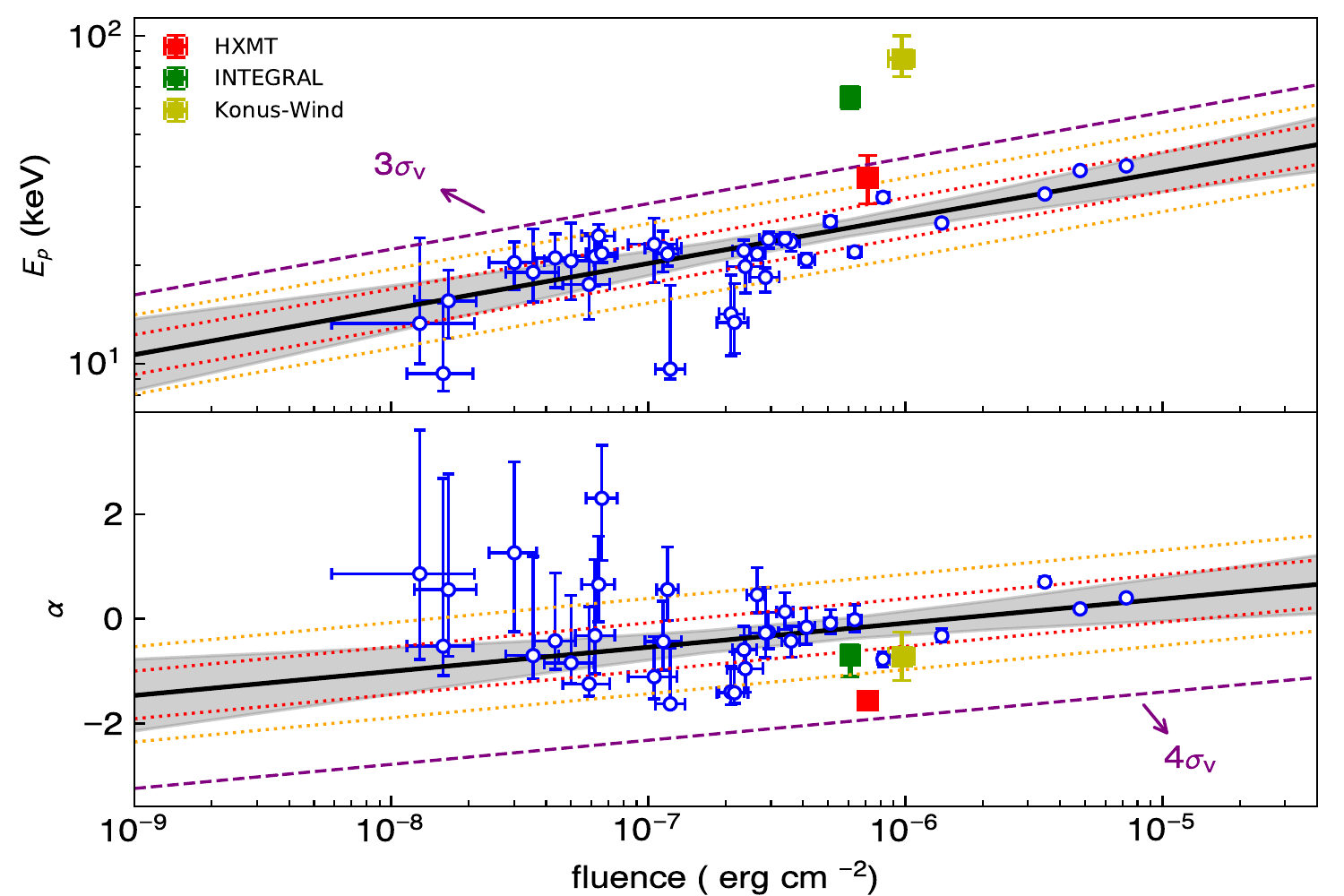}
\caption{Correlation plots for time-integrated spectral fitting. The same method of linear regression is imported from \citet{2018Tu}. The black solid lines represent the best-fitting results. The red and yellow dotted lines represent $1\sigma _v$ and $2\sigma _v$ regions of extra variability, respectively \citep{2005D'Agostini}, and the gray areas mark the 95\% confidence interval of fitting uncertainties. The black dashed line in the top-left panel is $T_{\rm bb} = T_{90}$. The red, green, and yellow data points in the bottom panels are the same as the corresponding value with identical color in Figure \ref{fig:dis}. The purple dashed lines indicate upper/lower limits of several times of $\sigma_v$ region that can contain the red points.}
\label{fig:cor}
\end{figure}

The distributions of $T_{\rm bb}$ and $T_{90}$ are presented in Figure \ref{fig:dis}, both being a Gaussian shape in the logarithmic scale (i.e., log-normal), which is similar to the findings in other magnetars \citep[e.g.][]{2015Collazzi}. The $T_{\rm bb}$ ($T_{90}$) distribution peaks at 0.175 (0.123) s, which are slightly longer than those of previous studies \citep{2020Lina}. The correlation between $T_{\rm bb}$ and $T_{90}$ is plotted in Figure \ref{fig:cor}. The corresponding Spearman’s rank correlation coefficient is 0.92, with a chance probability of $1.7 \times 10^{-14}$. A power-law fit to the trend results in $T_{90} \propto T_{\rm bb}^{0.91\pm 0.05}$.

We also calculated the waiting time between the neighboring bursts by measuring the difference of their starting time $T_{\rm bb}$ ($T_{90}$). The corresponding distributions are presented in Figure \ref{fig:dis}, which are both consistent with a log-normal shape peaking at 17.4 s. No significant correlation is observed between the burst duration and waiting time (Figure \ref{fig:cor}).

By plotting the distribution of MTV of our sample in Figure \ref{fig:dis}, we find that their typical variability is roughly a few tens of milliseconds. No apparent correlation between MTV and the burst duration (e.g., $T_{90}$) can be found (Figure \ref{fig:cor}).

\subsection{Time-integrated Spectral Fitting}\label{subsec:int}

For each burst, we extract the source spectra, background spectra, and generate the instrumental response matrix within the expand of \tbb following the standard procedures as described in \citet{2016Zhang,2018ZhangBB}. We choose three detectors: two NaI detectors with the smallest viewing angles, and one BGO detector with brightest peak count rate. The selected energy band is 8-300 keV for NaI detectors and 200-300 keV for the BGO detector. The total energy channel number is $\sim$150, which is the number of data points in spectral fitting. Such observed data are then fitted by various spectral models using a self-developed spectral fitting tool \textit{McSpecFit} \citep{2018ZhangBB}. Five frequently used models, namely the blackbody (BB), the power law (PL), the combination of two blackbodies (BB+BB), the PL with an exponential cutoff at higher energies (CPL), and the combination of a BB and a PL (BB+PL) are employed and compared. In particular, the CPL model is defined as \citep{2016Yu}
\begin{equation}
	N(E) = AE^{\alpha} {\rm exp} [-(\alpha +2 )E/E_p ],
\end{equation}
where A is the normalization factor, $\alpha$ is the power-law photon index, and $E_p$ is the peak energy of the energy density spectrum ($\nu F_{\nu}$) in units of keV.

To determine whether a certain spectral model can be accepted to explain the data well, regardless of its physical meaning, we required that (1) its reduced statistics (namely the PGSTAT value from \citet{1996Arnaud} divided by the degree of freedom, PGSTAT/dof) should be less than 1.2\footnote{Except for Burst \#10, which has PGSTAT/dof $\sim$ 1.9 due to strong spectral evolution effect.};(2) the best-fit parameters are fully constrained in the physical reasonable regions. According to such criteria, we find that 28, 18, 22, 32, and 11 bursts can be fitted with BB, BB+BB, PL, CPL, and BB+PL model, respectively. For comparison, previous studies \citep[e.g.][]{2008Israel,2011Lin,2020Lina,2012Horst} have shown that the BB+BB and CPL models are preferred in broadband-spectra fitting.

To further determine which acceptable model is the ``best" one to fit the data, we employ the Bayesian information criterion (BIC) to conduct the model comparison. A numerical value BIC is introduced as \citep{1978Schwarz}
\begin{equation}
	{\rm BIC} = -2\ln \mathcal{L}+k\ln n
\end{equation}
where $\mathcal{L}$ is the maximum likelihood, $k$ is the number of parameters of the model, and $n$ is the number of data points used in the fitting. A model with a smaller BIC value is considered to be better fitted with the data. Therefore, we identified the model whose BIC has the smallest value as the ``best" one. A BIC difference of $<2$, 2-6, 6-10, and $>10$ are regarded as weak evidence, positive evidence, strong evidence, and very strong evidence, respectively\citep{1961Jeffreys, 1995Kass, 1998Mukherjee}. Accordingly, we identified the ``best" model with the minimum BIC. For the cases when two BICs with a difference of $<2$, we consider that both models can equally explain the data. If one of the pair gives the minimum BIC, we consider both as ``best" models. After selecting the best model(s), we further obtain the burst fluence and calculate its flux and isotropic energy within the energy band of 8--200 keV using a distance of 6.6 kpc \citep{2020Zhou}\footnote{We found no significant difference in terms of the energetics calculation when the BICs of two models are close (e.g., $\Delta$BIC $<$ 2).}. The ``acceptable" and ``best" models (displayed in bold), their corresponding parameters, as well as the derived fluences, are all listed in Table \ref{table:int_spectra}.

For the 32 bursts whose time-integrated spectra can be well fitted by the CPL model, we plot the distributions of their peak energy $E_p$ and photon index $\alpha$ in Figure \ref{fig:dis}. Both distributions are Gaussian-like with a peak value of $\sim 22.4$ keV and $\sim -0.37$, respectively. The $E_p$ and $\alpha$ values of the FRB-associated X-ray burst observed by HXMT \citep{2020Li} are plotted in red for comparison.

Similarly, we plot the distributions of the low and high blackbody temperatures for the 18 bursts that can be fitted by the BB+BB model in Figure \ref{fig:dis}. Both temperatures are distributed as a Gaussian-like shape with peak values of $\sim$4.42 keV and $\sim$11.2 keV, respectively. There is no significant correlation between the high and low temperatures in our sample (Figure \ref{fig:cor}).

\begin{figure}
\centering
\includegraphics[angle=0,width=0.7\textwidth]{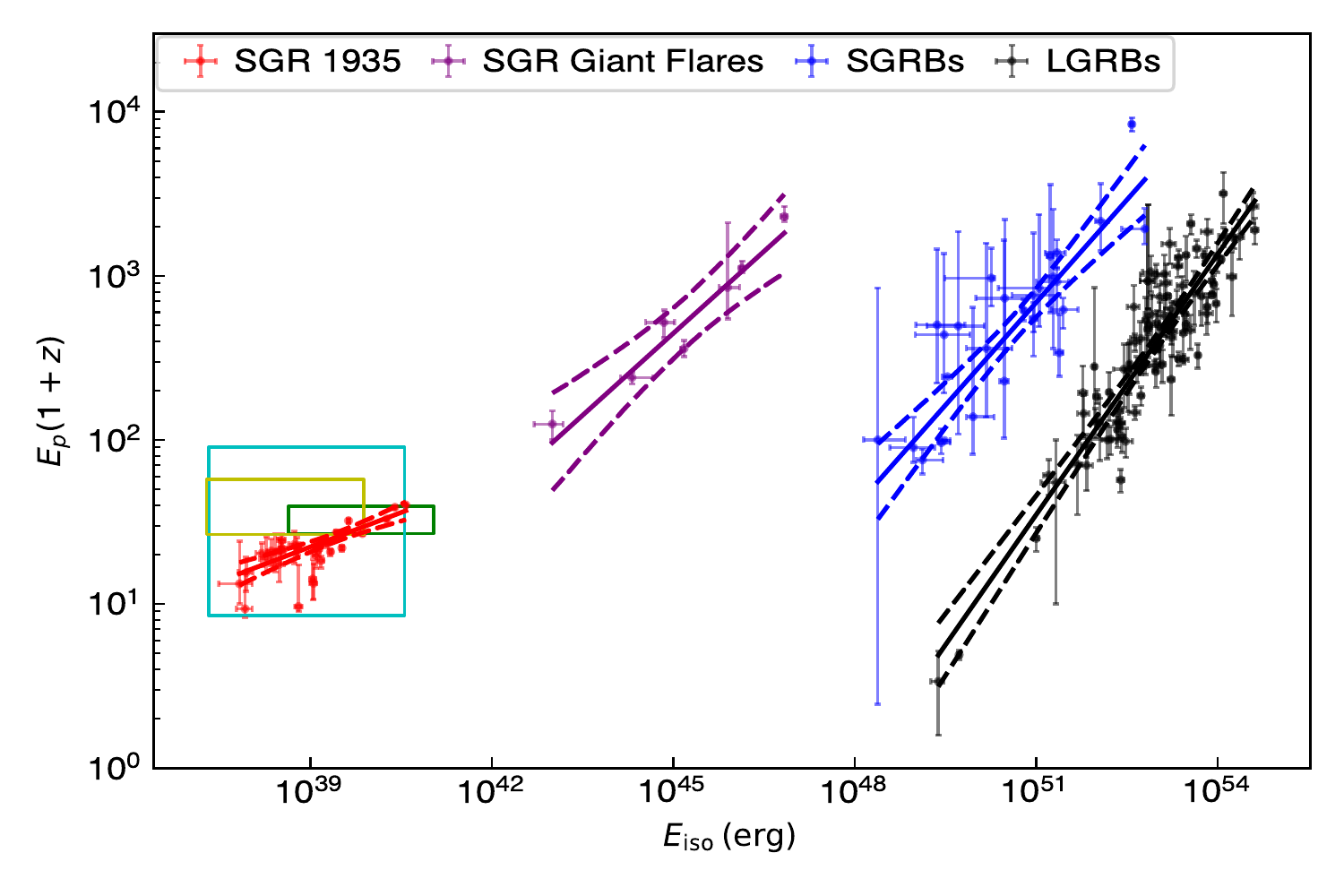}
\caption{$E_p-E_{\rm iso}$ correlation of SGR J1935+2154 bursts in our sample compared with that of long and short GRBs, as well as SGR giant flares \citep{1982Mazets,1999Mazets,2008Mazets,1999Hurley,2005Hurley,2001Aptekar,2006Ofek,2008Ofek,2007aFrederiks,2007bFrederiks,2007Tanaka,2020Yang}. The solid and dashed lines represent the best-fitting result and 95\% confidence intervals of fitting uncertainties for different populations. The green, yellow, and cyan boxes represent the range of $E_p$ and $E_{\rm iso}$ for the previous bursts of SGRs J1935+2154 \citep{2020Lina}, J0501+4516 \citep{2011Lin}, and J1550-5418 \citep{2012Horst}.}
\label{fig:EpEiso}
\end{figure}

We can further investigate correlations between the bursts' energetics properties and their temporal and spectral properties. Such correlations are presented in Figure \ref{fig:cor}. A brighter (more energetic) burst with higher flux (fluence) yields a smaller MTV, suggesting that MTV calculation is subject to a selection effect from observations. We find a tight positive correlation between $E_p$ and flux (fluence) with a slope of $0.20\pm 0.02$ ($0.14\pm 0.02$). Such correlation is consistent with what was previously reported by \citet{2020Lina,2020Linb}, \citeauthor{2020Younes} (\citeyear{2020Younes}; however, c.f., \citealt{2011Lin,2012Horst}). By converting fluence to isotropic energy ($E_{\rm iso}$), we compare the $E_p$-$E_{\rm iso}$ correlation of our sample to that of GRBs and giant flares of SGRs \citep{2020Yang,2020ZhangHM} and find they are located in different tracks and with different slopes (Figure \ref{fig:EpEiso}). We also plot the ranges of $E_p$ and $E_{\rm iso}$ for the previous X-ray bursts of SGRs J1935+2154 \citep{2020Lina}, J0501+4516 \citep{2011Lin} and J1550-5418 \citep{2012Horst} in Figure \ref{fig:EpEiso}, which are consistent with our results. On the other hand, no significant correlation is identified between photon index and flux (fluence) in our sample.

\subsection{Time-resolved Spectral Analysis}\label{subsec:res}

\begin{figure}
\centering
\includegraphics[angle=0,width=0.48\textwidth]{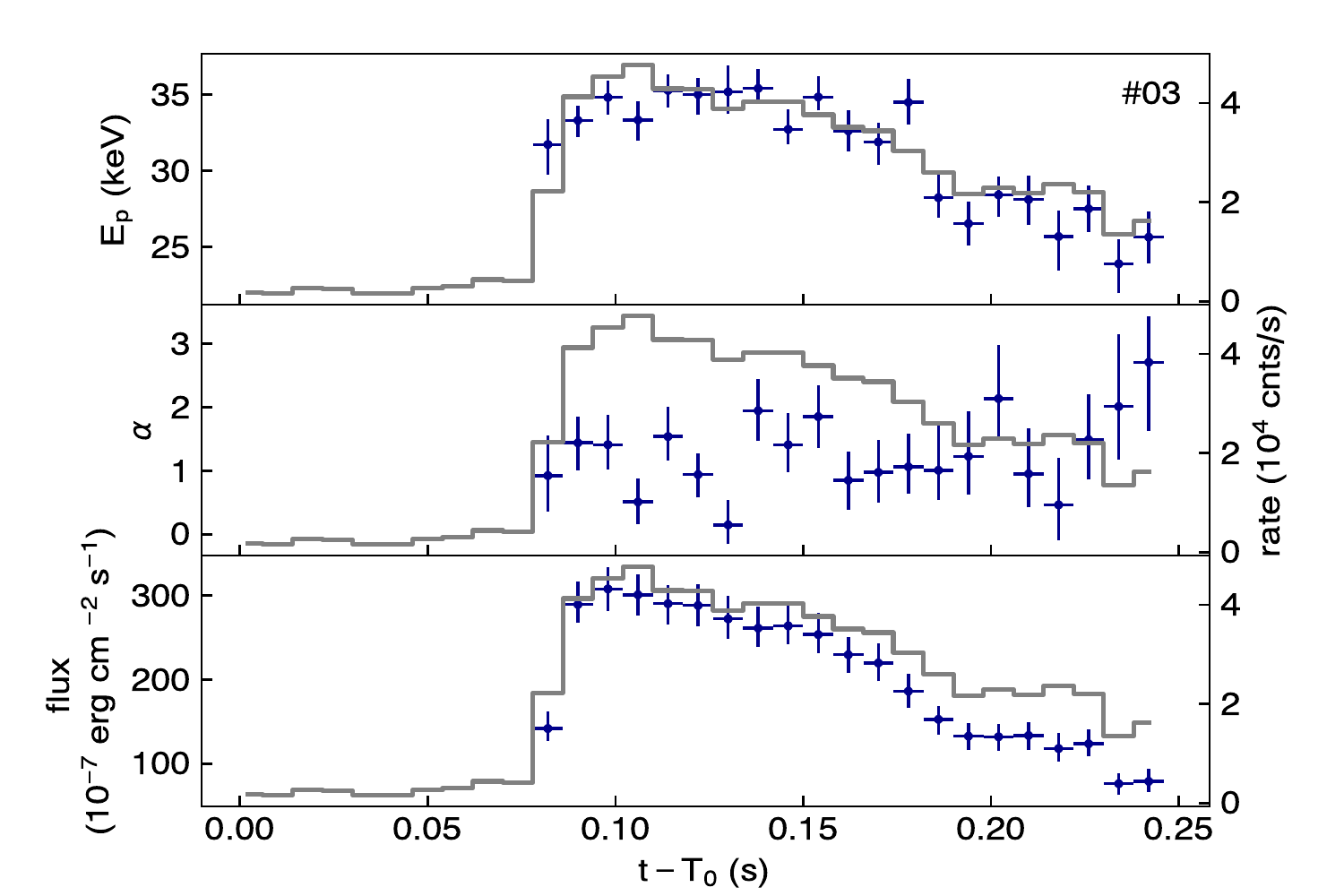}
\includegraphics[angle=0,width=0.48\textwidth]{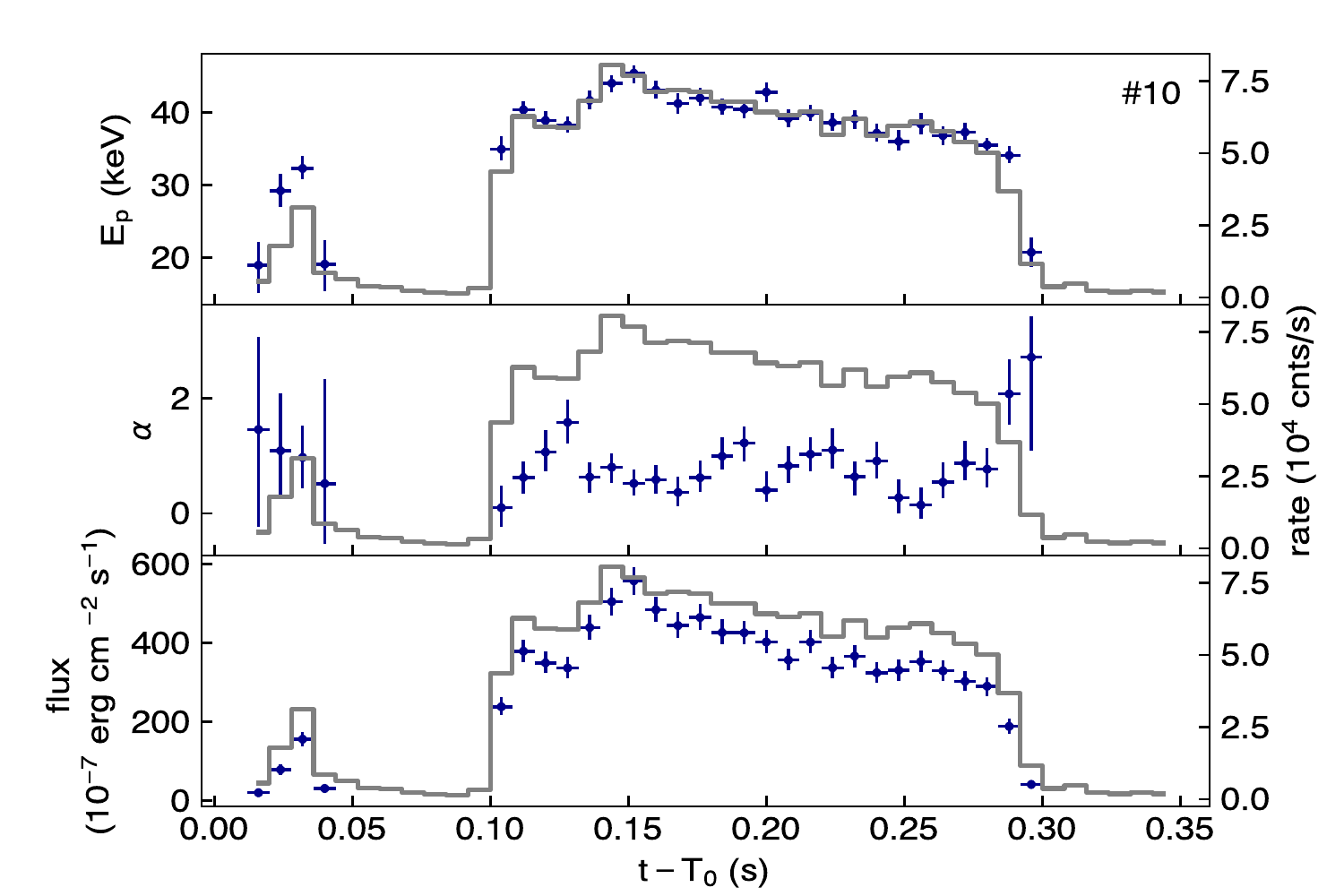}
\includegraphics[angle=0,width=0.48\textwidth]{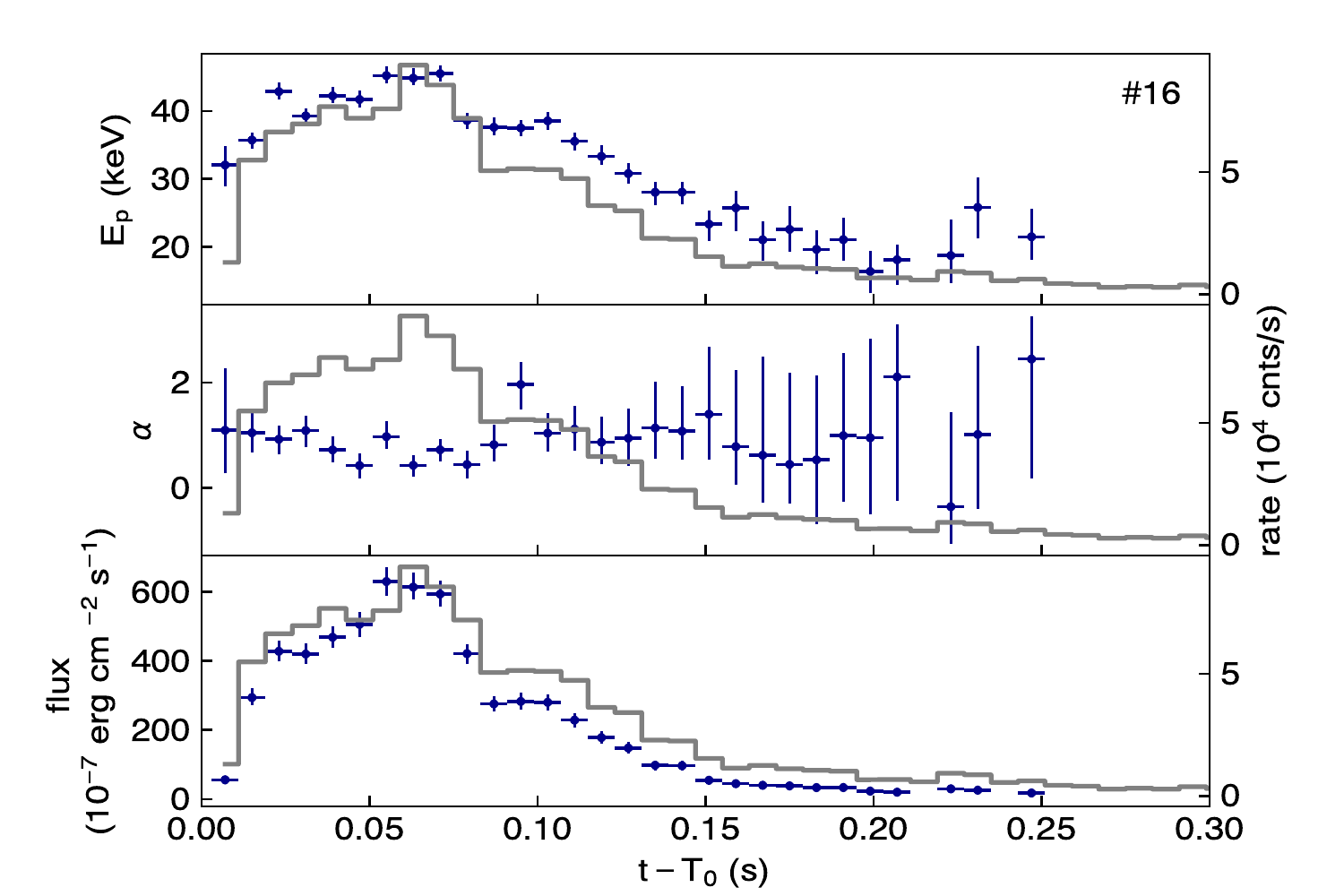}
\caption{Spectral evolution of the top three brightest bursts. The CPL parameters, $E_p$, $\alpha$, and the derived flux are plotted on top of the light curves.}
\label{fig:res_cpl_evolution}
\end{figure}

To perform the time-resolved spectral analysis, we divide the time \tbb of each burst into several segments with 8-ms width, extract spectral files (namely, the source and background spectra and the response matrix), and fit them with spectral models. Since almost all the bursts in our sample can be fitted by the CPL model, we thus only employ the CPL model to do the time-resolved spectral fitting. For some time segments of the weak bursts, CPL parameters are unconstrained due to very few photons. For simplicity, we ignore the spectra of those segments. In this approach, we finally obtain a sample of 157 time-resolved spectra. The time-resolved spectral analysis of the top three brightest bursts (Bursts \#03, \#10, \#16) are shown in Figure \ref{fig:res_cpl_evolution}. A spectral evolution is clearly observed in these cases, where $E_p$ always tracks the flux behavior, and peaks when the flux reaches its peak. Given that the data points provided by the three brightest bursts account for half of the total time-resolved spectra and that they are only special for their light curves, we consider them as a sub-sample in this study.

\begin{figure}
\centering
\includegraphics[angle=0,width=0.48\textwidth]{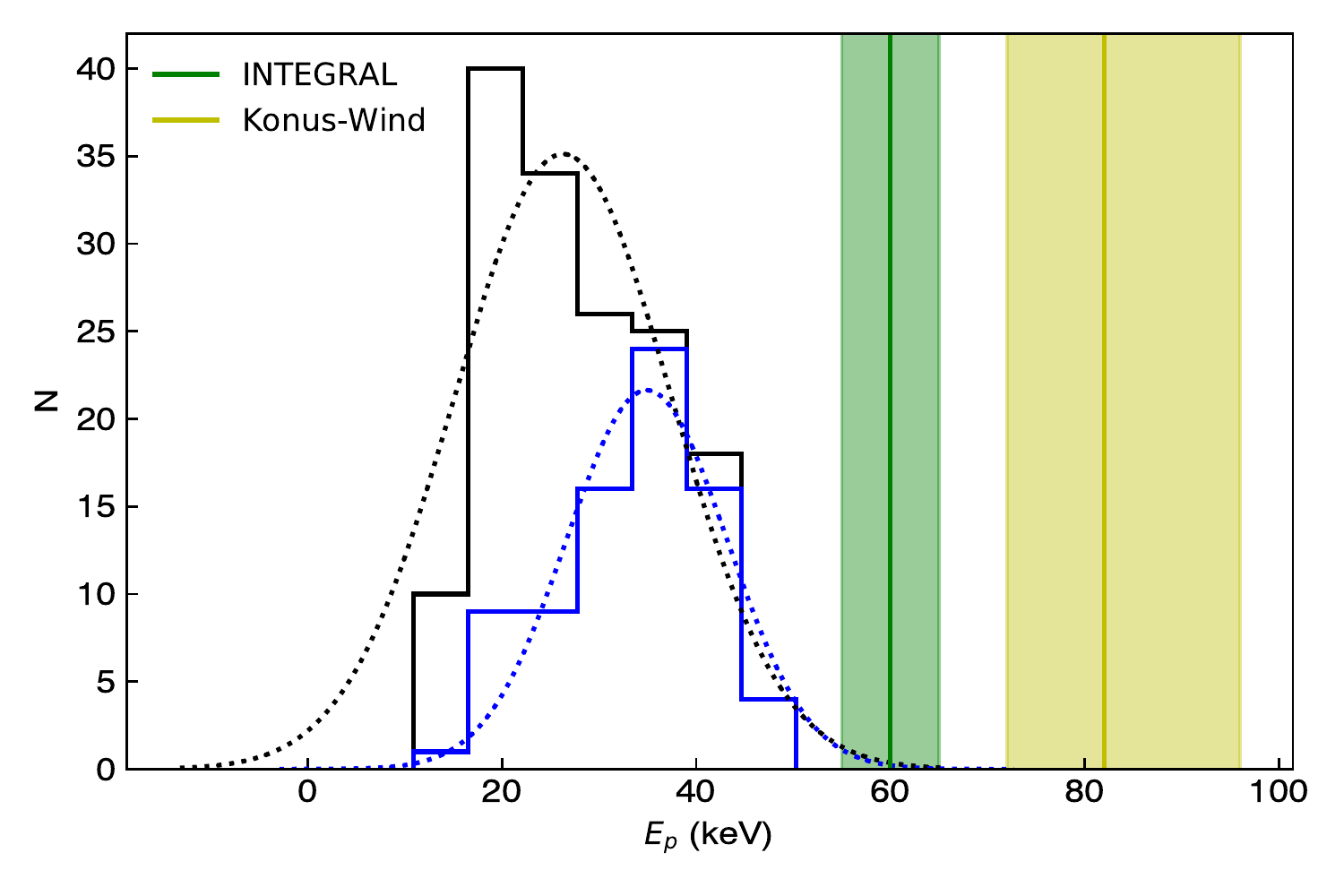}
\includegraphics[angle=0,width=0.48\textwidth]{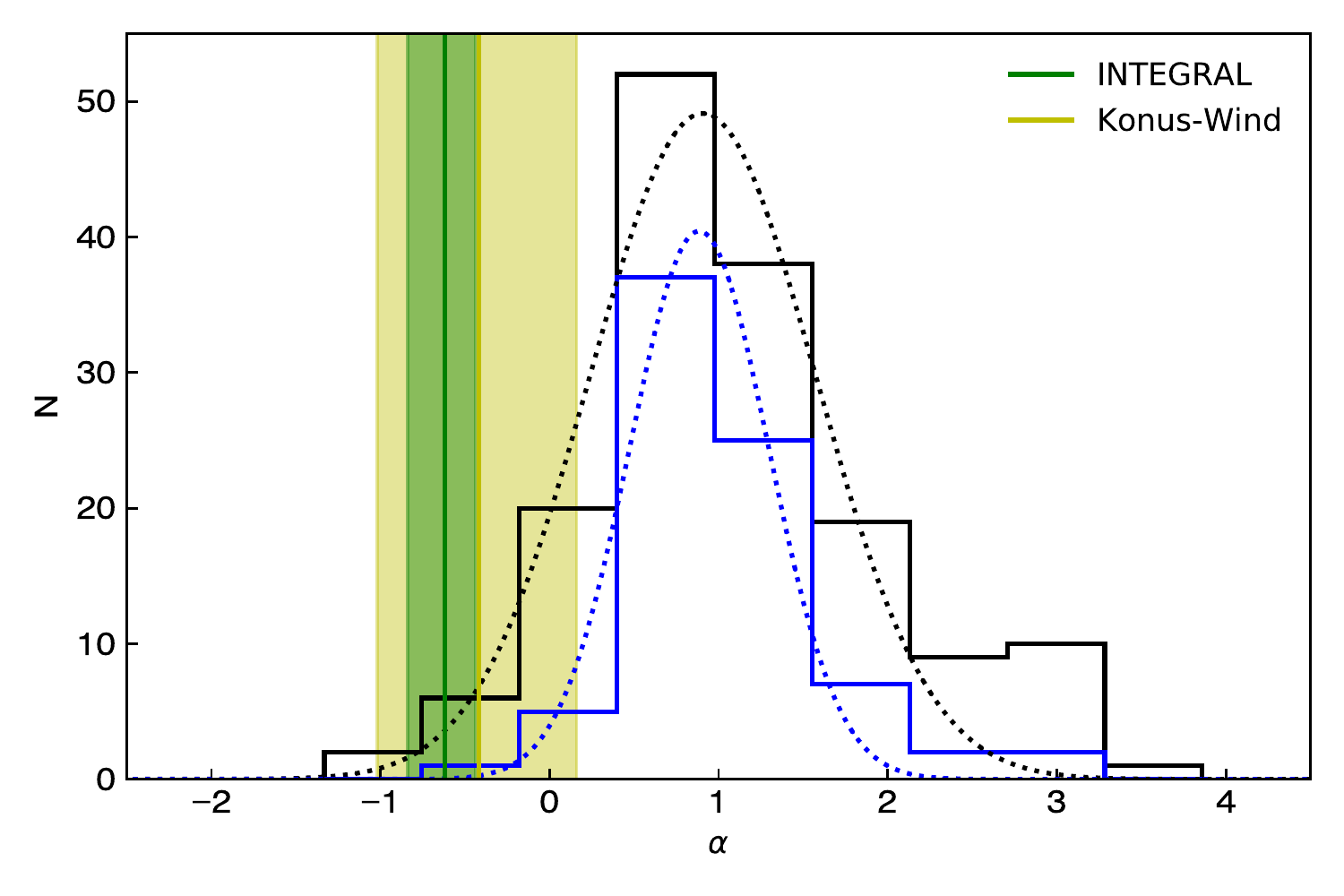}
\includegraphics[angle=0,width=0.48\textwidth]{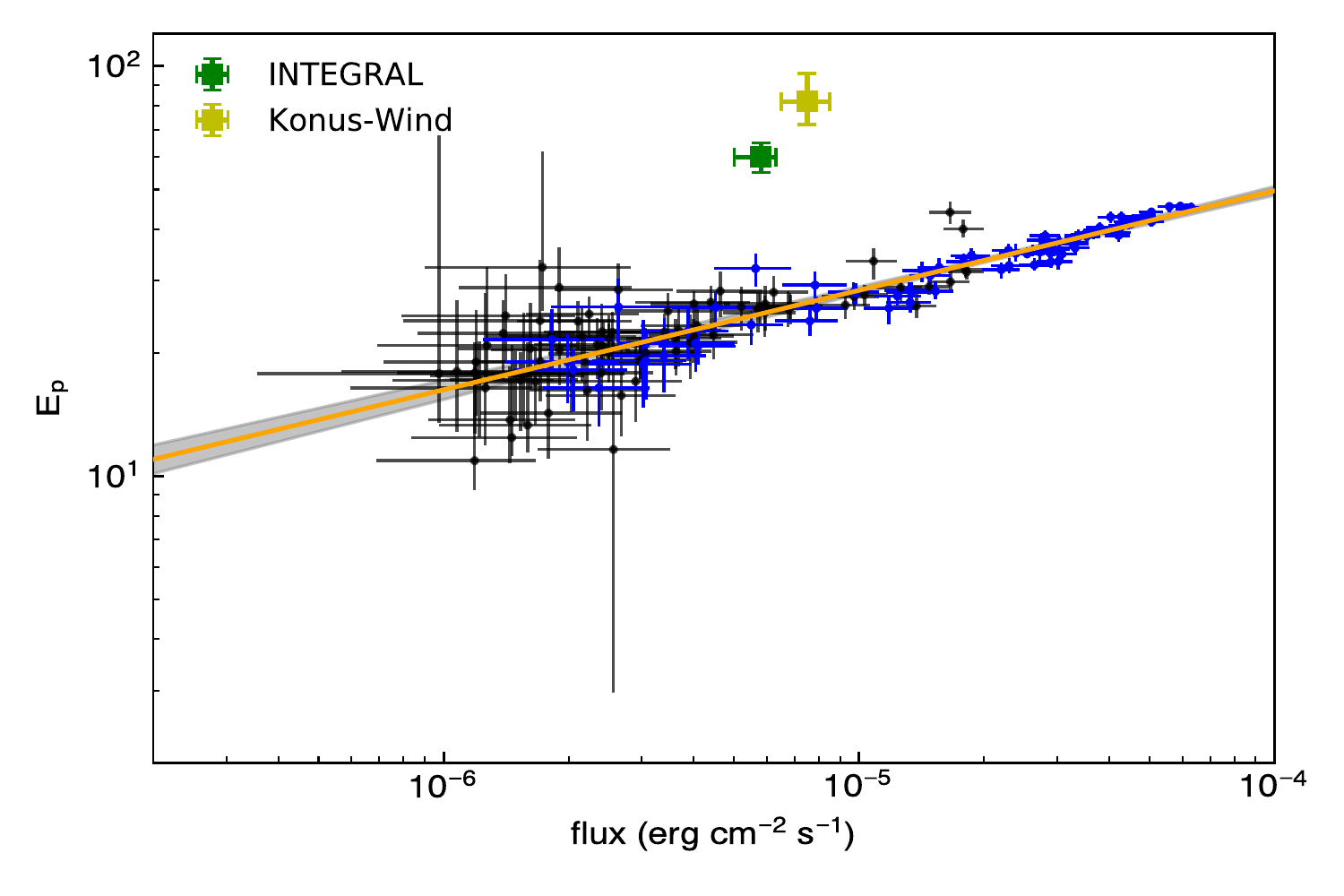}
\includegraphics[angle=0,width=0.48\textwidth]{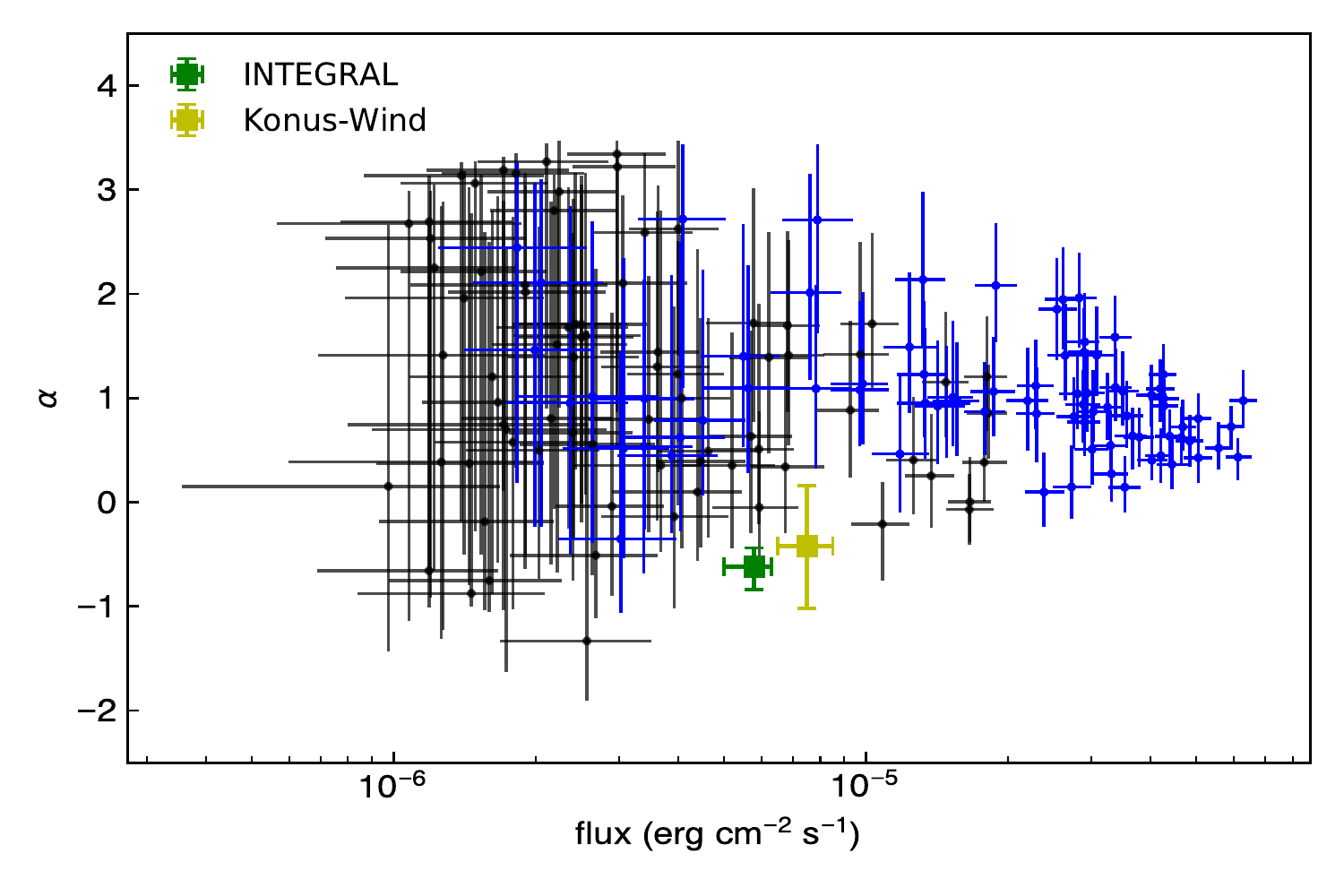}

\caption{Top row: distributions of $E_p$ (left) and photon index (right) of the CPL model fits for resolved spectra. The dotted curves are Gaussian fits to the histograms. Bottom row: the evolution of $E_p$ (left) and photon index (right) as a function of flux. The green/yellow vertical lines, corresponding shadow areas and data points, have the same meaning mentioned in Figure \ref{fig:dis} and \ref{fig:cor} but for spectra with narrower time intervals around the peak of the light curve of the FRB-associated burst. The contribution of three brightest bursts (\#03, 10, 16) are highlighted in blue. The yellow solid line is the best power-law fitting result. The gray area shows the 95\% confidence interval of fitting uncertainties.}
\label{fig:res_cpl}
\end{figure}

The distributions of the spectral fitting results and their correlations are shown in Figure \ref{fig:res_cpl}, where the contribution of the sub-sample of the brightest bursts is highlighted in blue. Both $E_p$ and photon index $\alpha$ for the whole sample (sub-sample) are consistent with Gaussian distributions with peak values of $\sim 26.3$ keV ($34.9$ keV) and $\sim 0.91$ ($0.88$), respectively. Similarly to the integrated spectra, we find no correlation between $\alpha$ and flux but a significant correlation between $E_p$ and flux with a slope of $0.24 \pm 0.01$.

\section{How Special Is the FRB-associated Burst?} \label{sec:special}

As listed in Table \ref{table:FRB-XRB}, the FRB-associated X-ray burst was observed by several facilities in different energy bands and exhibited slightly different properties at different time intervals. We compared those properties with those of the burst in our sample in the following aspects.

\begin{enumerate}

 \item Duration. The duration of the FRB-associated burst is $\sim 0.5-0.6$ s and is longer than 97\% of the bursts in our sample, although it still falls into the 3$\sigma$ region of the duration distribution (Figure \ref{fig:dis}).

 \item Light curve profile. The FRB-associated burst shows some complex features such as multiple spikes, multiple episodes, and a large flux. We found that some bursts (namely, \#1, \#9, \#11, etc.) exhibit similar features in our sample. The FRB-associated burst is not distinctively different from the bursts in our sample in terms of their light curves features.

 \item Spectral properties. As discussed in Section \ref{subsec:int}, most of the bursts in our sample can be well fitted by the CPL model with a typical $\alpha \sim -0.37$ and $E_p \sim 22.4$. Such values are consistent with previous studies, which suggest that SGR bursts are thermal-like. 
In order to compare the spectral properties of our sample with the FRB-associated burst, we overplot the HXMT, INTEGRAL, and Konus-Wind results with our sample in Figure \ref{fig:dis}. The INTEGRAL or Konus-Wind observation provides a high $E_p$ and a soft photon index. However, we note that emission below 10 keV is critical to determine the nonthermal or highly thermalized spectra, particularly the photon index \citep{2012Lin}. As the detectors on board HXMT are most sensitive to low-energy photons, we take the photon index value from the HXMT result, which is softer than those measurements of INTEGRAL and Konus-Wind. By comparing such a photon index as well as the peak energy with those of our 34-burst sample, we conclude that in terms of spectral type, the FRB-associated burst is special with a softer photon index and a harder peak energy.

 \item Energy. As shown in Figure \ref{fig:dis}, the average flux of FRB-associated burst is typical when compared to those of our sample. The fluence is slightly higher but is close to the 1$\sigma$ region. So in terms of the energetics properties, the FRB-associated burst is not unusual.

 \item Spectra-energy correlation. As shown in Figure \ref{fig:cor}, the FRB-associated burst becomes off-track from the correlations mainly due to its higher $E_p$, as well as a softer $\alpha$ obtained with HXMT.

 \item Time-resolved properties and correlation. We used the time-resolved spectral results around the peak of the burst (as listed Table \ref{table:FRB-XRB}) measured by INTEGRAL and Konus-Wind to compare with our 157 time-resolved spectra sample and find that the FRB-associated burst is still a special case, especially in terms of its high $E_p$ (Figure \ref{fig:res_cpl}).

\end{enumerate}

In summary, compared with our FRB-absent sample we find that the FRB-associated burst is longer and more energetic than most other bursts but in any case is consistent with the FRB-absent sample statistically. It distinguishes itself for its nonthermal spectrum and higher spectral peak energy. Unfortunately, we lack of the Fermi/GBM data of the FRB-associated burst because Fermi was occulted by the Earth. Therefore, our result may be subject to instrumental selection effect when we compare the observations from different missions (e.g., HXMT/INTEGRAL/Konus-Wind vs. Fermi).

\section{SGR burst rate and comparison with the FRB burst rate} \label{sec:frb}

\begin{figure}
\centering
\includegraphics[width=0.7\linewidth]{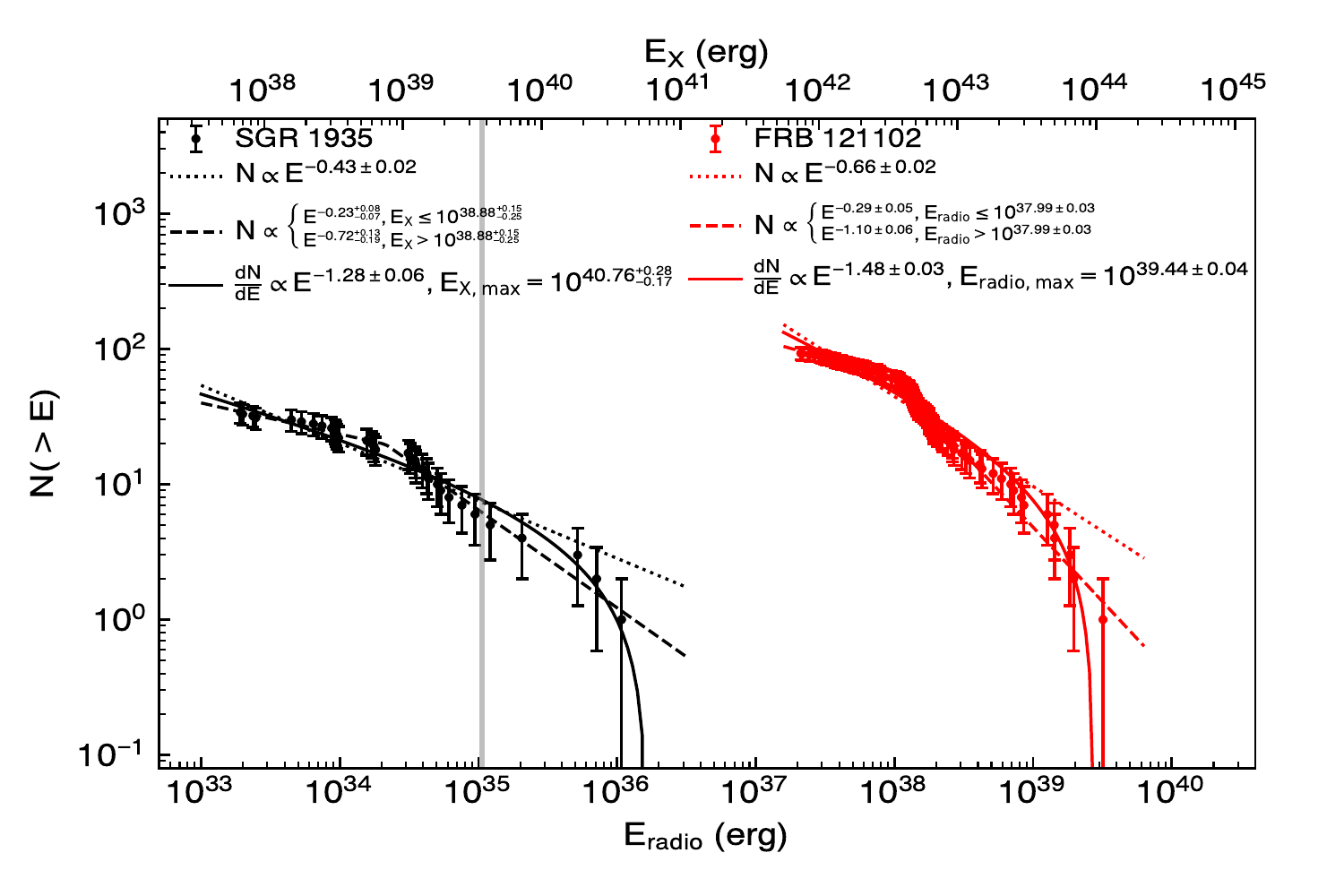}
\caption{Cumulative energy distribution of SGR J1935+2154 bursts and FRB 121102. The radio energy of \sgr and the X-ray (8-200 keV) energy of FRB 121102 are estimated based on the energy ratio in radio and X-ray band of the FRB-associated burst. Different colors represent different sources, \sgr (black) and FRB 121102 (red). Dotted, dashed, and solid lines are the best-fitted model curves, including the PL model, the broken PL model, and the differential PL model with a maximum energy.}
\label{fig:NE}
\end{figure}

In Figure \ref{fig:NE}, we present the cumulative energy distribution of 34 bursts from SGR J1935+2154. We fit the distribution with three models as follows.

\begin{enumerate}

 \item PL model. The best fit index for the PL model is $0.43\pm 0.02$.
 \item Broken PL model. The best-fit index for the lower and higher energies are $-0.23 _{-0.07}^{+0.08}$ and $-0.72 _{-0.19}^{+0.13}$, respectively, with the break energy being $E\sim 10^{38.9}$ erg.
 \item PL model with a maximum energy $E_{\rm max}$ in the form of $N(N>E)\propto (E^{\alpha}-E_{\rm max}^{\alpha})$. The best-fit parameter values are $\alpha =-0.28 \pm 0.06$ and $E_{\rm max} \sim 10^{40.8}$ erg.

\end{enumerate}
Several previous works have shown that magnetar burst energies follow a similar power-law distribution \citep{Gogus1999,Gavriil2004,Scholz2011,Wang2017,2019MNRAS.487.3672Z,Cheng2020,2020Lina}. By comparing the released energy of FRB 200428 and its associated X-ray burst, we calculate the radio-to-X-ray energy ratio as $\eta = E_{\rm radio}/E_{\rm X} \simeq 2.9 \times 10^{-5}$ \citep{2020Bochenek,2020Li}. Assume that such a ratio is typical for FRB-SGR burst associations, we can compare the cumulative energy distribution of SGR bursts with that of radio bursts from repeating FRB sources. In Figure \ref{fig:NE}, we plot the energy (both projected X-ray energies and radio energies) distributions of the X-ray bursts in our sample and the radio bursts from the repeater FRB 121102 obtained from \citet{2018Zhang}\footnote{We only use the C-band (4-8 GHz) data from the Green Bank Telescope observation \citep{2018Zhang}.}. One can see that the bursts from FRB 121102 are projected to be much more energetic than the bursts in our sample.

\begin{figure}
\centering
\includegraphics[angle=0,width=0.7\textwidth]{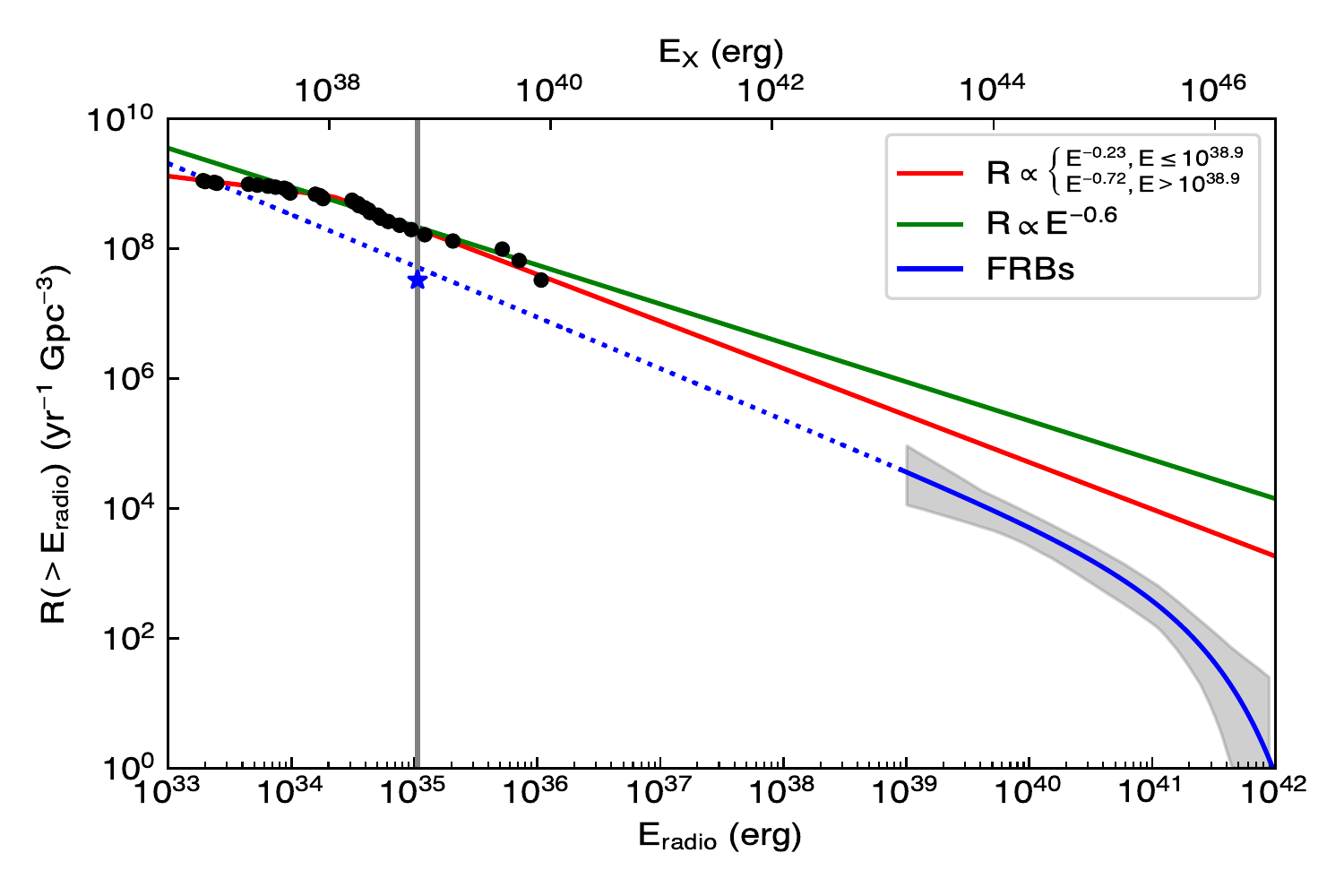}
\caption{Energy-dependent event rate densities of magnetar X-ray bursts and FRBs. Black circles represent the cumulative energy distribution of our sample scaled to the whole magnetar population. The red line shows its best fitted broken PL model. The green line is the rate density of giant flares derived from \citet{2019Beniamini}. The event rate density of FRBs \citep{2020Luo} and its extrapolation is shown by the blue solid and dotted line. The gray shaded area represents the 2$\sigma$ confidence region. The blue star is the event rate density inferred from FRB 200428.}
\label{fig:LF}
\end{figure}

Assuming that all FRBs are produced by magnetars, we investigate the event rate densities of FRBs and X-ray bursts from SGR J1935+2154. Based on an assumption that the burst energy distribution is $-dN/dE \propto E^{-1.6}$~$(E>10^{36})$~erg, \citet{2019Beniamini} found that the rate of the giant flares with energy $>10^{46}$~erg in the Milky Way is $\sim5$~kyr$^{-1}$ by modeling the rate and evolution of magnetars. We adopt this cumulative distribution as the rate of magnetar bursts in our Galaxy. The number density of the galaxy in the universe is 0.006 Mpc$^{-3}$ \citep{2010Mo}. Assuming that all galaxies have the same magnetar population as the Milky Way, the burst rate density of all magnetars can be derived as $5\times 10^{-3}\times (E_{\rm X}/10^{46})^{-0.6}\times 0.006\times 10^9$\,yr$^{-1}$ Gpc $^{-3}$. The magnetar burst rate density becomes $5.6\times10^8$\,yr$^{-1}$ Gpc $^{-3}$ when $E_{\rm X}=10^{38.9}$\,erg. In our sample, there are 17 bursts with $E>10^{38.9}$\,erg and the distribution of our sample suggests $-dN/dE \propto E^{-1.7}$ at higher energies. Therefore, we lift the accumulated energy distribution of \sgr in Figure \ref{fig:NE} by $3.3\times 10^7$ yr$^{-1}$ Gpc$^{-3}$ to approximately estimate the total burst rate density of magnetars (Figure \ref{fig:LF}).

By assuming the luminosity function (LF) of FRBs follows a Schechter function \citep{1976Schechter}, \citet{2020Luo} deduced the event rate density distribution of FRBs as
\begin{equation}
	R_{\rm FRB}(>L)=\phi ^\star \Gamma \left(\alpha +1, \frac{L}{L^\star}\right),\ L\ge 9.1\times 10^{41} {\rm \ erg\ s} ^{-1},
\end{equation}
where $\Gamma$ is the incomplete GAMMA function, the best-fitted parameters are $\phi ^{\star} =339$ Gpc$^{-3}$ yr$^{-1}$, $\alpha =-1.79$ and $L^{\star} = 2.9 \times 10^{44}$ erg s$^{-1}$. We simply assume that the average duration of FRBs is $\sim$1 ms to deduce the relation $R_{\rm FRB}(>E)$. We also plot the FRB LF curve and its extrapolation in Figure \ref{fig:LF}. Because one FRB, i.e. FRB 200428, was detected at $E_{\rm radio} = 10^{35}$ erg by STARE-2 during $\sim$ one year operation, we simply use the factor $3.3\times 10^7$ yr$^{-1}$ Gpc $^{-3}$ to obtain the all-sky event rate density of FRB, which is denoted as the blue star  adjacent to the blue dotted line in Figure \ref{fig:LF}. It is located at the $2\sigma$ confidence region of FRB event rate density derived from \citet{2020Luo}. Comparing this point with the estimated SGR burst event rate density curve, one can see that the difference between the FRB and magnetar burst event rate is about a factor of 150, i.e. every $\sim 150$ SGR burst would have one FRB associated. This is consistent with the results obtained in \cite{2020Linc} and \cite{2020Lu}.

\section{Summary} \label{sec:sum}

We systematically analyzed the FRB-absent bursts of SGR J1935+2154 just hours before the FRB 200428 event. After a comprehensive investigation of the burst's temporal and spectral properties, we find that the FRB-associated X-ray burst observed by HXMT only distinguishes itself in terms of its nonthermal $\alpha$ and spectral peak energy, but is otherwise consistent with the burst population. Since the FRB-associated burst was not detected by Fermi/GBM, potential instrumental selection effects may also play a role in the apparent differences. Future complete samples of FRB-associated and FRB-absent X-ray bursts from Galactic magnetars are needed to determine whether the FRB-associated bursts are truly atypical.

\citet{2020Younes} performed a similar analysis by focusing on the broadband (1-300 keV) time-integrated spectral fitting of the joint sample of NICER and GBM observations of 24 bursts. Our sample selection is based on the joint time coverage by both GBM and FAST, which provides 34 bursts (with a different identification method as discussed in Section \ref{sec:obs}) guaranteed to be not associated with any pulsed radio emission. Furthermore, we performed a time-resolved spectral analysis on most bursts with the finest time resolution allowed by statistics. Consequently, our results yield a slightly softer photon index, although the overall distributions of both the peak energy and photon index of our sample are consistent with those in \citet{2020Younes}. In addition to this, the comparison between FRB-associated and FRB-absent bursts in both approaches are in agreement.

We also compare the cumulative energy distribution of our burst sample with that of the FRB burst sample of FRB 121102. We further compare the event rate density of the X-ray bursts with the event rate density of FRBs assuming that all FRBs originate from magnetars. Using the FRB 200428 and its associated X-ray burst as a calibrator, we found that the event rate density of FRBs is lower than the event rate density of magnetar bursts by a factor of $\sim 150$, suggesting that only a small fraction of magnetar bursts can produce FRBs. This strengthens the observational and theoretical evidence of such a discrepancy, as discussed earlier \citep{2020Linc,2020Lu}. As discussed by \cite{2020Linc}, there could be three possibilities for such a discrepancy: either beaming or narrow spectra of FRB emission with most outside the GHz band, or the uniqueness of the FRB-associated X-ray burst. This Letter suggests that the last possibility may not be the sole reason for the discrepancy. Since the second option (narrow spectra) is very contrived \citep{2020Linc}, our result suggests that FRB beaming remains an attractive possibility to account for missing FRBs in the majority of SGR bursts.

\acknowledgments
We acknowledge support by the Fundamental Research Funds for the Central Universities (14380035). This work is supported by National Key Research and Development Programs of China (2018YFA0404204), the National Natural Science Foundation of China (grant Nos. 11833003, U1831207 and 11703002), and the Program for Innovative Talents, Entrepreneur in Jiangsu.

%%%%%%%%%%%%%%%%%%%%%%%%%%%%%%%%%%%
\clearpage
\begin{table}
\centering
\caption{Properties of the FRB 200428-associated X-Ray Burst}
\label{table:FRB-XRB}
\begin{center}
\begin{tabular}{ccccccc}
\hline
\hline
Instrument & Energy Band & Burst Duration &  \multicolumn{4}{c}{Spectral Fitting} \\
&&&Time Interval\footnote{The times are relative to $T_0=$ 2020 April 28 14:34:24.0 UTC, when the burst triggered HXMT and INTEGRAL.}& $E_p$ & $\alpha$ & Flux\footnote{The fluxes are calculated in the energy ranges listed in the second column, respectively.}\\
 & (keV) & (ms) & (ms) & (keV) & & ($10^{-7}$ erg cm$^{-2}$ s$^{-1}$) \\
\hline
HXMT\footnote{Results from \citet{2020Li}.} & 1-250 & 530 & [-200, 1000] & $36.9\pm6.2$ & $-1.56\pm0.06$ & $5.95^{+0.34}_{-0.32}$ \\
\hline
\multirow{2}{*}{INTEGRAL}\footnote{Results from \citet{2020Mereghetti}.} & \multirow{2}{*}{20-200} & \multirow{2}{*}{600} & [190, 790] & $65\pm5$ & $-0.7^{+0.2}_{-0.4}$ & $10.2\pm0.5$ \\
& & & [395, 536]\footnote{The main pulse contains three peaks in the light curve.} & $60\pm5$ & $-0.62^{+0.18}_{-0.22}$ & $58^{+5}_{-8}$\footnote{Peak flux on a 10 ms scale around the second peak of the light curve.}\\
\hline
\multirow{2}{*}{Konus-Wind}\footnote{Results from \citet{2020Ridnaia}.} & \multirow{2}{*}{20-500} & \multirow{2}{*}{484} & [436, 692]\footnote{The first peak of the light curve is not included.} & $85^{+15}_{-10}$ & $-0.72^{+0.47}_{-0.46}$ & $20.0\pm2.3$\footnote{Calculated from the fluence $9.7\pm1.1\times10^{-7}$ erg cm$^{-2}$, which is measured over the spectral interval and scaled to the 484 ms interval, i.e., burst duration.} \\
& & & [436, 500] & $82^{+14}_{-10}$ & $-0.42^{+0.58}_{-0.60}$ & $75\pm10$\footnote{Peak flux on the 16 ms interval corresponding to the second peak in the light curve.} \\   
\hline
\end{tabular}
\end{center}
\leftline{\bf{Notes.}}
\end{table}

\newpage
\begin{table}
\centering
\caption{Burst Time, Duration, and Minimum Time Variability (MTV) of Each SGR J1935+2154 Burst}
\label{table:time}
\begin{center}
\begin{tabular}{c c c c c}%
\hline%
\hline%
ID&Burst Time (UTC 2020 April 28))&$T_{\rm bb}$ (s)&$T_{\rm 90}$ (s)&MTV (s)\\%
\hline%
1&00:19:44.192&0.138&0.080$_{-0.017}^{+0.016}$&0.020\\%
2&00:23:04.728&0.028&0.021$_{-0.007}^{+0.053}$&0.028\\%
3&00:24:30.296&0.252&0.122$_{-0.001}^{+0.002}$&0.004\\%
4&00:25:43.945&0.054&0.076$_{-0.022}^{+0.085}$&0.054\\%
5&00:37:36.153&0.108&0.095$_{-0.019}^{+0.085}$&0.008\\%
6&00:39:39.513&0.244&0.194$_{-0.037}^{+0.062}$&0.098\\%
7&00:40:33.072&0.228&0.190$_{-0.006}^{+0.005}$&0.008\\%
8&00:41:32.136&0.390&0.222$_{-0.008}^{+0.008}$&0.016\\%
9&00:43:25.169&0.374&0.174$_{-0.006}^{+0.004}$&0.008\\%
10&00:44:08.202&0.340&0.154$_{-0.001}^{+0.001}$&0.002\\%
11&00:44:09.302&0.156&0.112$_{-0.001}^{+0.003}$&0.004\\%
12&00:45:31.097&0.042&0.030$_{-0.004}^{+0.004}$&0.004\\%
13&00:46:00.009&0.312&0.208$_{-0.019}^{+0.021}$&0.012\\%
14&00:46:00.609&0.220&0.126$_{-0.006}^{+0.007}$&0.012\\%
15&00:46:06.408&0.040&0.019$_{-0.006}^{+0.009}$&0.010\\%
16&00:46:20.176&0.854&0.166$_{-0.005}^{+0.005}$&0.002\\%
17&00:46:23.504&0.842&0.742$_{-0.015}^{+0.017}$&0.122\\%
18&00:46:43.208&0.226&0.128$_{-0.032}^{+0.018}$&0.010\\%
19&00:47:24.961&0.206&0.152$_{-0.014}^{+0.024}$&0.172\\%
20&00:47:57.528&0.104&0.084$_{-0.014}^{+0.005}$&0.038\\%
21&00:48:44.824&0.538&0.382$_{-0.024}^{+0.020}$&0.048\\%
22&00:48:49.272&0.302&0.112$_{-0.011}^{+0.018}$&0.006\\%
23&00:49:00.273&0.154&0.120$_{-0.017}^{+0.030}$&0.154\\%
24&00:49:01.121&0.186&0.151$_{-0.006}^{+0.010}$&0.186\\%
25&00:49:01.936&0.306&0.181$_{-0.043}^{+0.045}$&0.088\\%
26&00:49:06.472&0.026&0.022$_{-0.006}^{+0.026}$&0.026\\%
27&00:49:16.592&0.312&0.234$_{-0.004}^{+0.003}$&0.012\\%
28&00:49:22.392&0.124&0.078$_{-0.006}^{+0.013}$&0.060\\%
29&00:49:27.280&0.090&0.082$_{-0.017}^{+0.073}$&0.034\\%
30&00:49:46.142&0.046&0.036$_{-0.011}^{+0.034}$&0.046\\%
31&00:49:46.680&0.368&0.150$_{-0.014}^{+0.022}$&0.074\\%
32&00:50:01.012&0.080&0.047$_{-0.016}^{+0.025}$&0.012\\%
33&00:50:01.358&0.156&0.095$_{-0.006}^{+0.006}$&0.012\\%
34&00:50:21.969&0.022&0.019$_{-0.010}^{+0.026}$&0.022\\%
\hline
\end{tabular}
\end{center}
\end{table}

\clearpage
\begin{sidewaystable}\tiny
\setlength{\tabcolsep}{2.2pt}
\centering
\caption{Time-integrated Spectral Fit Parameters and Fluence of Each SGR J1935+2154 Burst}
\label{table:int_spectra}
\begin{center}
\begin{tabular}{c c c c c c c c c c c c c c c c c c c c}%
\hline%
\hline%
\multirow{3}{*}{ID}&\multicolumn{3}{c}{BB}&\multicolumn{4}{c}{BB+BB}&\multicolumn{3}{c}{PL}&\multicolumn{4}{c}{CPL}&\multicolumn{4}{c}{BB+PL}&Fluence\\%
 &kT&\multirow{2}{*}{BIC}&\multirow{2}{*}{$\frac{\rm PGSTAT}{\rm dof}$}&kT$_1$&kT$_2$&\multirow{2}{*}{BIC}&\multirow{2}{*}{$\frac{\rm PGSTAT}{\rm dof}$}&\multirow{2}{*}{$\alpha$}&\multirow{2}{*}{BIC}&\multirow{2}{*}{$\frac{\rm PGSTAT}{\rm dof}$}&\multirow{2}{*}{$alpha$}&E$_{\rm p}$&\multirow{2}{*}{BIC}&\multirow{2}{*}{$\frac{\rm PGSTAT}{\rm dof}$}&\multirow{2}{*}{$\alpha$}&kT&\multirow{2}{*}{BIC}&\multirow{2}{*}{$\frac{\rm PGSTAT}{\rm dof}$}&\multirow{2}{*}{$\left( \begin{array}{c} 10^{-8}\\ {\rm \ erg} {\rm \ cm}^{-2}\end{array} \right)$}\\%
 &(keV)&&&(keV)&(keV)&&&&&&&(keV)&&&&(keV)&&&\\%
\hline%
\hline%
1&$\mathbf{6.41_{-0.70}^{+0.51}}$\footnote{The fitting results of best models are marked in bold.}&\textbf{124.6}&\textbf{114.5/152}&...&...&...&...&$-2.63_{-0.13}^{+0.12}$&139.1&129.1/152&$\mathbf{-0.43_{-0.55}^{+0.77}}$&$\mathbf{22.46_{-3.43}^{+2.89}}$&\textbf{122.8}&\textbf{107.7/151}&...&...&...&...&$11.42_{-1.72}^{+2.00}$\\%
2&$\mathbf{5.20_{-0.61}^{+0.96}}$&\textbf{53.4}&\textbf{43.4/153}&...&...&...&...&$-2.79_{-0.30}^{+0.19}$&63.2&53.1/153&$1.26_{-1.51}^{+1.74}$&$20.35_{-3.53}^{+3.19}$&58.5&43.4/152&...&...&...&...&$3.01_{-0.61}^{+0.66}$\\%
3&...&...&...&$6.09_{-0.46}^{+0.46}$&$10.57_{-0.47}^{+0.57}$&182.8&162.7/151&...&...&...&$\mathbf{0.71_{-0.09}^{+0.10}}$&$\mathbf{32.94_{-0.35}^{+0.24}}$&\textbf{179.7}&\textbf{164.6/152}&...&...&...&...&$\mathbf{348.79_{-7.21}^{+6.89}}$\footnote{The fluences of three top brightest bursts are bolded.}\\%
4&$\mathbf{3.63_{-0.67}^{+3.87}}$&\textbf{86.7}&\textbf{76.6/153}&...&...&...&...&$\mathbf{-3.27_{-1.27}^{+0.53}}$&\textbf{88.1}&\textbf{78.0/153}&$0.86_{-1.64}^{+2.75}$&$13.27_{-3.27}^{+10.89}$&91.7&76.6/152&...&...&...&...&$1.29_{-0.70}^{+0.81}$\\%
5&$\mathbf{6.68_{-1.27}^{+0.91}}$&\textbf{91.9}&\textbf{81.9/152}&...&...&...&...&$-2.84_{-0.27}^{+0.17}$&96.2&86.1/152&$\mathbf{-0.84_{-0.40}^{+1.30}}$&$\mathbf{20.59_{-4.91}^{+6.29}}$&\textbf{92.2}&\textbf{77.1/151}&...&...&...&...&$5.00_{-1.06}^{+1.14}$\\%
6&$4.83_{-0.30}^{+0.53}$&145.6&135.6/152&$3.61_{-0.39}^{+0.42}$&$10.51_{-1.62}^{+2.18}$&126.5&106.4/150&$-2.66_{-0.11}^{+0.09}$&133.5&123.4/152&$\mathbf{-1.40_{-0.24}^{+0.49}}$&$\mathbf{14.18_{-3.60}^{+4.46}}$&\textbf{123.3}&\textbf{108.2/151}&$-2.79_{-0.34}^{+0.18}$&$6.38_{-0.95}^{+2.02}$&132.0&111.8/150&$20.94_{-2.13}^{+2.61}$\\%
7&$6.38_{-0.24}^{+0.26}$&164.2&154.1/152&$\mathbf{4.48_{-0.21}^{+0.45}}$&$\mathbf{11.58_{-1.19}^{+2.12}}$&\textbf{132.3}&\textbf{112.1/150}&...&...&...&$-0.42_{-0.31}^{+0.28}$&$23.47_{-1.48}^{+1.22}$&136.3&121.2/151&$-2.41_{-0.18}^{+0.20}$&$6.36_{-0.45}^{+0.49}$&146.4&126.3/150&$35.88_{-2.83}^{+2.89}$\\%
8&...&...&...&$4.78_{-0.17}^{+0.32}$&$11.04_{-0.42}^{+0.91}$&141.5&121.4/150&...&...&...&$\mathbf{-0.32_{-0.12}^{+0.14}}$&$\mathbf{26.87_{-0.65}^{+0.63}}$&\textbf{138.7}&\textbf{123.5/151}&$-2.57_{-0.10}^{+0.07}$&$7.69_{-0.17}^{+0.22}$&159.8&139.7/150&$138.49_{-4.48}^{+4.55}$\\%
9&...&...&...&$\mathbf{4.57_{-0.22}^{+0.34}}$&$\mathbf{10.17_{-1.04}^{+1.82}}$&\textbf{159.3}&\textbf{139.2/150}&...&...&...&$-0.02_{-0.23}^{+0.28}$&$21.93_{-0.86}^{+0.80}$&163.9&148.8/151&$-2.59_{-0.24}^{+0.18}$&$5.92_{-0.16}^{+0.39}$&178.3&158.1/150&$63.51_{-3.59}^{+3.97}$\\%
10&...&...&...&$\mathbf{7.22_{-0.28}^{+0.28}}$&$\mathbf{13.96_{-0.43}^{+0.40}}$&\textbf{306.2}&\textbf{286.1/150}&...&...&...&$0.40_{-0.05}^{+0.06}$&$40.13_{-0.25}^{+0.26}$&308.8&293.8/149&...&...&...&...&$\mathbf{723.51_{-11.81}^{+12.49}}$\\%
11&$7.32_{-0.19}^{+0.17}$&163.9&153.8/152&$4.53_{-0.57}^{+0.54}$&$9.84_{-0.87}^{+1.03}$&134.3&114.1/150&...&...&...&$\mathbf{-0.08_{-0.20}^{+0.26}}$&$\mathbf{27.12_{-0.99}^{+0.93}}$&\textbf{129.2}&\textbf{114.1/151}&$-2.92_{-0.29}^{+0.17}$&$7.87_{-0.26}^{+0.25}$&143.0&122.9/150&$51.07_{-2.70}^{+2.92}$\\%
12&$\mathbf{6.51_{-0.54}^{+0.52}}$&\textbf{87.1}&\textbf{77.1/153}&...&...&...&...&$-2.58_{-0.16}^{+0.11}$&113.6&103.5/153&$0.66_{-0.71}^{+0.92}$&$24.53_{-2.40}^{+2.10}$&91.5&76.3/152&...&...&...&...&$6.41_{-0.91}^{+0.98}$\\%
13&$5.95_{-0.34}^{+0.39}$&137.1&127.0/152&$4.17_{-0.33}^{+1.09}$&$9.88_{-1.06}^{+9.90}$&129.4&109.2/150&$-2.46_{-0.08}^{+0.07}$&151.7&141.6/152&$\mathbf{-0.59_{-0.42}^{+0.45}}$&$\mathbf{22.13_{-2.25}^{+1.73}}$&\textbf{124.5}&\textbf{109.3/151}&$-2.42_{-0.43}^{+0.16}$&$6.15_{-0.42}^{+0.93}$&129.7&109.6/150&$23.54_{-2.33}^{+2.87}$\\%
14&$6.33_{-0.28}^{+0.27}$&133.4&123.3/152&$\mathbf{4.76_{-0.45}^{+0.44}}$&$\mathbf{11.31_{-1.73}^{+1.99}}$&\textbf{119.4}&\textbf{99.2/150}&...&...&...&$\mathbf{-0.25_{-0.32}^{+0.38}}$&$\mathbf{23.95_{-1.44}^{+1.28}}$&\textbf{117.6}&\textbf{102.5/151}&...&...&...&...&$29.30_{-2.30}^{+2.68}$\\%
15&$\mathbf{6.07_{-0.79}^{+0.68}}$&\textbf{82.2}&\textbf{72.2/152}&...&...&...&...&$-2.72_{-0.21}^{+0.17}$&89.2&79.1/152&$\mathbf{-0.42_{-0.54}^{+1.30}}$&$\mathbf{21.02_{-3.97}^{+3.87}}$&\textbf{83.8}&\textbf{68.7/151}&...&...&...&...&$4.34_{-0.75}^{+0.85}$\\%
16&...&...&...&$6.42_{-0.23}^{+0.34}$&$13.53_{-0.36}^{+0.54}$&200.1&179.9/151&...&...&...&$\mathbf{0.19_{-0.06}^{+0.07}}$&$\mathbf{38.83_{-0.35}^{+0.34}}$&\textbf{197.4}&\textbf{182.2/152}&...&...&...&...&$\mathbf{478.40_{-9.20}^{+8.67}}$\\%
17&$\mathbf{4.72_{-0.22}^{+0.31}}$&\textbf{119.1}&\textbf{109.0/153}&$\mathbf{4.18_{-0.14}^{+0.49}}$&$\mathbf{12.44_{-2.73}^{+5.07}}$&\textbf{119.2}&\textbf{99.0/151}&$-2.57_{-0.08}^{+0.08}$&142.9&132.8/153&$\mathbf{-0.27_{-0.48}^{+0.86}}$&$\mathbf{18.36_{-1.83}^{+1.22}}$&\textbf{118.7}&\textbf{103.5/152}&...&...&...&...&$28.56_{-3.76}^{+3.67}$\\%
18&$6.37_{-0.23}^{+0.20}$&157.4&147.3/153&$4.82_{-0.28}^{+0.77}$&$10.27_{-1.27}^{+4.58}$&149.4&129.2/151&...&...&...&$\mathbf{0.14_{-0.34}^{+0.36}}$&$\mathbf{23.99_{-0.99}^{+1.16}}$&\textbf{146.5}&\textbf{131.4/152}&$-2.52_{-0.60}^{+0.21}$&$6.58_{-0.36}^{+0.32}$&153.4&133.2/151&$34.03_{-2.48}^{+2.56}$\\%
19&$\mathbf{5.48_{-0.55}^{+0.87}}$&\textbf{101.4}&\textbf{91.3/153}&...&...&...&...&$-2.57_{-0.21}^{+0.14}$&109.5&99.4/153&$-0.32_{-0.73}^{+1.45}$&$21.24_{-4.00}^{+3.39}$&104.1&89.0/152&...&...&...&...&$6.20_{-1.00}^{+1.20}$\\%
20&$\mathbf{5.58_{-0.33}^{+0.37}}$&\textbf{101.8}&\textbf{91.8/152}&...&...&...&...&$-2.63_{-0.11}^{+0.08}$&147.1&137.0/152&$0.56_{-0.59}^{+0.82}$&$21.63_{-1.80}^{+1.13}$&105.0&89.9/151&...&...&...&...&$11.87_{-1.11}^{+1.20}$\\%
21&$5.37_{-0.31}^{+0.45}$&125.5&115.4/153&$4.65_{-0.25}^{+0.55}$&$17.67_{-4.20}^{+8.40}$&120.7&100.5/151&$-2.43_{-0.09}^{+0.08}$&124.7&114.6/153&$\mathbf{-0.95_{-0.45}^{+0.63}}$&$\mathbf{19.79_{-3.34}^{+2.65}}$&\textbf{117.4}&\textbf{102.2/152}&$\mathbf{-2.23_{-0.34}^{+0.24}}$&$\mathbf{5.34_{-0.41}^{+1.07}}$&\textbf{117.2}&\textbf{97.1/151}&$23.82_{-3.56}^{+4.10}$\\%
22&...&...&...&$\mathbf{5.13_{-0.18}^{+0.31}}$&$\mathbf{15.15_{-0.83}^{+0.84}}$&\textbf{160.3}&\textbf{140.2/150}&...&...&...&$-0.77_{-0.14}^{+0.13}$&$32.17_{-1.29}^{+1.04}$&168.5&153.4/151&...&...&...&...&$81.81_{-3.91}^{+4.63}$\\%
23&$\mathbf{4.68_{-0.50}^{+1.62}}$&\textbf{87.7}&\textbf{77.6/152}&...&...&...&...&$\mathbf{-2.62_{-0.30}^{+0.17}}$&\textbf{86.8}&\textbf{76.8/152}&$\mathbf{-1.24_{-0.23}^{+1.48}}$&$\mathbf{17.46_{-3.82}^{+7.53}}$&\textbf{87.6}&\textbf{72.5/151}&...&...&...&...&$5.88_{-1.23}^{+1.20}$\\%
24&$4.17_{-0.29}^{+0.44}$&145.8&135.7/152&$3.41_{-0.30}^{+0.73}$&$10.44_{-2.35}^{+5.45}$&146.3&126.1/150&$\mathbf{-2.66_{-0.17}^{+0.13}}$&\textbf{141.6}&\textbf{131.5/152}&$\mathbf{-1.62_{-0.05}^{+1.06}}$&$\mathbf{9.64_{-0.65}^{+7.69}}$&\textbf{142.1}&\textbf{127.0/151}&...&...&...&...&$12.17_{-1.53}^{+1.68}$\\%
25&$5.55_{-0.53}^{+0.89}$&98.1&88.0/153&...&...&...&...&$-2.40_{-0.16}^{+0.12}$&97.8&87.8/153&$\mathbf{-1.11_{-0.42}^{+0.77}}$&$\mathbf{23.14_{-5.47}^{+4.65}}$&\textbf{95.8}&\textbf{80.6/152}&$-2.37_{-0.88}^{+0.22}$&$6.47_{-1.30}^{+2.76}$&102.3&82.1/151&$10.54_{-2.20}^{+2.37}$\\%
26&$\mathbf{4.33_{-0.63}^{+0.89}}$&\textbf{41.9}&\textbf{31.8/152}&...&...&...&...&$-3.14_{-0.52}^{+0.26}$&46.1&36.0/152&$0.56_{-1.27}^{+2.21}$&$15.55_{-3.60}^{+3.74}$&46.7&31.6/151&...&...&...&...&$1.67_{-0.44}^{+0.47}$\\%
27&$5.42_{-0.18}^{+0.22}$&149.8&139.7/152&$\mathbf{4.73_{-0.17}^{+0.25}}$&$\mathbf{15.51_{-2.33}^{+2.59}}$&\textbf{124.5}&\textbf{104.4/150}&...&...&...&$-0.16_{-0.32}^{+0.37}$&$20.78_{-1.10}^{+0.89}$&134.8&119.6/151&$-2.23_{-0.17}^{+0.24}$&$5.28_{-0.25}^{+0.27}$&130.1&109.9/150&$41.24_{-2.76}^{+3.29}$\\%
28&$\mathbf{4.26_{-0.30}^{+0.39}}$&\textbf{97.3}&\textbf{87.2/152}&...&...&...&...&$-2.80_{-0.19}^{+0.14}$&120.8&110.7/152&...&...&...&...&...&...&...&...&$6.28_{-0.77}^{+0.83}$\\%
29&$\mathbf{6.02_{-0.91}^{+0.93}}$&\textbf{99.5}&\textbf{89.4/152}&...&...&...&...&$-2.76_{-0.35}^{+0.21}$&102.9&92.8/152&$-0.70_{-0.44}^{+1.90}$&$19.00_{-3.56}^{+6.74}$&102.2&87.1/151&...&...&...&...&$3.56_{-0.77}^{+0.92}$\\%
30&$\mathbf{3.19_{-0.45}^{+0.83}}$&\textbf{71.6}&\textbf{61.5/152}&...&...&...&...&$\mathbf{-3.29_{-0.81}^{+0.34}}$&\textbf{71.8}&\textbf{61.7/152}&$-0.52_{-0.56}^{+3.21}$&$9.34_{-1.10}^{+5.95}$&76.3&61.2/151&...&...&...&...&$1.59_{-0.44}^{+0.49}$\\%
31&$4.60_{-0.24}^{+0.31}$&155.6&145.5/152&$\mathbf{4.28_{-0.25}^{+0.29}}$&$\mathbf{19.17_{-4.30}^{+4.85}}$&\textbf{149.1}&\textbf{129.0/150}&$-2.60_{-0.10}^{+0.08}$&153.4&143.3/152&$\mathbf{-1.41_{-0.20}^{+0.78}}$&$\mathbf{13.36_{-2.62}^{+4.20}}$&\textbf{147.6}&\textbf{132.5/151}&$\mathbf{-2.56_{-0.30}^{+0.26}}$&$\mathbf{4.97_{-0.55}^{+0.98}}$&\textbf{148.2}&\textbf{128.1/150}&$21.62_{-3.14}^{+2.78}$\\%
32&$\mathbf{5.65_{-0.42}^{+0.49}}$&\textbf{91.7}&\textbf{81.7/152}&...&...&...&...&$-2.56_{-0.14}^{+0.12}$&125.4&115.4/152&$2.31_{-1.18}^{+1.02}$&$21.72_{-1.38}^{+2.05}$&96.6&81.5/151&...&...&...&...&$6.59_{-0.87}^{+0.97}$\\%
33&$\mathbf{5.63_{-0.23}^{+0.22}}$&\textbf{101.1}&\textbf{91.0/152}&$4.82_{-0.41}^{+0.84}$&$9.46_{-2.86}^{+8.42}$&102.6&82.5/150&...&...&...&$\mathbf{0.46_{-0.34}^{+0.52}}$&$\mathbf{21.64_{-0.99}^{+0.92}}$&\textbf{99.8}&\textbf{84.7/151}&...&...&...&...&$26.51_{-2.39}^{+1.94}$\\%
34&$\mathbf{4.10_{-0.63}^{+1.21}}$&\textbf{55.2}&\textbf{45.2/152}&...&...&...&...&$-2.83_{-0.76}^{+0.30}$&58.7&48.7/152&...&...&...&...&...&...&...&...&$1.34_{-0.49}^{+0.58}$\\%
\hline
\hline
\end{tabular}%
\end{center}
\leftline{\bf{Notes.}}
\end{sidewaystable}

\clearpage

\clearpage

\clearpage

\clearpage

\end{document}